\documentclass[twocolumn,showpacs,preprintnumbers,amsmath,amssymb,prd,letterpaper,floatfix,nofootinbib,superscriptaddress]{revtex4}  

\usepackage[latin1]{inputenc}
\usepackage{graphics}
\usepackage{graphicx}
\usepackage{epsfig}
\usepackage{color}
\usepackage{longtable}
\usepackage{hyperref}
\usepackage{latexsym}
\usepackage{amsmath}
\usepackage{amsthm}
\usepackage{amsfonts}
\usepackage{amssymb}
\usepackage{dsfont}
\usepackage{bm}
\usepackage{graphics,psfrag}
\usepackage{graphicx,psfrag}
\usepackage[nice]{nicefrac}

\graphicspath{}
\newcommand{\be}{\begin{equation}}
\newcommand{\ee}{\end{equation}}
\newcommand{\bea}{\begin{eqnarray}}
\newcommand{\eea}{\end{eqnarray}}
\newcommand{\eq}[1]{(\ref{#1})}

\newcommand{\ot}{\ensuremath{\frac{1}{2}}}

\newcommand{\E}{\mathrm{e}}

\renewcommand{\imath}{\mathrm{i}}

\newcommand{\myparagraph}[1]{~\\ \noindent\ensuremath{\mathbf{#1}}:~}
\newlength{\mylenC}
\begin{document}
%
\title{QCD with Two Light Dynamical Chirally Improved Quarks: Baryons}

\author{Georg P.~Engel}
\email{georg.engel@mib.infn.it}
\affiliation{Institut f\"ur Physik, FB Theoretische Physik, Universit\"at
Graz, A--8010 Graz, Austria}
\affiliation{Dipartimento di Fisica, Universit\`a di Milano-Bicocca and INFN, sezione di Milano-Bicocca, I--20126 Milano, Italy}

\author{C. B. Lang}
\email{christian.lang@uni-graz.at}
\affiliation{Institut f\"ur Physik, FB Theoretische Physik, Universit\"at Graz, A--8010 Graz, Austria}

\author{Daniel Mohler}
\email{dmohler@fnal.gov}
\affiliation{Fermi National Accelerator Laboratory, Batavia, 60510-5011Illinois, USA}
\affiliation{TRIUMF, 4004 Wesbrook Mall Vancouver, BC V6T 2A3, Canada}

\author{Andreas Sch\"afer}
\email{andreas.schaefer@physik.uni-regensburg.de}
\affiliation{Institut f\"ur Theoretische Physik, Universit\"at
Regensburg, D--93040 Regensburg, Germany}

\collaboration{BGR [Bern-Graz-Regensburg] Collaboration}

\date{\today}

\begin{abstract}
We present a study of baryon ground states and low lying excitations of non-strange
and strange baryons. The results
are based on seven gauge field ensembles with two dynamical light Chirally Improved (CI)
quarks corresponding to pion masses between 255 and 596 MeV and a strange valence quark with mass
fixed by the $\Omega$ baryon. The lattice spacing varies between 0.1324 and 0.1398 fm.  
Given in lattice units, the bulk of our results are for size $16^3\times 32$, for 
two ensembles with light pion masses (255 and 330 MeV) we also use $24^3\times 48$ 
lattices and perform an infinite volume extrapolation. 
We derive energy levels for the spin 1/2 and 3/2 channels for both parities.
In general, our results in the infinite volume limit compare well with experiment.
We analyze the flavor symmetry content by 
identifying the singlet/octet/decuplet contributions of the resulting eigenstates. The 
ground states compositions agree with quark model expectations. In some cases
the excited states, however, disagree and we discuss possible reasons. 
\end{abstract}


\maketitle

\newpage

\section{Introduction}

Restricting to strong interactions, almost all of the hadrons are resonances. 
For lattice studies, due to the finiteness of the lattice volumes the smallest momenta come in units $2\pi/L$. 
Moreover, for unphysically heavy pion masses decay channels are often not
open or the resulting phase space is small, leading to energy levels in the
vicinity of the resonance energy. This motivates the identification of the low
energy levels with masses of corresponding resonances. 
Eventually, towards physical pion masses and larger lattices, the interpretation becomes invalid and the observed energy levels show a
more intricate pattern, related in the elastic channel to two-hadron states \cite{Luscher:1990ux,Luscher:1991cf}.
Recent work, where correlators of only single hadron operators were studied 
\cite{Bulava:2010yg,Engel:2010my,Engel:2012qp}, found no clear signal of possibly coupling two-hadron states
(with the possible exception of $s$ wave channels). It was concluded that for a 
full study one should include such interpolators explicitly. In \cite{Lang:2011mn,Lang:2012sv} 
it was demonstrated in meson correlation studies, that neglect of two-meson interpolators
may obscure the obtained energy level picture in some cases. Attempts towards including
meson-baryon interpolators are discussed in \cite{Gockeler:2012yj,Hall:2012wz}
and a recent study including $\pi N$ interpolators in the negative
parity nucleon sector demonstrated significant effects in the observed energy spectrum \cite{Lang:2012db}.

The present work is a continuation of a study of single baryon correlators, with more ensembles and
larger statistics as compared to \cite{Engel:2010my}. Like before we
see no obvious signal of coupling meson-baryon channels (with a few possible exceptions where the meson-baryon
system is in $s$ wave, as will be discussed). We therefore identify the lowest energy levels
with baryon ground states and excitations. 

We use two mass identical light quarks with the Chirally Improved (CI) fermion action
\cite{Gattringer:2000js,Gattringer:2000qu,Engel:2010my}. The strange quark is considered as valence quark, its mass
fixed by setting the $\Omega$-mass to its physical value. The pion masses for the seven ensembles of 200--300 gauge
configurations each range from 255 MeV to 596 MeV, with lattice size $16^3\times 32$ and lattice 
spacing between 0.1324 and 0.1398 fm. For two ensembles with light pion masses
also lattices of size $12^3\times 24$ and $24^3\times 48$ were used to allow extrapolation to infinite volume. 

Other recent studies aiming at light and strange baryon excitations, some of them with 2+1 dynamical quarks, include
\cite{WalkerLoud:2008bp,WalkerLoud:2008pj,Bulava:2009jb,Bulava:2010yg,Bulava:2011xj,%
Edwards:2011jj,Mahbub:2010rm,Mahbub:2010vu,Menadue:2011pd,Mahbub:2012ri,Alexandrou:2012xk,%
Arthur:2012yc}. In \cite{Edwards:2012fx} excited spectra for non-strange and strange
baryon are derived from anisotropic lattices and standard improved Wilson fermions. 
See also recent reviews \cite{Bulava:2011np,Lin:2011ti,Fodor:2012gf} and references therein. 

In Sect. \ref{setup} we discuss the setup for our simulations and remark
on the methods used for the data analysis. Results from the $16^3\times 32$
lattices for light and strange baryons are presented in Sections
\ref{sec:baryons:light} and \ref{sec:baryons:strange} respectively. In Sect.
\ref{sec:results_vol} the infinite volume extrapolation and uncertainties with
regard to the strange quark mass chosen in our simulations are discussed. We
conclude with a summary in Sect. \ref{summary}.

\section{\label{setup}Setup of the Simulation and Analysis}

The CI fermion action \cite{Gattringer:2000js,Gattringer:2000qu}
results from a parametrization of a general fermion action
connecting each site along  gauge link paths  to other sites up to distance
three (in lattice units). 
This truncated ansatz is used to algebraically solve the Ginsparg-Wilson
equation. The action consists of several hundred terms and obeys the
Ginsparg-Wilson relation approximately. It was used in quenched
\cite{Gattringer:2003qx,Burch:2006cc} and dynamical simulations
\cite{Lang:2005jz}.   It was found that the
small eigenvalues fluctuate predominantly towards the inside of the
Ginsparg-Wilson unit circle \cite{Gattringer:2008vj}.
Exceptionally small eigenvalues are suppressed, which allows to simulate smaller
pion masses on coarse lattices.
For further improvement of the fermion action one level of
stout smearing of the gauge fields  \cite{Morningstar:2003gk} was included in its
definition. The parameters are adjusted such that the value of the plaquette is
maximized ($\rho=0.165$ following \cite{Morningstar:2003gk}). For the pure gauge
field part of the action we use the tadpole-improved L\"uscher-Weisz  gauge
action \cite{Luscher:1984xn}. For a given gauge coupling we use the same assumed
plaquette value for the different values of the bare quark mass parameter.

The lattice spacing $a$ is defined as discussed in \cite{Engel:2011aa}, using the
static potential with a Sommer parameter $r_0=0.48$ fm and setting the scale at 
the physical pion mass for each value of $\beta_{LW}$. This value of the
Sommer parameter may be slightly too small for $n_f=2$, as has been argued recently
\cite{Fritzsch:2012wq,Bali:2012qs}, where a value near 0.5 fm is preferred.

All parameters as well as details of the implementation in the Hybrid Monte
Carlo simulation \cite{Duane:1987de,Lang:2005jz} and various quality check are given in
\cite{Engel:2010my,Engel:2011aa}. For reference we summarize the parameters of 
the used gauge field ensembles in Table \ref{tab:ensembles}.

\begin{table*}[t]
\begin{ruledtabular}
\begin{tabular}{c c c c c c c c c c}
set&	$\beta_{LW}$	&$m_0$	&$m_s$& configs. &$m_\pi$ [MeV]	&$L^3\times T \,[a^4]$	&$m_{\pi}L$ 	& $a$ [fm]	\\
\hline
A50&	4.70& 		 -0.050	&-0.020	&200 &596(5) 	& $16^3\times 32$	&6.40			&0.1324(11)	\\
A66&	4.70& 		 -0.066	&-0.012	&200 &255(7) 	& $16^3\times 32$	&2.72			&0.1324(11)	\\
B60&	4.65& 		 -0.060	&-0.015	&300 &516(6) 	& $16^3\times 32$	&5.72			&0.1366(15)	\\
B70&	4.65& 		 -0.070	&-0.011	&200 &305(6) 	& $16^3\times 32$	&3.38			&0.1366(15)	\\
C64&	4.58& 		 -0.064	&-0.020	&200 &588(6) 	& $16^3\times 32$	&6.67			&0.1398(14)	\\
C72&	4.58& 		 -0.072	&-0.019	&200 &451(5) 	& $16^3\times 32$	&5.11			&0.1398(14)	\\
C77&	4.58& 		 -0.077	&-0.022	&300 &330(5) 	& $16^3\times 32$	&3.74			&0.1398(14)	\\
\hline
LA66&	4.70& 		 -0.066	&-0.012	&~97 &	& $24^3\times 48$	&4.08			&0.1324(11)	\\
SC77&	4.58& 		 -0.077	&-0.022	&600  &	& $12^3\times 24$	&2.81			&0.1398(14)	\\
LC77&	4.58& 		 -0.077	&-0.022	&153  & 	& $24^3\times 48$	&5.61			&0.1398(14)	\\
\end{tabular}
\end{ruledtabular}
\caption[Parameters of the simulation]{\label{tab:ensembles}
\noindent 
Parameters of the simulation:  Ensemble names are given in the first row. We show
the gauge couplings $\beta_{LW}$, the light quark mass parameter $m_0$,  the
strange quark mass parameter $m_s$, the number of configurations analyzed
(``configs.''), the pion mass and the volume $L^3\times T$ in lattice units. The
dimensionless product of the pion mass with the spatial extent of the lattice,
$m_\pi L$,  enters finite volume corrections.  We also give the lattice spacing
$a$ as discussed in \cite{Engel:2011aa}. The three ensembles LA66, SC77 and LC77
are separated from the others by a horizontal line,  since they are used only for
a discussion of finite volume effects.  For these ensembles we use the pion
masses of A66 and C77, respectively.
}
\end{table*}

In each baryon channel with given quantum numbers the eigenenergy levels are
determined with the so-called variational method
\cite{Luscher:1990ck,Michael:1985ne}. One uses interpolators with the correct
symmetry properties and computes the cross-correlation matrix $C_{ik}(t)=\langle 
O_i(t) O_k(0)^\dagger\rangle$. One then  solves the generalized eigenvalue
problem
\begin{equation}
C(t) \vec u_n(t)=\lambda_n(t) C(t_0) \vec u_n(t)
\end{equation}
in order to approximately recover the energy eigenstates $|n\rangle$.
The exponential decay of the eigenvalues 
\begin{equation}\label{eq:eigenvalues}
\lambda_n(t)=\E^{-E_n\,(t-t_0)} (1+\mathcal{O}(\E^{-\Delta E_n(t-t_0)} ))
\end{equation}
allows us to obtain the energy values, where $\Delta E_n$ is the distance to
other spectral values. In \cite{Blossier:2009kd} it was shown that for $t_0\leq t
\leq 2 t_0$ the value of $\Delta E_n$ is the distance to the first neglected
eigenenergy. In an actual computation the statistical fluctuations limit the
values of $t_0$ and one estimates the fit range by identifying  plateaus of the
effective energy. The eigenvectors serve as  fingerprints of the states,
indicating their content in terms of the lattice interpolators. 

\begin{figure}[t]
\noindent\includegraphics[width=\columnwidth,clip]{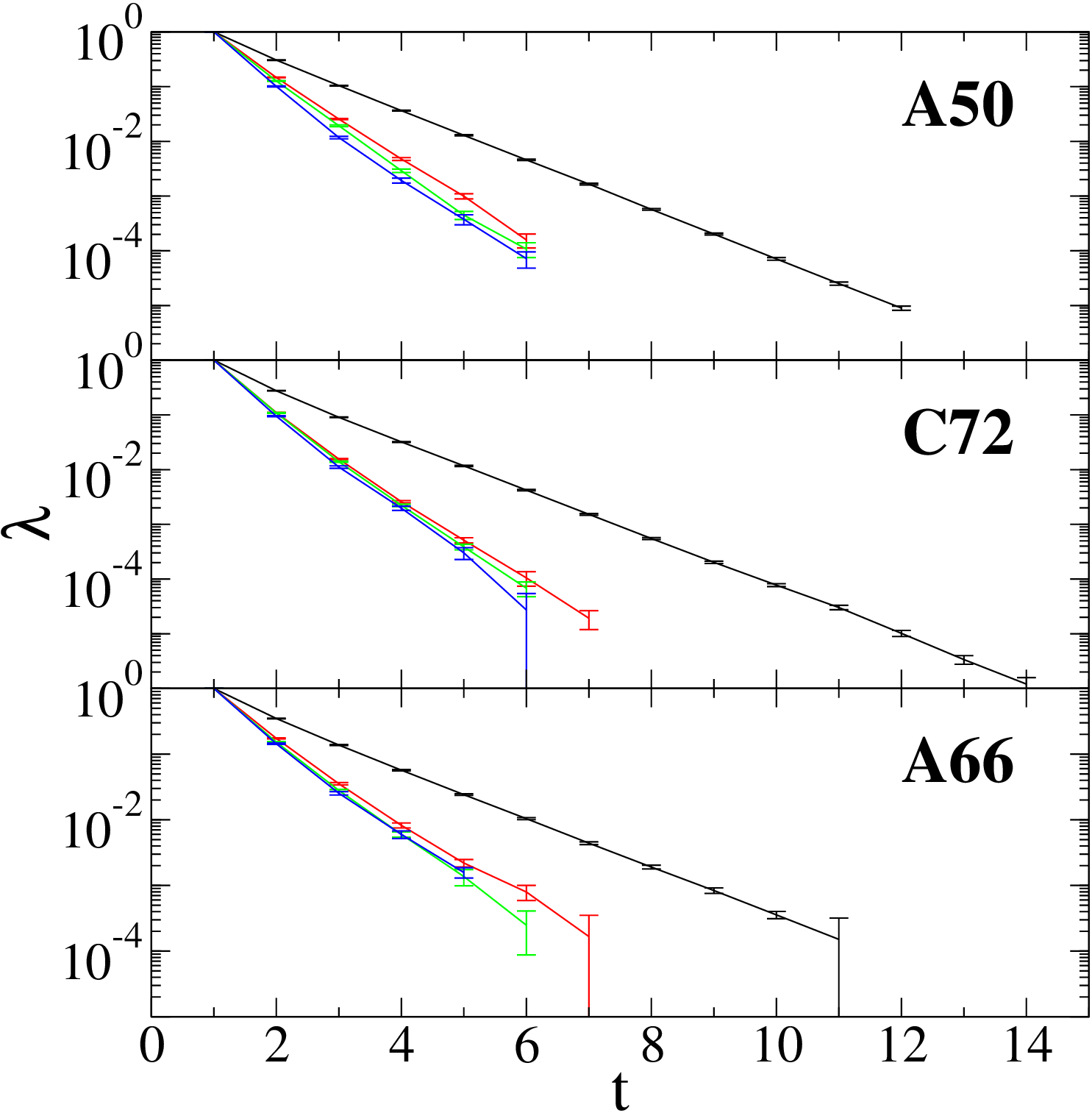}
\caption{
Eigenvalues for the four lowest states in the $\Sigma$ $1/2^+$ channel for ensembles A50, C72 and A66 (top to bottom) which covers the whole range of pion masses considered.
}
\label{fig:sigma_1half_eigvals}
\end{figure}

The quality of
the results depends on the statistics and the set of lattice operators.
The dependence on $t_0$ is studied. Larger values of $t_0$ increase
the noise and reduce the possible fit range, although the results are 
consistent. In the final analysis we use $t_0=1$ (with the origin at 0).
The statistical error is
determined with single-elimination jack-knife. For the fits to the
eigenvalues \eq{eq:eigenvalues} we use single
exponential behavior but check the stability with double exponential fits; we
take the correlation matrix for the correlated fits from the complete sample
\cite{Engel:2011aa}. 
As an example we show eigenvalues for the four lowest states in the $\Sigma$ $1/2^+$ channel for three ensembles in Fig.~\ref{fig:sigma_1half_eigvals}.
In general, we find very good agreement between the eigenstates of all considered ensembles. 
This suggests the interpretation of a signal with physical origin and in some cases serves to justify a fit relying on only few points. 

The set of interpolators used should be capable to approximate the eigenstates.
On the other hand, too large a set may add statistical noise. In practice one
tries to reduce the number of interpolators to a sufficient subset. We analyze
the dependence of the energy levels on the choice of interpolators and fit ranges
for the eigenvalues. For the final result, we make a reasonable choice of
interpolators and fit range and discuss the associated systematic
error.  For the extrapolation towards the physical pion mass we fit to the
leading order chiral behavior, i.e.,  linear in $m_\pi^2$.

The Dirac and flavor structure is motivated by the quark model 
\cite{Isgur:1978xj,Isgur:1978wd}, see also \cite{Glozman:1995fu}. Within the
relativistic quark model there have been many determinations of the
hadron spectrum, based on confining potentials and different assumptions
on the hyperfine interaction (see, e.g., \cite{Capstick:1986bm,Glozman:1997ag,Loring:2001kx}).
The singlet, octet and decuplet attribution \cite{Glozman:1995fu} of the states
has been evaluated based on such model calculations, e.g., in \cite {Melde:2008yr}
(see also the summary in \cite{Beringer:1900zz}).
We use sets of up to 24 interpolating fields in each quantum channel, combining
quark sources of different smearing widths, different Dirac structure and  octet
and decuplet flavor structure. In Appendix \ref{sec:app_interpol} (Tables \ref{tab:baryon:interpol:1} to \ref{tab:baryon:interpol:3}) we summarize
the structure and numbering of the baryon interpolators used in this study.

\section{Results for Light Baryons}\label{sec:baryons:light}

\subsection{Nucleon}
\begin{figure}[h!]
\noindent\includegraphics[width=\columnwidth,clip]{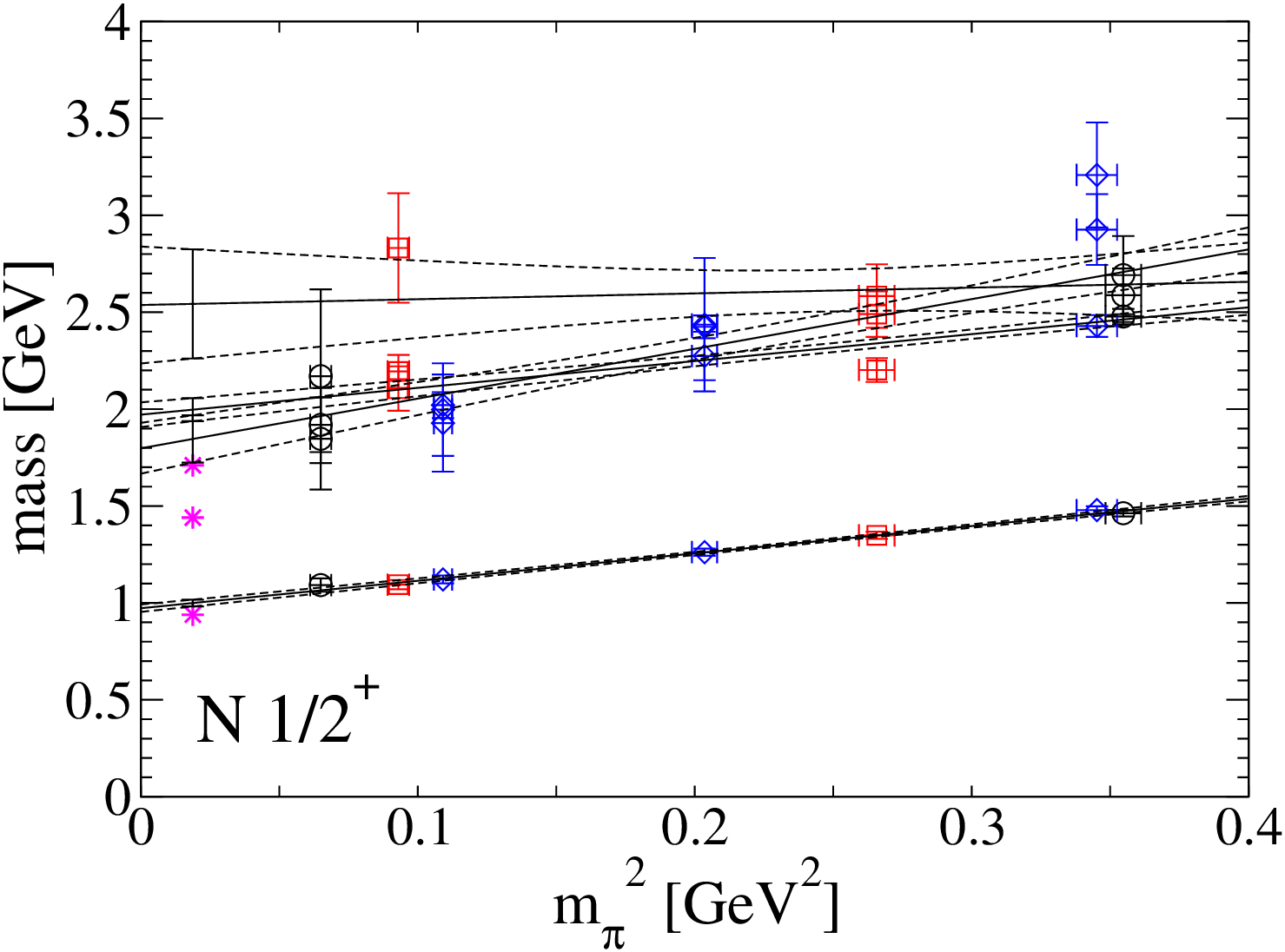}\\
\noindent\includegraphics[width=\columnwidth,clip]{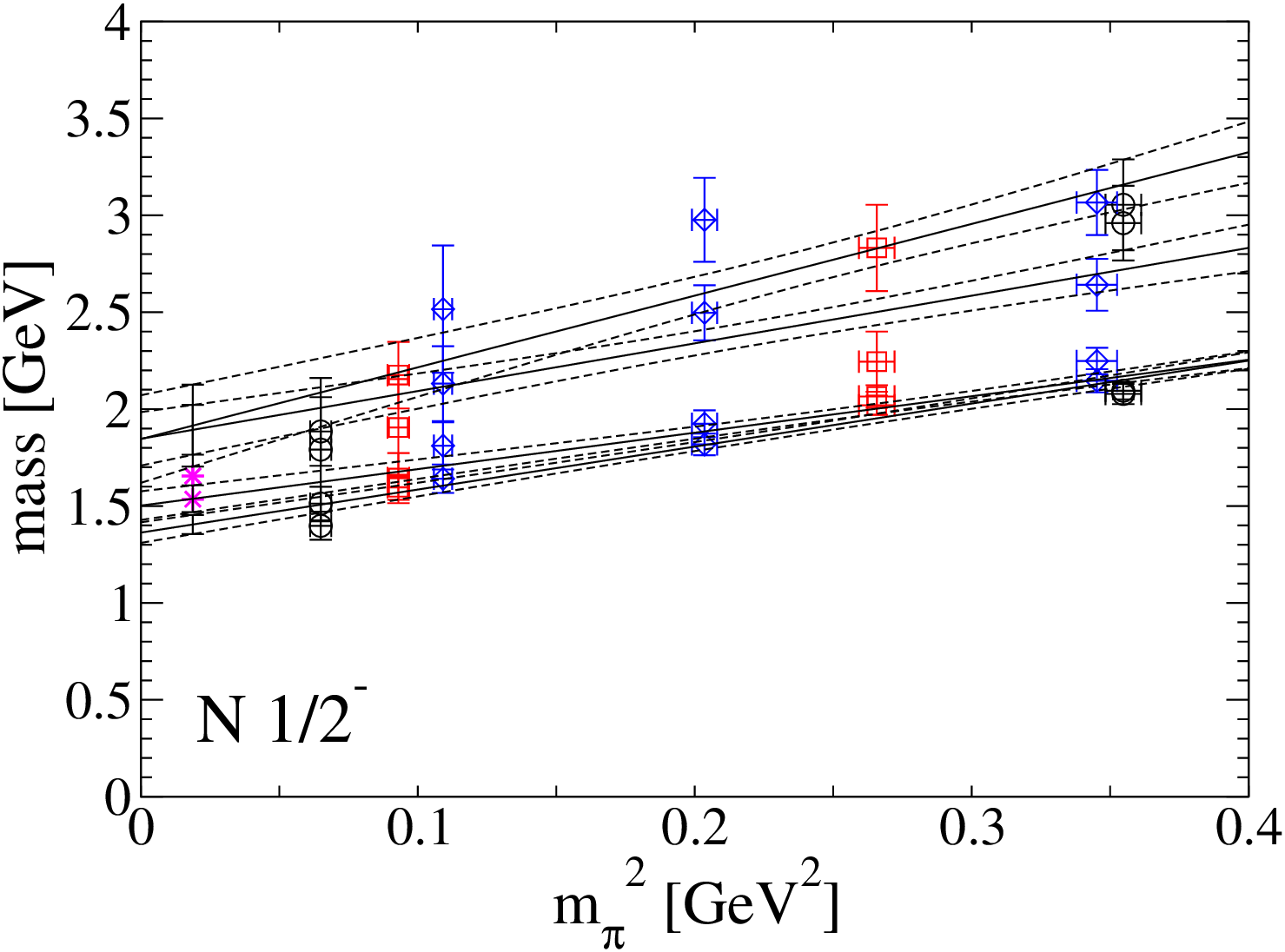}
\caption[Energy levels for nucleon spin 1/2]{
Energy levels for nucleon spin 1/2, positive (upper) and negative parity (lower).
Black, red and blue (color online) denote a value of $\beta$
equal to 4.70, 4.65 and 4.58, respectively.
The solid lines give the mean values of the fits in $m_\pi^2$, 
the dashed ones indicate the region of one $\sigma$.}
\label{fig:nucleon_1half}
\end{figure}

\begin{figure}
\noindent\includegraphics[width=\columnwidth,clip]{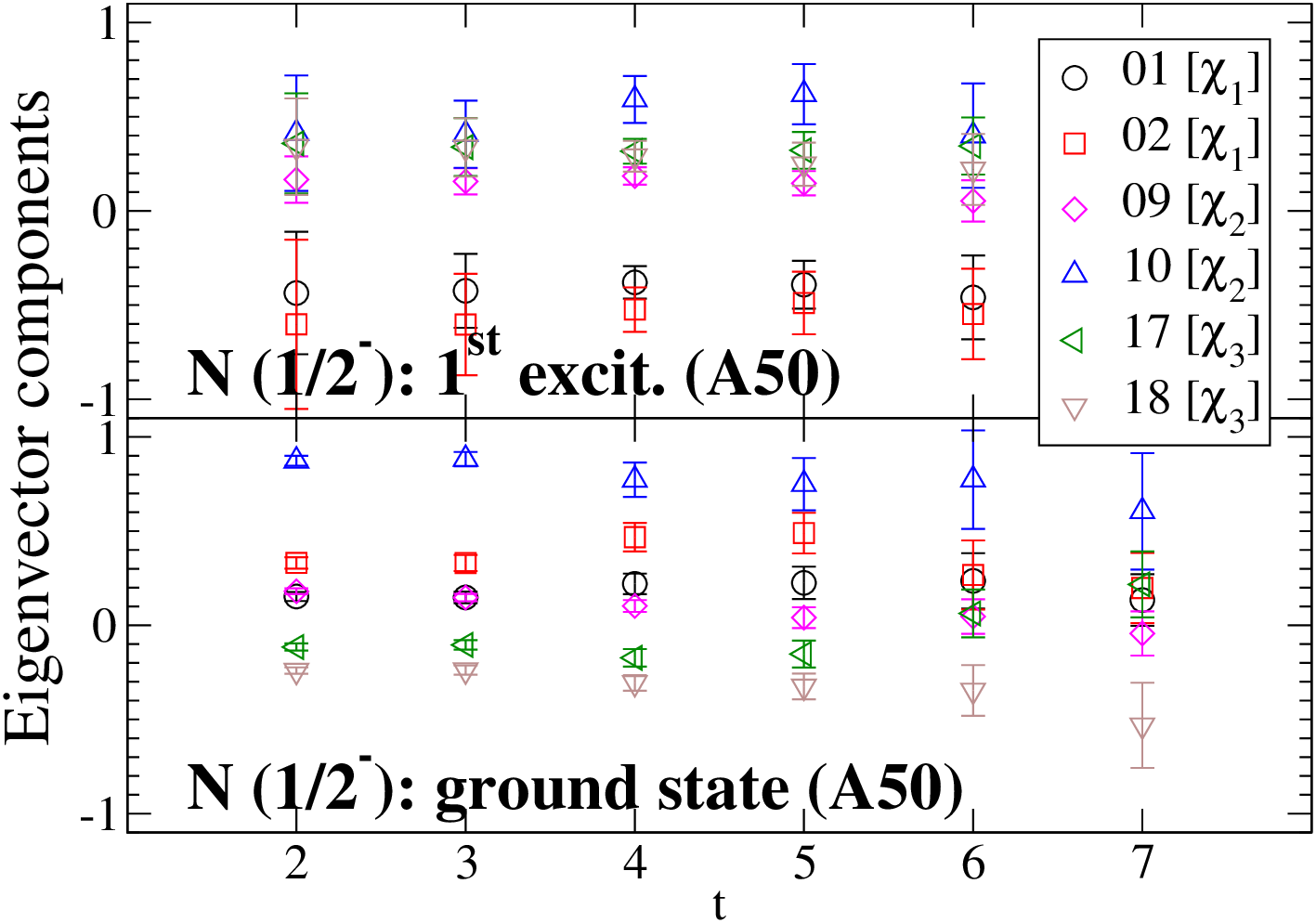}\\
\noindent\includegraphics[width=\columnwidth,clip]{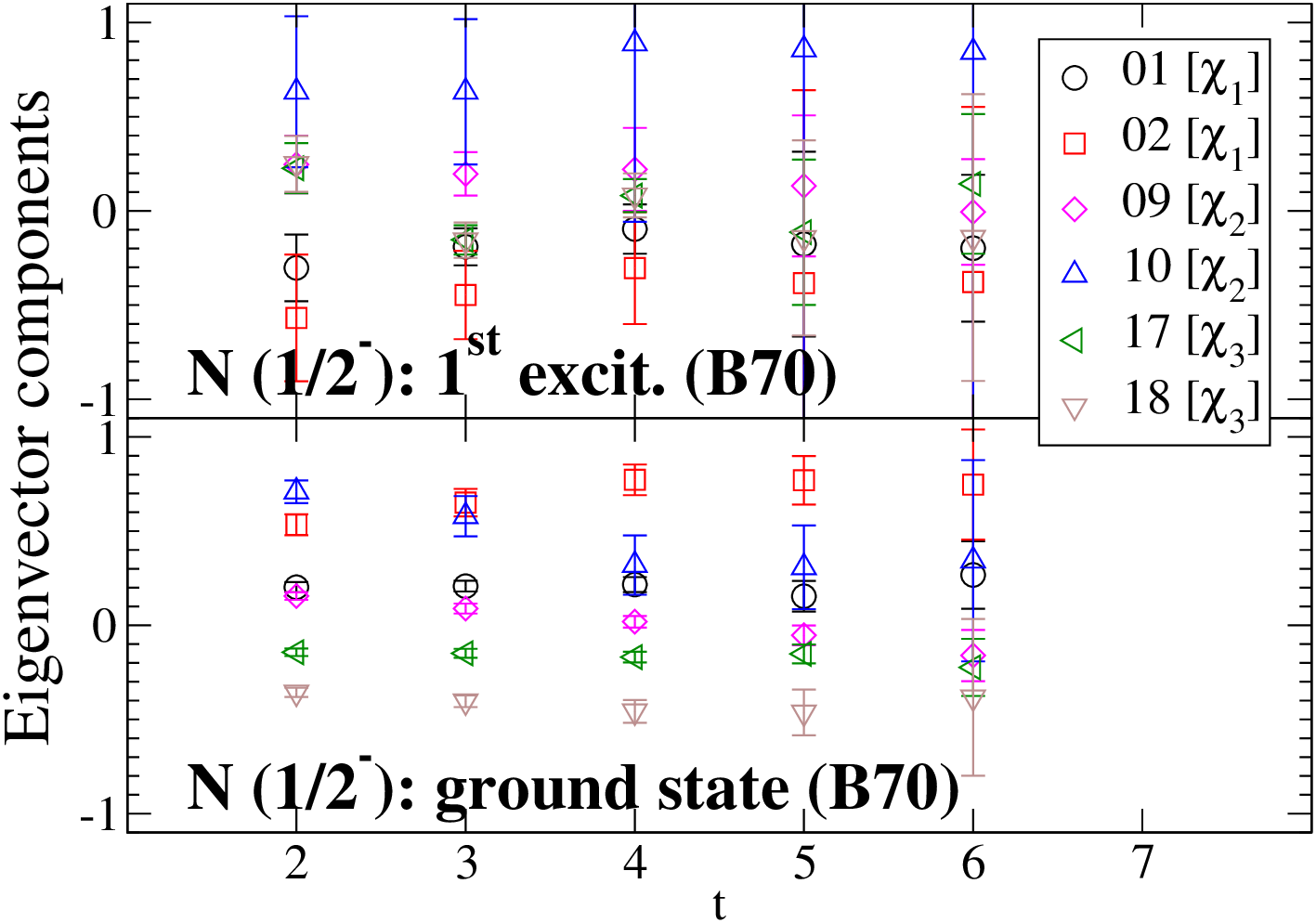}
\caption[Eigenvectors for nucleon spin 1/2 negative parity]{
Eigenvectors for nucleon spin 1/2 negative parity ground state and first excitation, 
ensemble A50 (upper) and B70 (lower).
Note the different composition of the states at the different pion masses.}
\label{fig:nucleon_1halfneg_vectors}
\end{figure}

\myparagraph{N:\;\;I(J^P)=\ot(\ot^+)}
The nucleon (spin 1/2 and positive parity) ground state is the lightest baryon. 
We use a $n_{\text{op}}\times n_{\text{op}}$ correlation matrix
with $n_{\text{op}}=6$
interpolators covering three Dirac structures and different levels of quark
smearing, (1,2,9,10,19,20) (see Appendix A), and extract the four lowest
eigenstates. 
For the ground state the leading order chiral extrapolation yields a mass value
roughly 7\% larger than the experimental $N$ (see Fig.~\ref{fig:nucleon_1half}). 
Part of the deviation is caused by finite volume effects, which will be
discussed in Section \ref{sec:results_vol}. The remaining small deviation might
be caused by systematic errors from scale setting (using $r_0=0.48$ fm), or a
curvature due to higher order terms in the chiral extrapolation 
(for a discussion on the latter, see, e.g., \cite{Bali:2012qs}).
Within the basis used in the variational method, the ground state is dominated by the first Dirac structure, with a
contribution of the third one (cf., Table \ref{tab:baryon:interpol:2}).  
We stress that all Dirac structures used here generate independent field operators which are not related by Fierz transformations.

The first excitation in the nucleon channel should be the ``Roper resonance $N(1440)$'',
notorious because it lies below the ground state in the corresponding negative parity
channel. This ``reverse level ordering'' differs from the expectations of most simple quark
models (see, e.g., \cite{Isgur:1977ef,Isgur:1978wd}). 
However, in our simulation, the first
excitation is $\mathcal{O}$(500 MeV) higher than the experimental value.
The levels are ordered conventionally
with alternating parity. 
This is also the case in lattice simulation with
quenched and dynamical results of other groups (e.g.,\cite{Cohen:2009zk,Mahbub:2009cf,Bulava:2010yg}). Towards physical pion masses, the first
excitation was reported to bend down significantly \cite{Mahbub:2010rm}, however, still all
lattice results are closer to the $N(1710)$ than to the Roper resonance $N(1440)$, with large 
error bars.

At present it is unclear to us, what the reason for this behavior may be, although there
are several suspects.
Finite volume effects could shift the energy level up. For the ground state this shift is comparatively small (as discussed in Section \ref{sec:results_vol}).
This could be significantly larger for the excited state, which is generally expected to have larger physical size.
(E.g.,  in quark models it is considered as a radial excitation.)
Unfortunately, the signal of this state is too weak in our study to allow for a reliable analysis of finite volume effects. 

Another interpretation may be that the used interpolators may not couple strongly enough to the Roper resonance and thus represent the
physical content poorly and we might even miss the physical Roper state altogether.
We observe a similar problem in the corresponding $\Lambda$ sector \cite{Engel:2012qp}. There the  first observed excitation 
is dominated by singlet interpolators (first Dirac structure)
matching the $\Lambda(1810)$ (singlet in the quark model).
The Roper-like  $\Lambda(1600)$ (octet in the quark model) seems to be missing. 

Furthermore, the energy levels of the $p$-wave scattering state $\pi N$ also could influence the situation dramatically. 
Inclusion of such baryon-meson interpolator may be necessary for a better representation of the physical state. The resulting
energy spectrum is related to the scattering phase shift in this channel \cite{Gockeler:2012yj,Hall:2012wz}. 
In small boxes and for broad resonances, the resulting energy levels are shifted
significantly with regard to noninteracting levels and the resonance mass has to
be extracted from the phase shift data. As the experimental Roper state is broad
this shift might be significant.

After chiral extrapolation, we obtain two close excitations within roughly 1800-2000 MeV.
One of those has a $\chi^2/$d.o.f.~of the fit of larger than three (see Table \ref{tab:chi2baryons_pospar}), 
which may suggests a non-linear dependence on $m_\pi^2$.
However, an extrapolation using only data with pion masses below 350 MeV misses the experimental Roper resonance as well.

In several of our ensembles the excited energy levels overlap with each other within error bars.
At light pion masses, the first excitation is dominated by a combination of interpolators of the second Dirac structure;
the second excitation is dominated by the first Dirac structure, with some contribution from the third one.
Towards heavier quark masses, this level ordering interchanges.

Finally, we note that the results in  the nucleon positive parity channel do not deviate significantly from the corresponding
quenched simulations \cite{Burch:2006cc}.

\myparagraph{N:\;\;I(J^P)=\ot(\ot^-)}
In general, we find somewhat low energy levels in the negative parity baryon channels, compared to experiment.
This is also true for the nucleon spin 1/2 negative parity channel.
We use again the set of interpolators (1,2,9,10,19,20), and find that the chiral extrapolation of
the ground state comes out too low and that of the first excitation ends up near the experimental ground state mass value (see Fig.~\ref{fig:nucleon_1half}).

The two lowest states are usually identified with the N(1535) and N(1650). However, in that channel the $N\pi$ state is
in $s$ wave. A naive estimate of its energy (neglecting the interaction energy) at values of the pion mass above 300 MeV puts it close to the observed lowest energy level. Towards small pion masses the $N\pi$ energy level should fall more steeply than the nucleon mass towards the physical point. 
This suggests a (avoided) level crossing of the (negative parity) nucleon and the $N\pi$ state with related energy level shifts, when
moving from larger to smaller pion energies.

Indeed, our results are compatible with such a picture.
In \cite{Engel:2010my} we analyzed only a subset of the configurations available in this work.
There, we argued that the eigenvectors show no indication for a level crossing in the range of pion masses between roughly 300 and 600 MeV.
In the present work, we can monitor the eigenvectors down to pion masses of 250 MeV.
Furthermore, we use a larger basis (at the cost of introducing additional noise.)
We use the same quark smearing structures for different Dirac structures, 
such that the eigenvectors give information about the content of the state without the need of additional normalization of the interpolators. 
We find indeed a significant change in the eigenvectors towards lighter pion masses.
The eigenvectors are shown for ensembles A50 and B70 in Fig.~\ref{fig:nucleon_1halfneg_vectors}.
In particular, the ground state is dominated by interpolator 2 ($\chi_1$) around $m_\pi=300$ MeV, 
and by interpolator 10 ($\chi_2$) above $m_\pi=500$ MeV.
For the first excitation, interpolator 10 contributes stronger at lighter pion masses compared to heavier ones.
This trend is observed also in the other ensembles and at partially quenched data.
However, the picture does not clearly support an (avoided) level crossing scenario, a unique conclusion is missing. 

The observed behavior towards smaller quark masses was also discussed in
\cite{Mahbub:2012ri} for the 2+1 flavor situation. A recent simulation including
(for the first time) also $\pi N$ interpolators \cite{Lang:2012db} demonstrated
significant changes on the spectrum. In that light we may interpret the states
obtained in the present study  (with only 3-quark interpolators), as effective
superpositions of resonance and meson baryon states.

We postpone further discussion of the content of the states to 
Section \ref{sec:results_vol}, where finite volume effects will be discussed. 

\begin{figure}
\noindent\includegraphics[width=\columnwidth,clip]{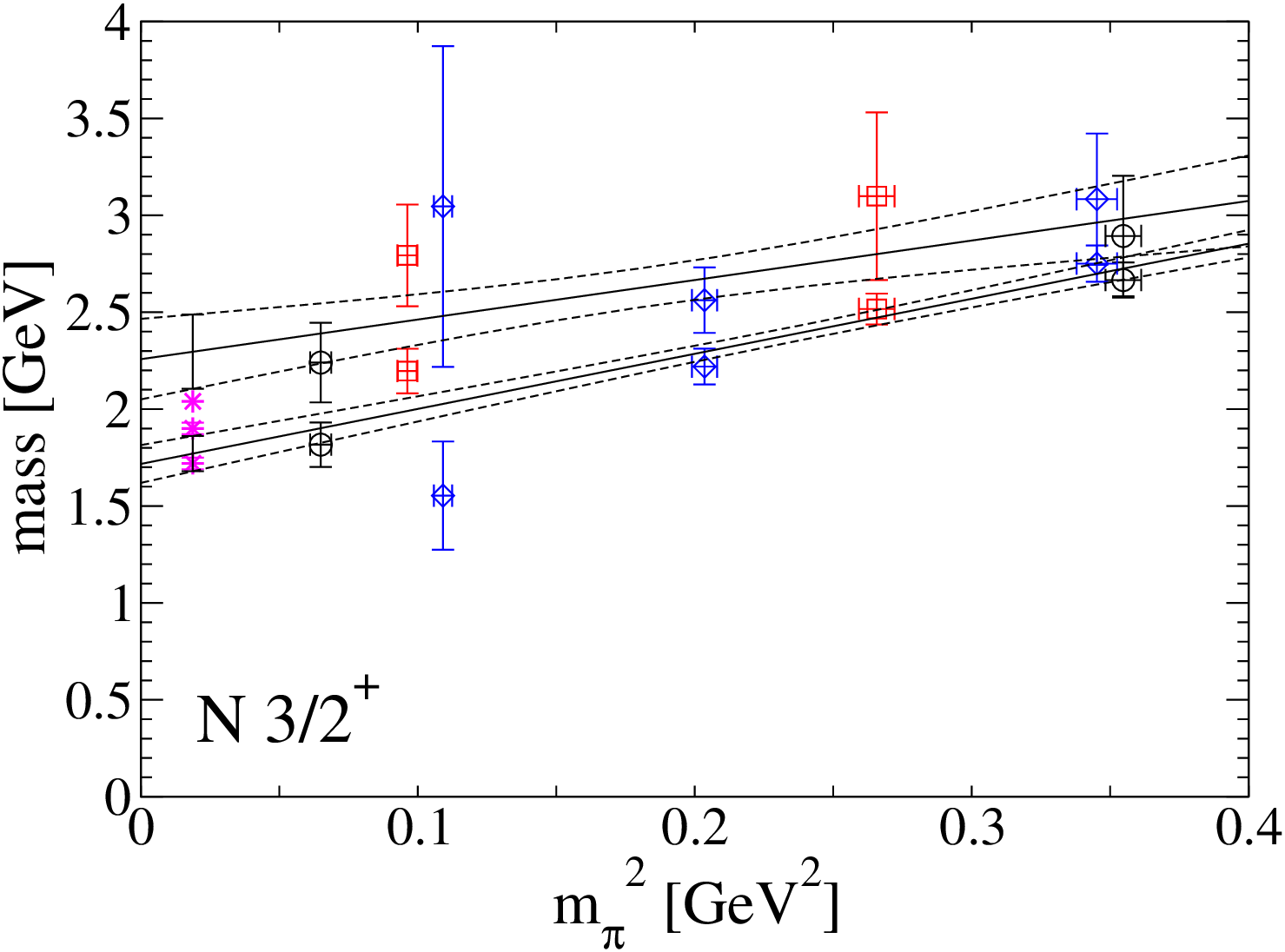}\\
\noindent\includegraphics[width=\columnwidth,clip]{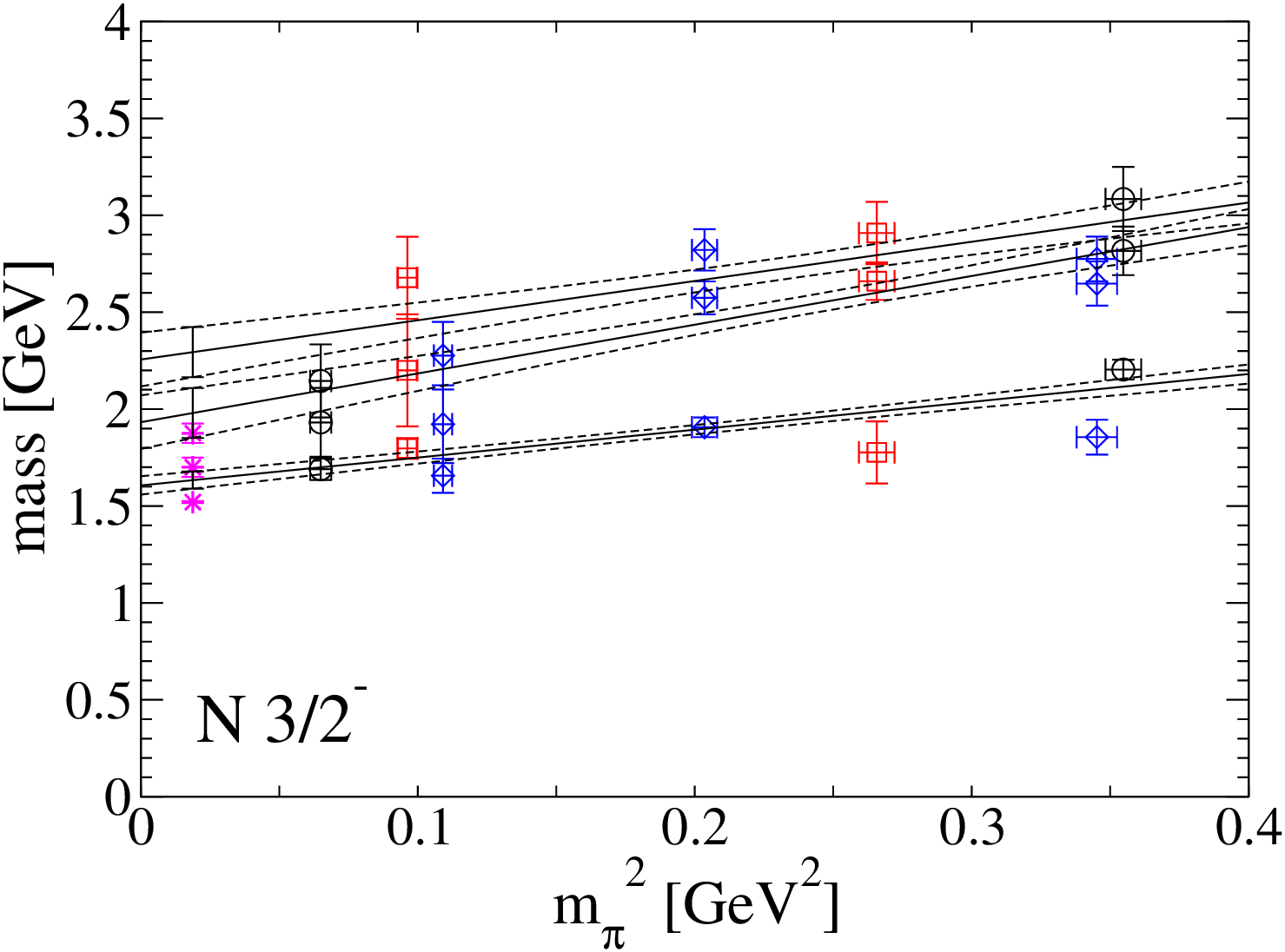}
\caption[Energy levels for nucleon spin 3/2]{
Energy levels for nucleon spin 3/2, positive (upper) and negative parity (lower).}
\label{fig:nucleon_3half}
\end{figure}

\myparagraph{N:\;\;I(J^P)=\ot(\frac{3}{2}^+)}
In the nucleon spin 3/2 positive parity channel, three states are known experimentally:
The $N(1720)$, $N(1900)$ and $N(2040)$, where the latter needs confirmation \cite{Beringer:1900zz}.
We use interpolators (1,4,5), respectively (1,2,3,4) in A66 and B70.
The signal is rather noisy and the effective mass plateaus appear to fall towards large time separations.
Sizable deviations from the chiral fit are observed in ensembles B70 and C77.
Nevertheless, the chiral extrapolation of the ground state agrees well with the experimental $N(1720)$ (see Fig.~\ref{fig:nucleon_3half}).
The first excitation overshoots the $N(1900)$ by about 2$\sigma$, which thus cannot be confirmed from this study.

\myparagraph{N:\;\;I(J^P)=\ot(\frac{3}{2}^-)}
In this channel, experimentally, $N(1520)$, $N(1700)$ and $N(1875)$ are established.
Using interpolators (1,2,3,4), three states can be extracted in our simulation (see Fig.~\ref{fig:nucleon_3half}).
The ground state extrapolates to a value between the $N(1520)$ and the $N(1700)$, the first and second
excitation come out higher than $N(1700)$ or $N(1875)$.

\subsection{Delta}
\begin{figure}
\noindent\includegraphics[width=\columnwidth,clip]{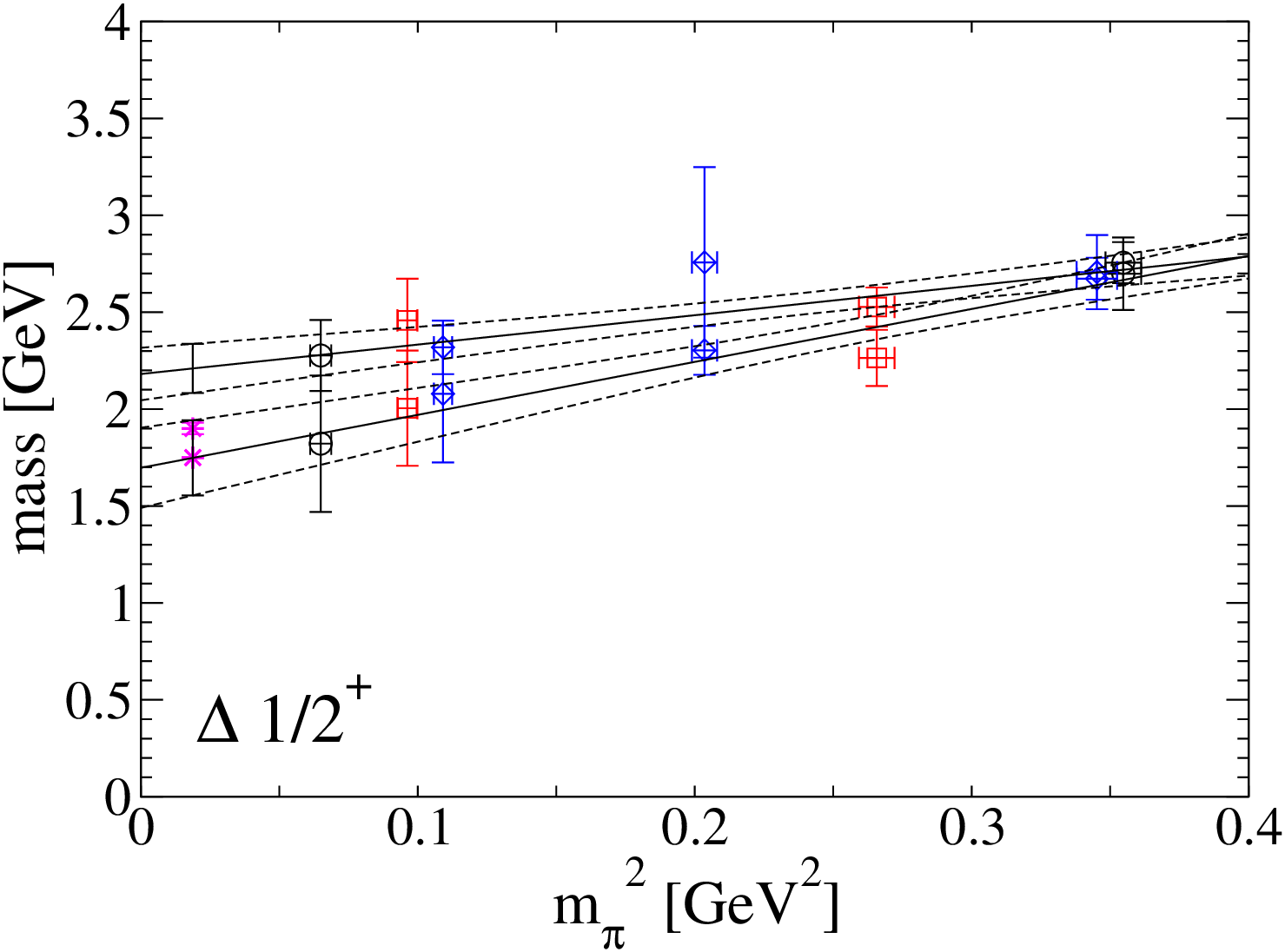}\\
\noindent\includegraphics[width=\columnwidth,clip]{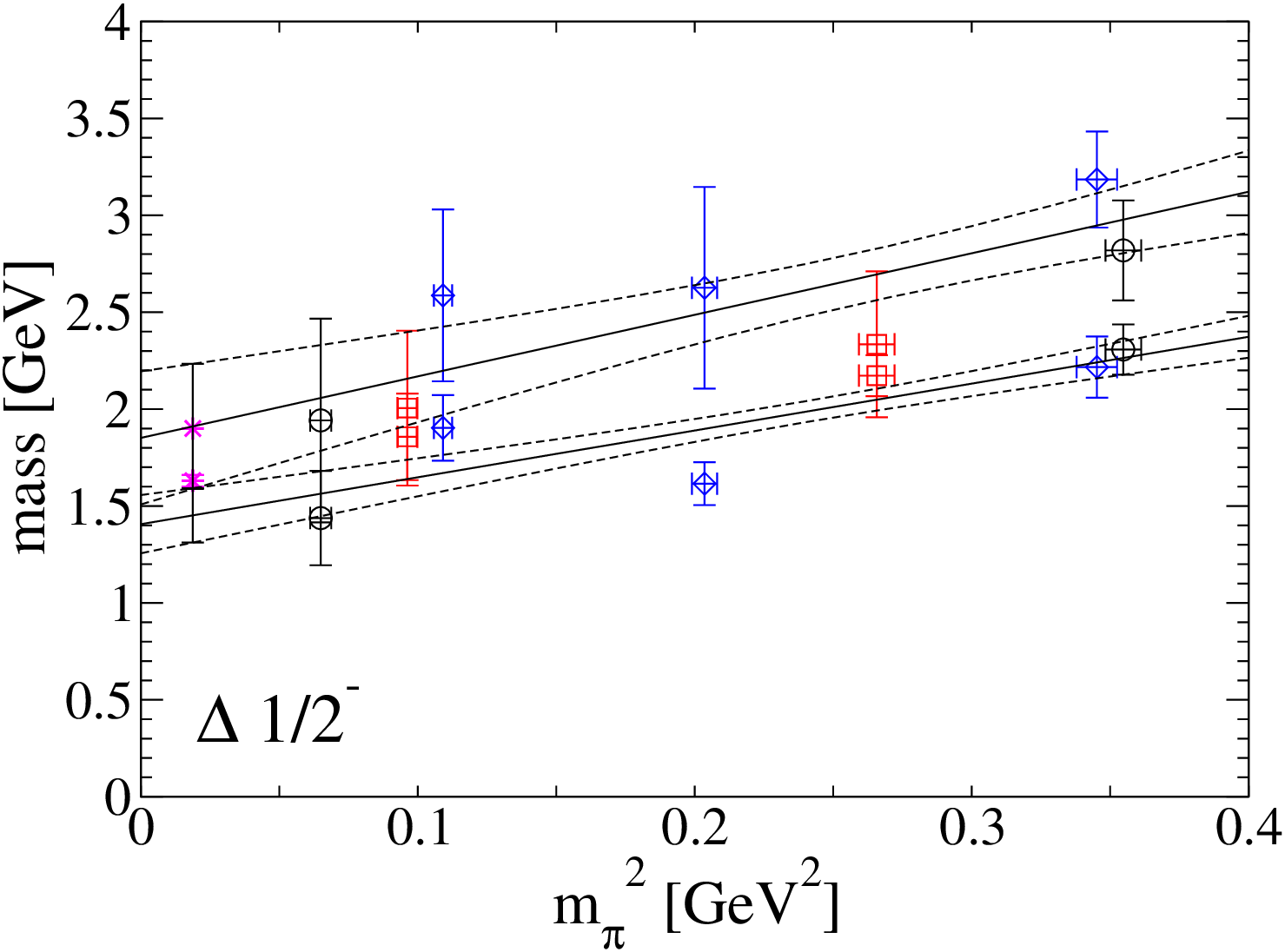}
\caption[Energy levels for $\Delta$ spin 1/2]{
Energy levels for $\Delta$ spin 1/2, positive (upper) and negative parity (lower).}
\label{fig:delta_1half}
\end{figure}
\begin{figure}
\noindent\includegraphics[width=\columnwidth,clip]{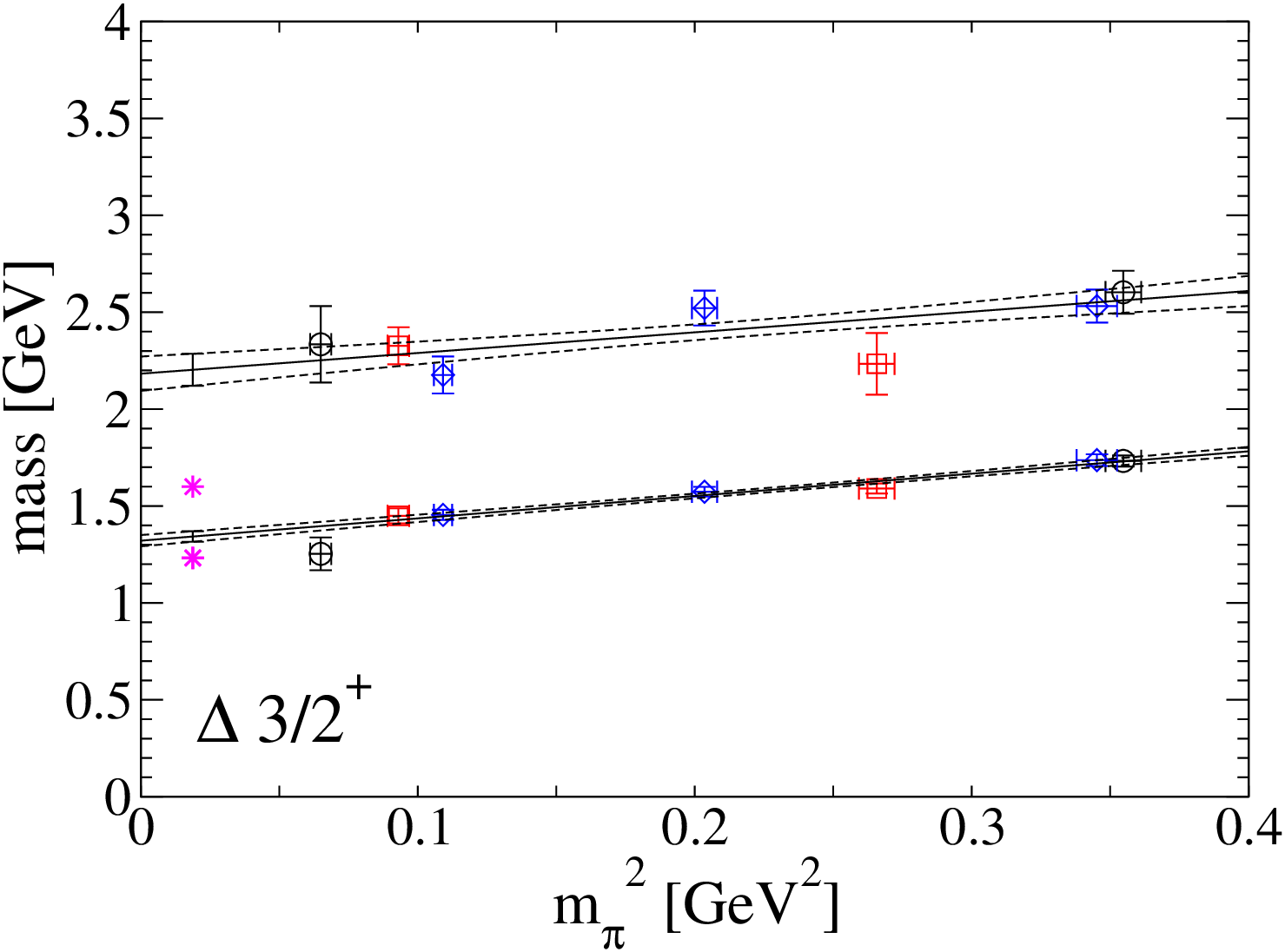}\\
\noindent\includegraphics[width=\columnwidth,clip]{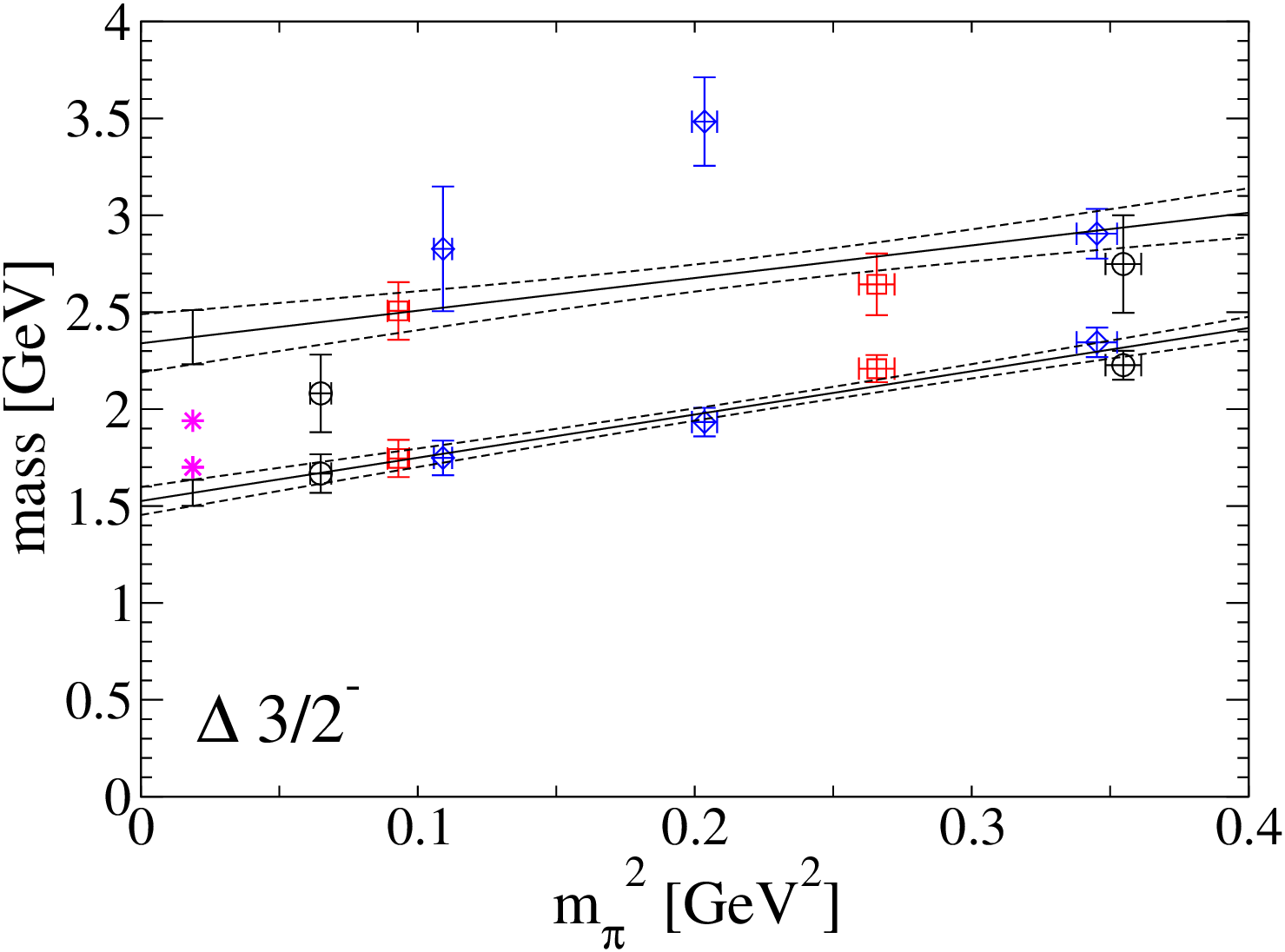}
\caption[Energy levels for $\Delta$ spin 3/2]{
Energy levels for $\Delta$ spin 3/2, positive (upper) and negative parity (lower).}
\label{fig:delta_3half}
\end{figure}
\myparagraph{\Delta :\;\;I(J^P)=\frac{3}{2}(\ot^+)}
Experimentally, the ground state $\Delta(1750)$ still needs confirmation, while $\Delta(1910)$ is well established.
In our simulation, using interpolators (1,4,5), we find two states, where the second eigenvalue decreases slower with the
pion mass than the first one.
The resulting crossing of the eigenvalues complicates the analysis and one has to follow the eigenvector composition in order
to properly assign the state.
However, the plateaus can be fitted and energy levels extracted, albeit with sizable error bars.
The chiral extrapolation of the ground state is compatible with both $\Delta(1750)$ and $\Delta(1910)$ within the error bars, the first excitation comes out higher (see Fig.~\ref{fig:delta_1half}).

\myparagraph{\Delta :\;\;I(J^P)=\frac{3}{2}(\ot^-)}
In the negative parity channel, $\Delta(1620)$ is established, while $\Delta(1900)$ needs confirmation.
Using interpolators (1,2,3,4) we extract two states in this channel.
The chiral extrapolation of the ground state hits the experimental $\Delta(1620)$ within $1.2 \sigma$ (see Fig.~\ref{fig:delta_1half}).
The excitation extrapolates to the $\Delta(1900)$, however, with a large associated error.

\myparagraph{\Delta :\;\;I(J^P)=\frac{3}{2}(\frac{3}{2}^+)}
The $\Delta(1232)$ is the lowest resonance of all spin 3/2 baryons.
We find a good signal of two states, the chiral extrapolations of both come out too high compared to the experimental $\Delta(1232)$ and the $\Delta(1600)$ (see Fig.~\ref{fig:delta_3half}).
Finite volume effects are a possible origin of the discrepancy, as will be discussed in Section \ref{sec:results_vol}. A possible
$p$-wave energy of a coupling $N\pi$ state would lie between the two observed levels and is not seen.
Note that the partially quenched data of this channel are used to set the strange quark mass parameter \cite{Engel:2011aa}. 

\myparagraph{\Delta :\;\;I(J^P)=\frac{3}{2}(\frac{3}{2}^-)}
We find a good signal in the $J^P=3/2^-$ $\Delta$ channel in all seven ensembles (see Fig.~\ref{fig:delta_3half}).
However, like in other negative parity baryon channels, the chiral extrapolation of the ground state comes out rather low compared to experiment.
The results for the first excitation are inconclusive, the $\chi^2/$d.o.f.~ of the chiral extrapolation fit is larger than three.

\section{Results for Strange Baryons}\label{sec:baryons:strange}

\subsection{Lambda}
$\Lambda$ baryons come as flavor singlets or octets, or as mixtures of them. Lattice simulations  in this channel are of particular interest, as for years no state was observed in the vicinity of the  prominent low-lying $\Lambda(1405)$ (see, e.g., \cite{Takahashi:2009ik,Takahashi:2009bu}). 
Only recent results show a level ordering compatible with experiment \cite{Menadue:2011pd}.
Our results for the $\Lambda$ baryons have been discussed elsewhere \cite{Engel:2012qp}. Here a few observations are summarized for completeness.

We include interpolators of flavor singlet and octet type and three Dirac structures in all four $J^P=\frac{1}{2}^\pm$ and $\frac{3}{2}^\pm$ channels.
In both 1/2 channels 
and in the $\frac{3}{2}^+$ channel we find ground states
extrapolating to the experimental values, whereas 
the $\frac{3}{2}^-$ ground state comes out too high. 
We confirm the $\Lambda(1405)$ and also find two low-lying excitations
in the $\ot^-$ channel. Our results 
suggest that the $\Lambda(1405)$ is dominated by flavor singlet 3-quark 
content, but at $m_\pi\approx 255$ MeV octet interpolators
contribute roughly 15-20\%, which may increase towards physical pion masses. 
The Roper-like (octet) state $\Lambda(1600)$ may  couple too weak to
our 3-quark interpolator basis.
We analyze the volume dependence and find that only the spin $\ot^+$
ground state shows a clear exponential dependence as expected 
for bound states.  For all other discussed states, the volume dependence is
either fairly flat or obscured by the statistical error.  

\subsection{Sigma}
\begin{figure}
\noindent\includegraphics[width=\columnwidth,clip]{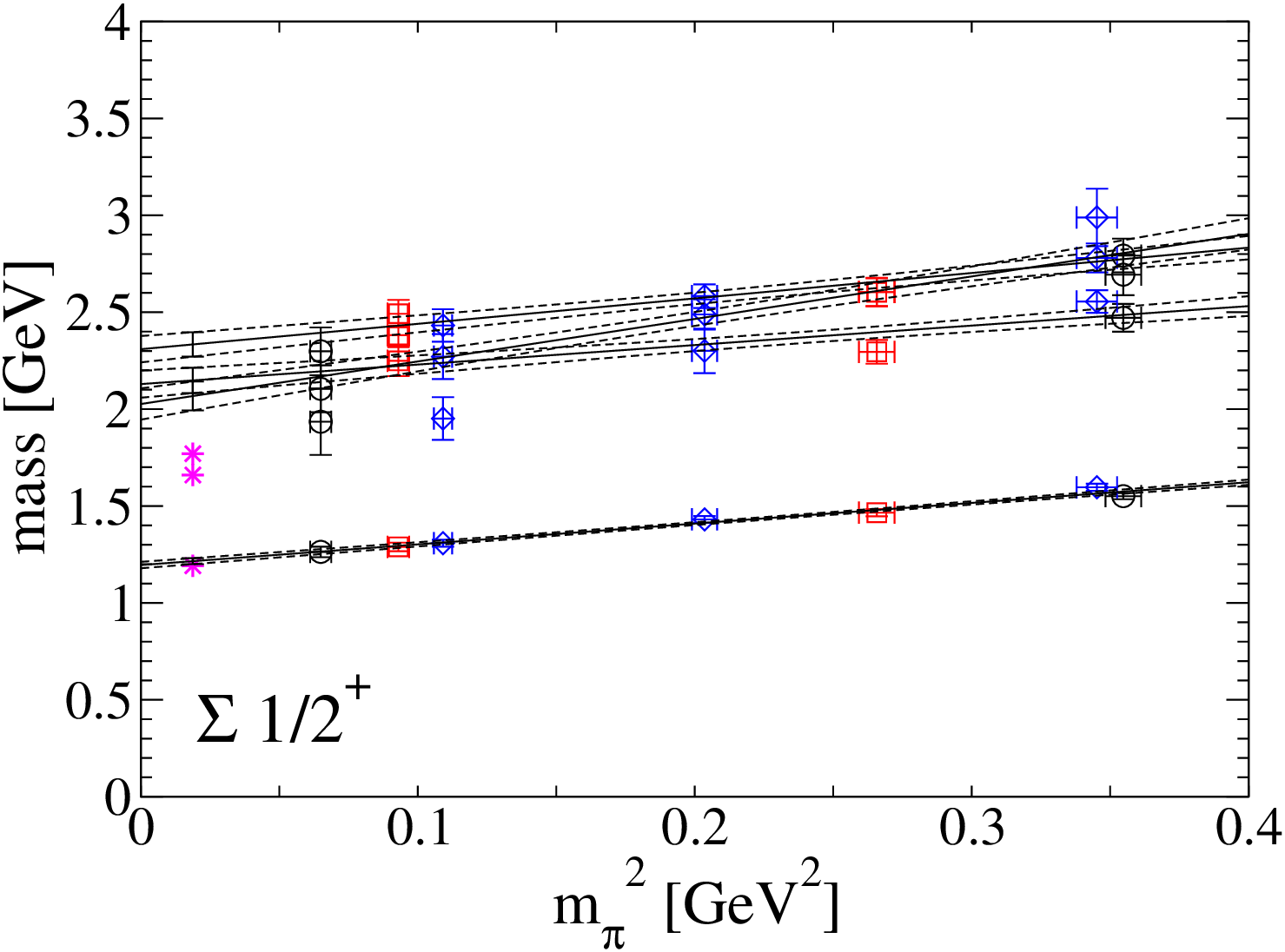}\\
\noindent\includegraphics[width=\columnwidth,clip]{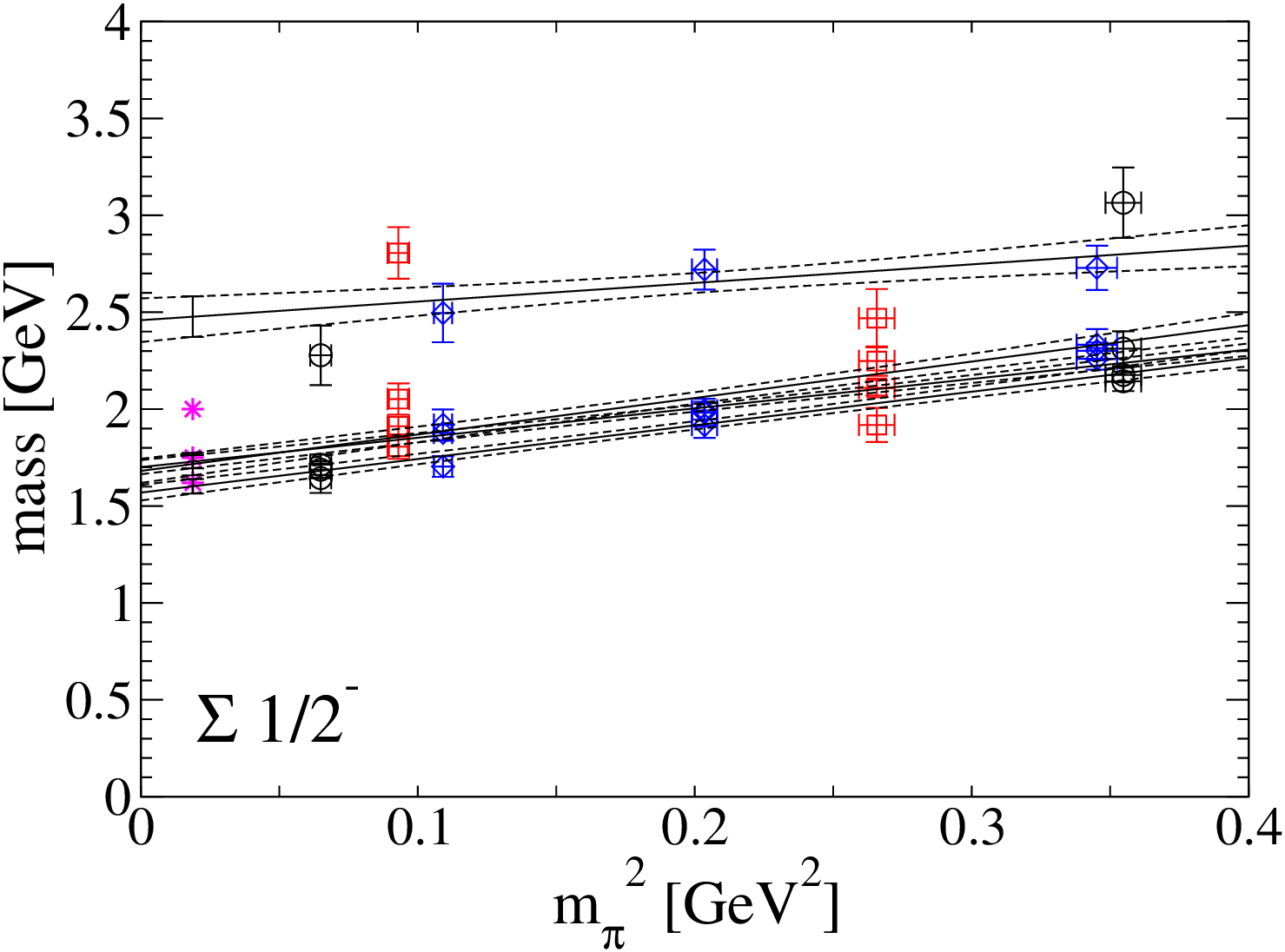}
\caption[Energy levels for $\Sigma$ spin 1/2]{
Energy levels for $\Sigma$ spin 1/2, positive (upper) and negative parity (lower).}
\label{fig:sigma_1half}
\end{figure}

\begin{figure}
\noindent\includegraphics[width=\columnwidth,clip]{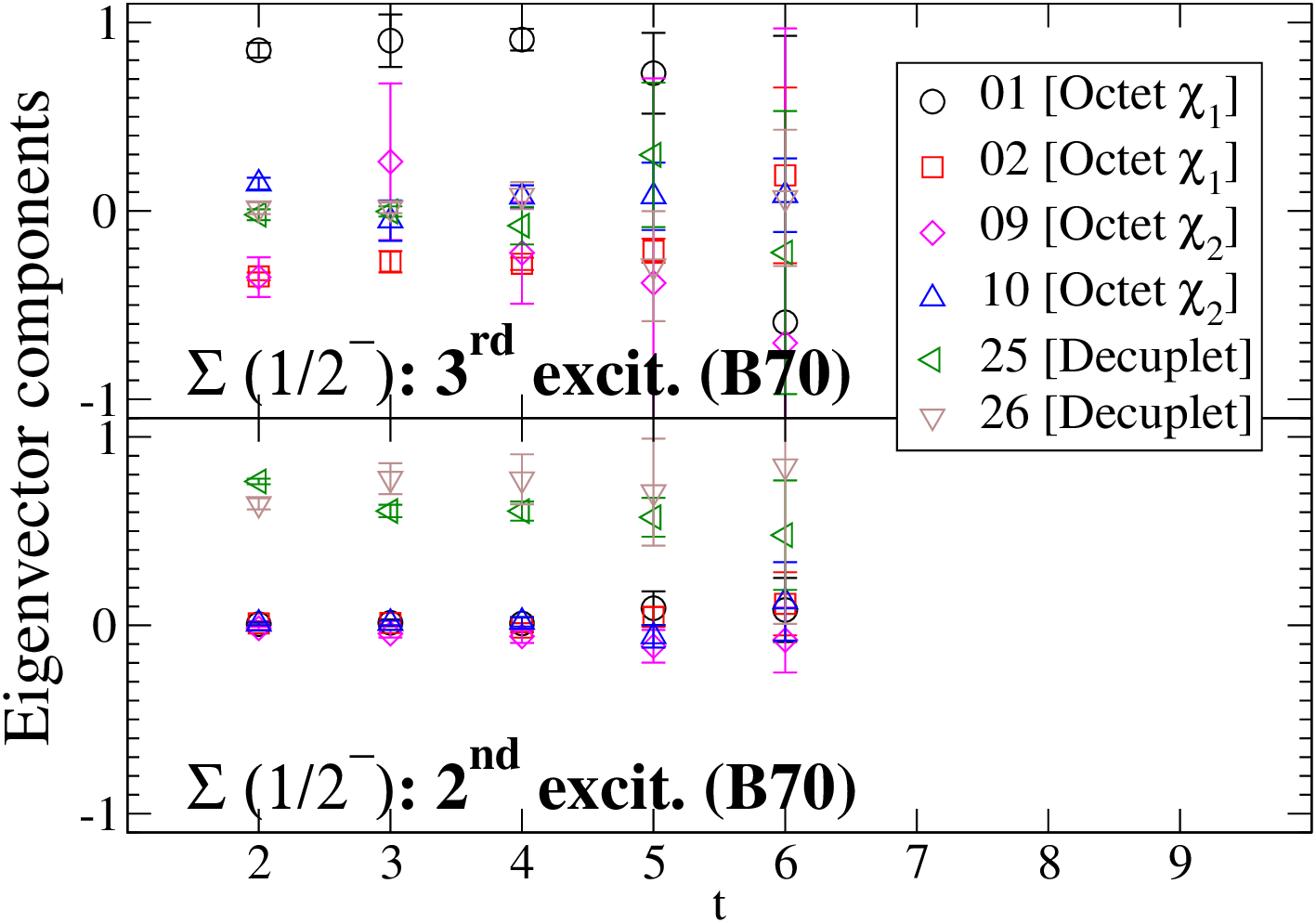}\\
\noindent\includegraphics[width=\columnwidth,clip]{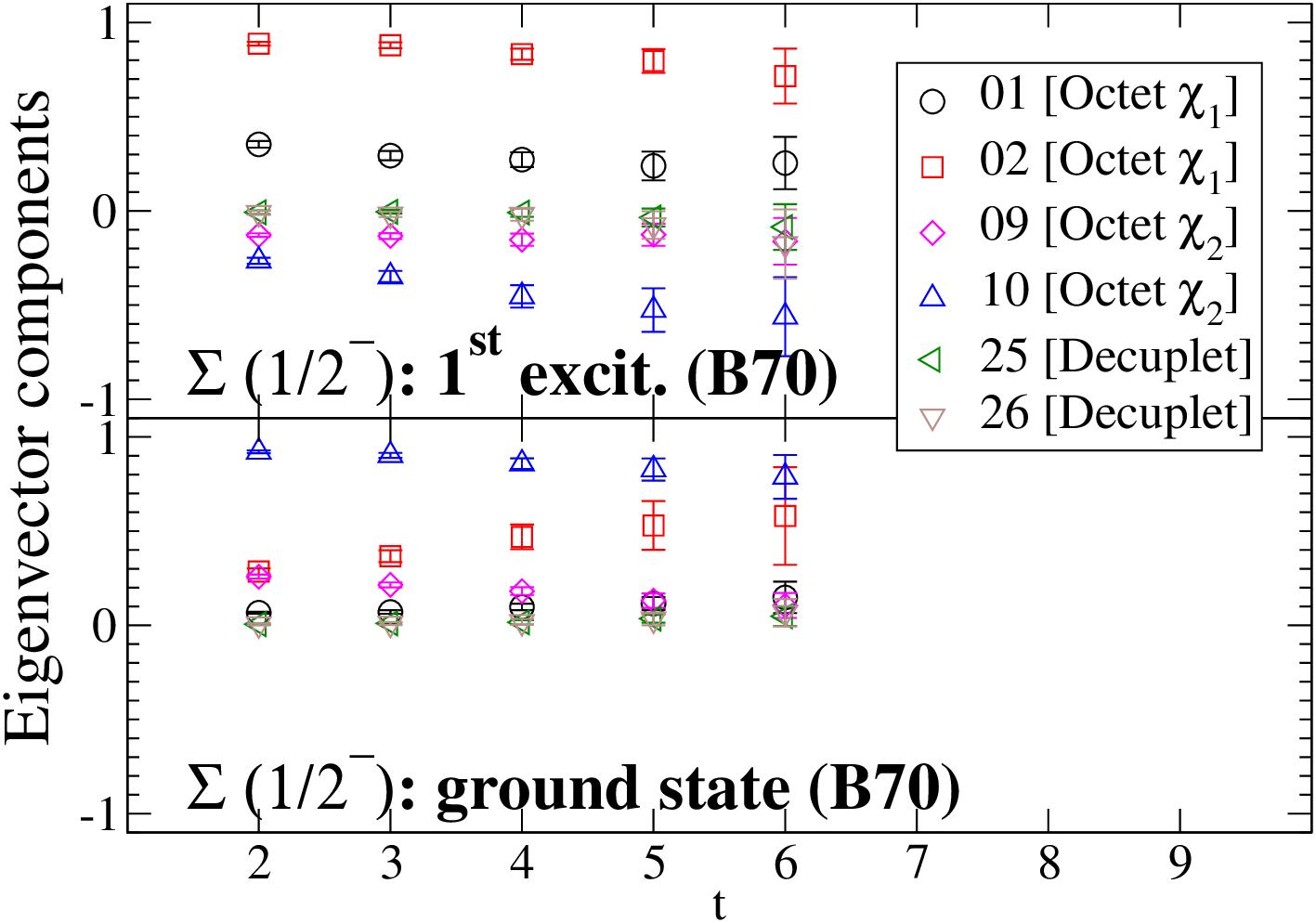}
\caption[Eigenvectors for $\Sigma$ spin 1/2 negative parity]{
Eigenvectors for $\Sigma$ spin 1/2 negative parity ground state and first excitation (upper)  
and second and third excitation (lower) for ensemble B70. 
Note the dominance of decuplet interpolators for the second excitation, which is a low lying state (see Fig.~\ref{fig:sigma_1half}). 
Details are discussed in the text.
}
\label{fig:sigma_1halfneg_vectors}
\end{figure}

\begin{figure}
\noindent\includegraphics[width=\columnwidth,clip]{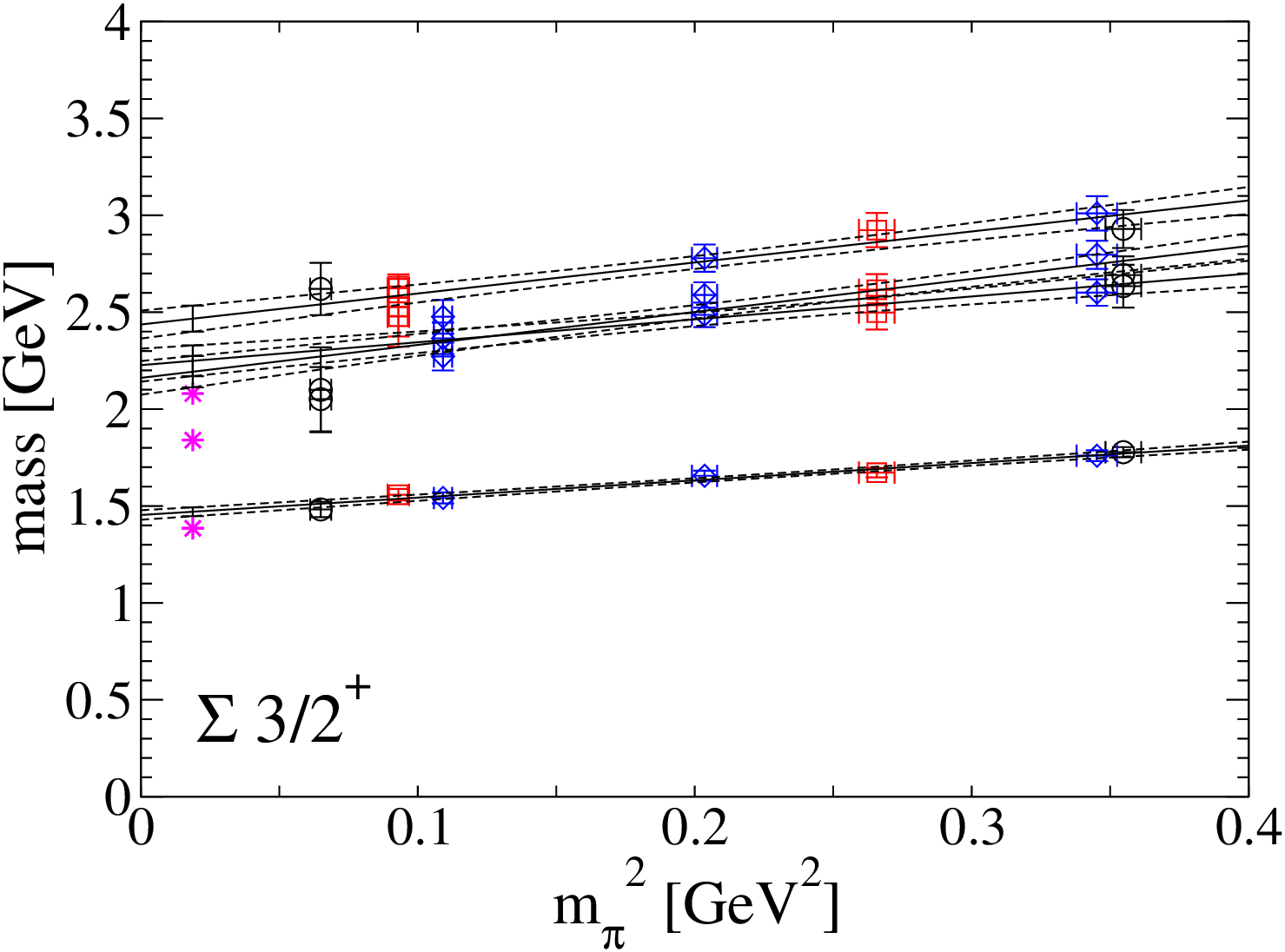}\\
\noindent\includegraphics[width=\columnwidth,clip]{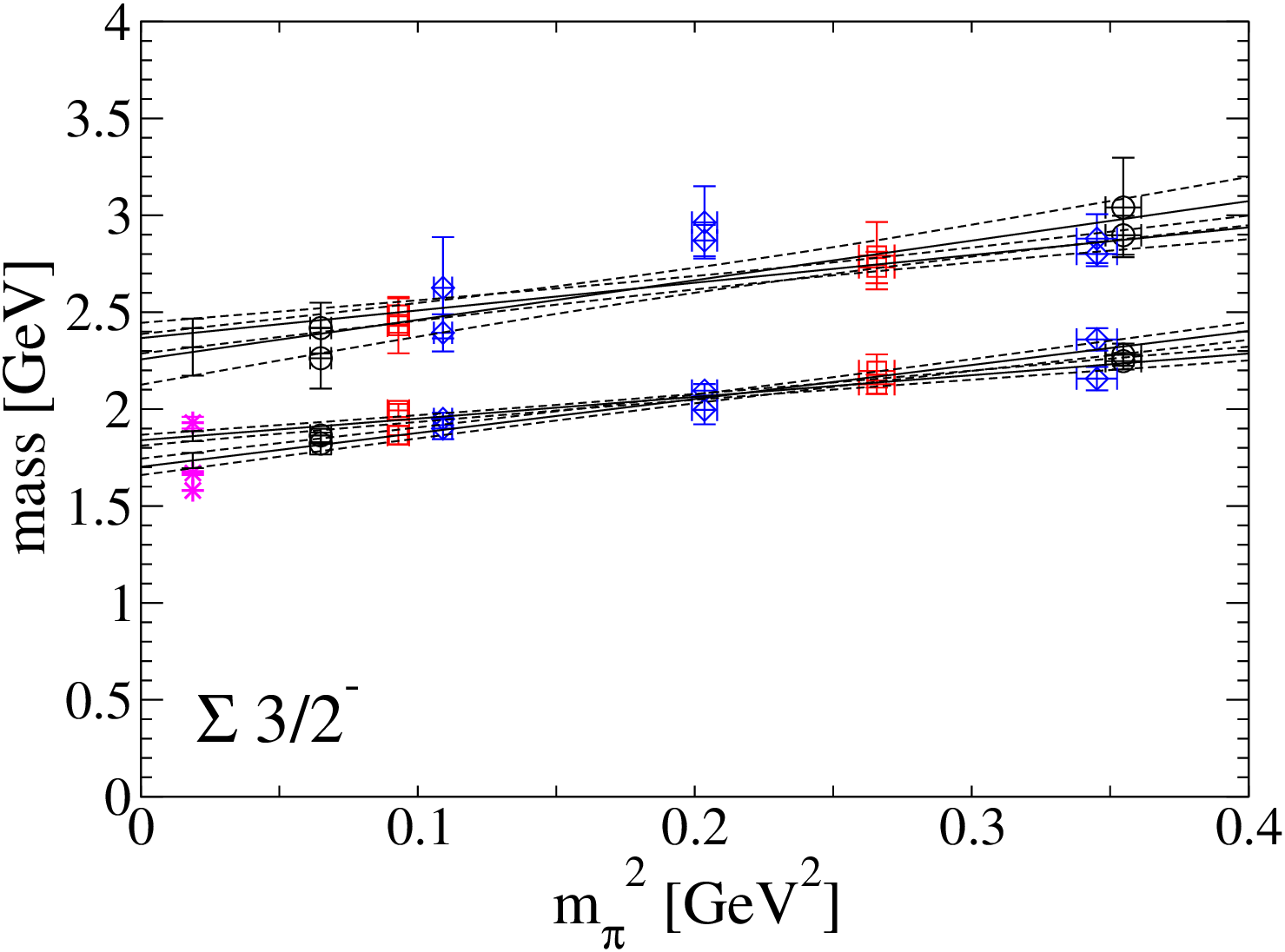}
\caption[Energy levels for $\Sigma$ spin 3/2]{
Energy levels for $\Sigma$ spin 3/2, positive (upper) and negative parity (lower).}
\label{fig:sigma_3half}
\end{figure}
\myparagraph{\Sigma: \;\;I(J^P)=1(\ot^+)}
The $\Sigma$(1189) ground state marks one of the lowest energy levels of the spin 1/2 baryons.
At the $SU(3)$ flavor symmetric point, the octet and decuplet irreducible representations are orthogonal.
Towards physical quark masses, $SU(3)_f$ is broken and hence octet and decuplet are allowed to mix.
We use the set (1,2,9,10,25,26), which includes octet interpolators with Dirac structures $\chi_1$ and $\chi_2$ 
and decuplet interpolators in the basis. We use the four lowest levels for our analysis.
The eigenvalues for three ensembles are shown in Fig.~\ref{fig:sigma_1half_eigvals}.
The ground state signal is fairly good and the chiral extrapolation results in a value close to the experimental $\Sigma(1189)$ (see Fig.~\ref{fig:sigma_1half}). 
The first excitation comes out too high compared to the experimental $\Sigma(1660)$.
Note the poor $\chi^2/$d.o.f.~of the corresponding chiral extrapolation, with a value larger than four (see Table \ref{tab:chi2baryons_pospar}).
The energy levels of the second and third excitations appear close to the first excitation in our simulations. 

Monitoring the eigenvectors, we analyze the octet/decuplet content of the states.
Within the finite basis employed, 
the ground state and the first excitation are strongly dominated by octet $\chi_1$. 
Of the second and third excitation, one is dominated by decuplet and
the other by octet $\chi_2$ interpolators.
The mixing of octet and decuplet interpolators is found to be negligible in the range of pion masses considered. 
As we will see, this holds for most $\Sigma$ and $\Xi$ channels discussed here.

\myparagraph{\Sigma: \;\;I(J^P)=1(\ot^-)}
In the $\Sigma$ spin $1/2$ negative parity channel, the Particle Data Group \cite{Beringer:1900zz} lists two low nearby states, $\Sigma(1620)$ and $\Sigma(1750)$, 
and one higher lying resonance, the  $\Sigma(2000)$.
Of those, only $\Sigma(1750)$ is established.

Again the set of interpolators (1,2,9,10,25,26) is used to extract four lowest states from our simulations.
We find three low nearby states, all of which extrapolate close to the experimental $\Sigma(1620)$ and $\Sigma(1750)$ (see Fig.~\ref{fig:sigma_1half}). 
Hence, our results confirm the $\Sigma(1620)$ and $\Sigma(1750)$ and even might suggest the existence of a third low lying resonance.
However, as discussed for the N$(\ot^-)$ (and like in the case of the $\Lambda(\ot^-)$) there are several $s$ wave baryon-meson channels ($N\overline K$, $\Lambda \pi$, $\Sigma \pi$), which,
for our values of the pion mass, have energies close to the ground state. We cannot exclude such contributions, although we
did not include them in the interpolators.

The eigenvectors of all four states are shown for ensemble B70 in Fig.~\ref{fig:sigma_1halfneg_vectors}.
Within the employed basis, the ground state is dominated by octet $\chi_2$,
the first excitation by octet $\chi_1$, the second excitation by decuplet and
the third excitation again  by octet $\chi_1$ interpolators.
We want to emphasize the existence of a low lying state in this channel which is dominated by decuplet interpolators.
This result also agrees with a recent quark model calculation \cite{Melde:2008yr}.
Again, the mixing of octet and decuplet interpolators appears to be negligible in the range of pion masses considered. 

\myparagraph{\Sigma: \;\;I(J^P)=1(\frac{3}{2}^+)}
The Particle Data Group lists $\Sigma(1385)$, $\Sigma(1840)$ and $\Sigma(2080)$, where only the lightest is established.
We use interpolators (2,3,10,11,12) and extract four energy levels (see Fig.~\ref{fig:sigma_3half}).
The chiral extrapolations come out high compared to the experimental values.
From the eigenvectors we find that 
the lowest two states are strongly dominated by decuplet,
the second excitation by octet and 
the third excitation again by decuplet interpolators.

\myparagraph{\Sigma: \;\;I(J^P)=1(\frac{3}{2}^-)}
In this channel, three states are known experimentally: $\Sigma(1580)$, $\Sigma(1670)$ and $\Sigma(1940)$, where the lightest one needs confirmation.
Using interpolators (2,3,10,11,12) we can extract four states.
We find two low lying states and two higher excitations (see Fig.~\ref{fig:sigma_3half}).
In general, the corresponding energy levels are high compared to experiment, thus not confirming the $\Sigma(1580)$.
However, the mixing of octet and decuplet might increase towards light pion masses, complicating the chiral behavior.
Analyzing the eigenvectors, we find that
of the two low lying states, one is dominated by octet and the other one by decuplet interpolators. Of the third and fourth state, one is dominated by octet and the other by decuplet interpolators.
Compared to the other $\Sigma$ channels, there appears a measurable mixing of octet and decuplet interpolators.
We remark the importance of decuplet interpolators for low-lying states 
in this channel.

\subsection{Xi}
\begin{figure}
\noindent\includegraphics[width=\columnwidth,clip]{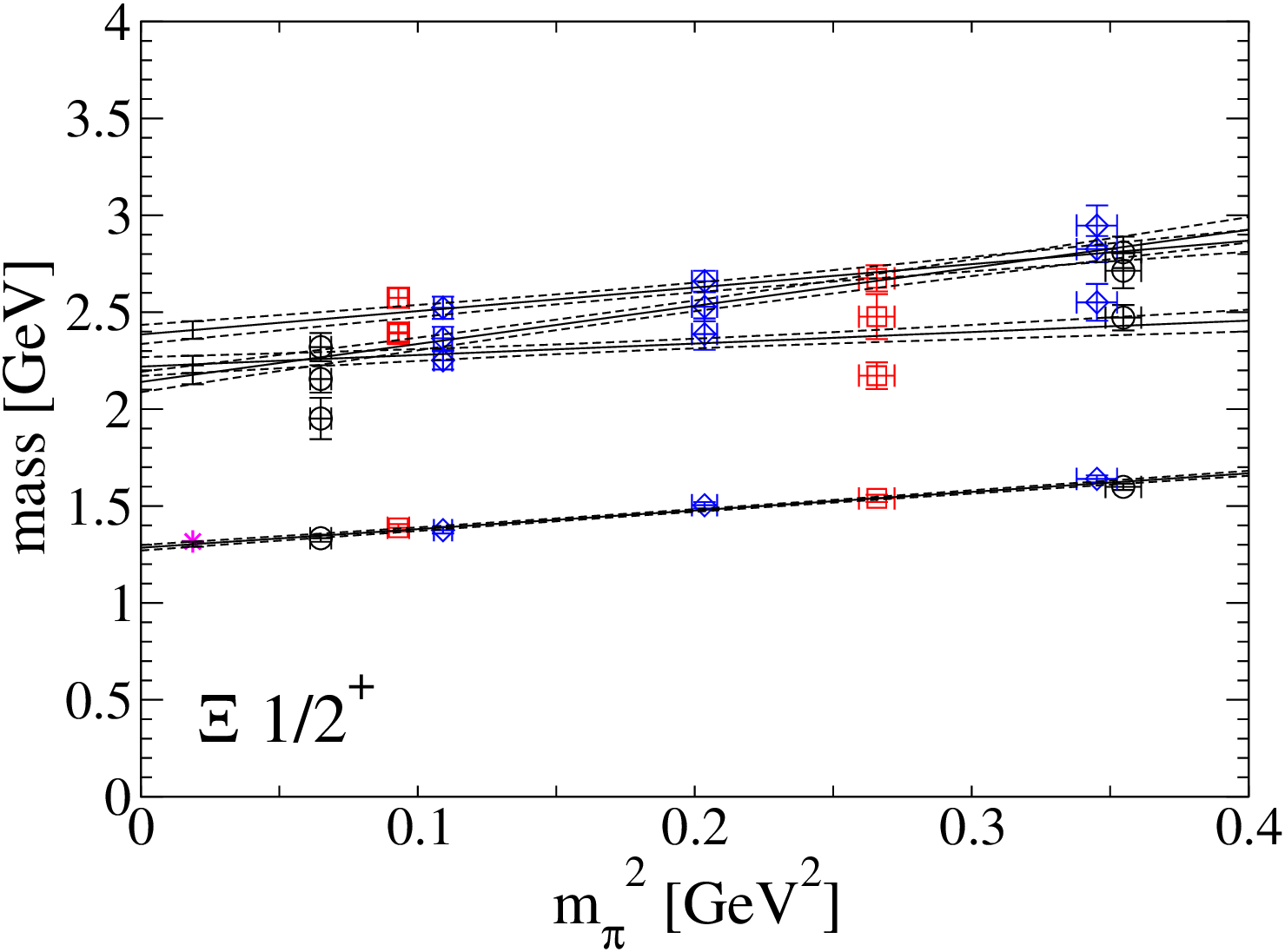}\\
\noindent\includegraphics[width=\columnwidth,clip]{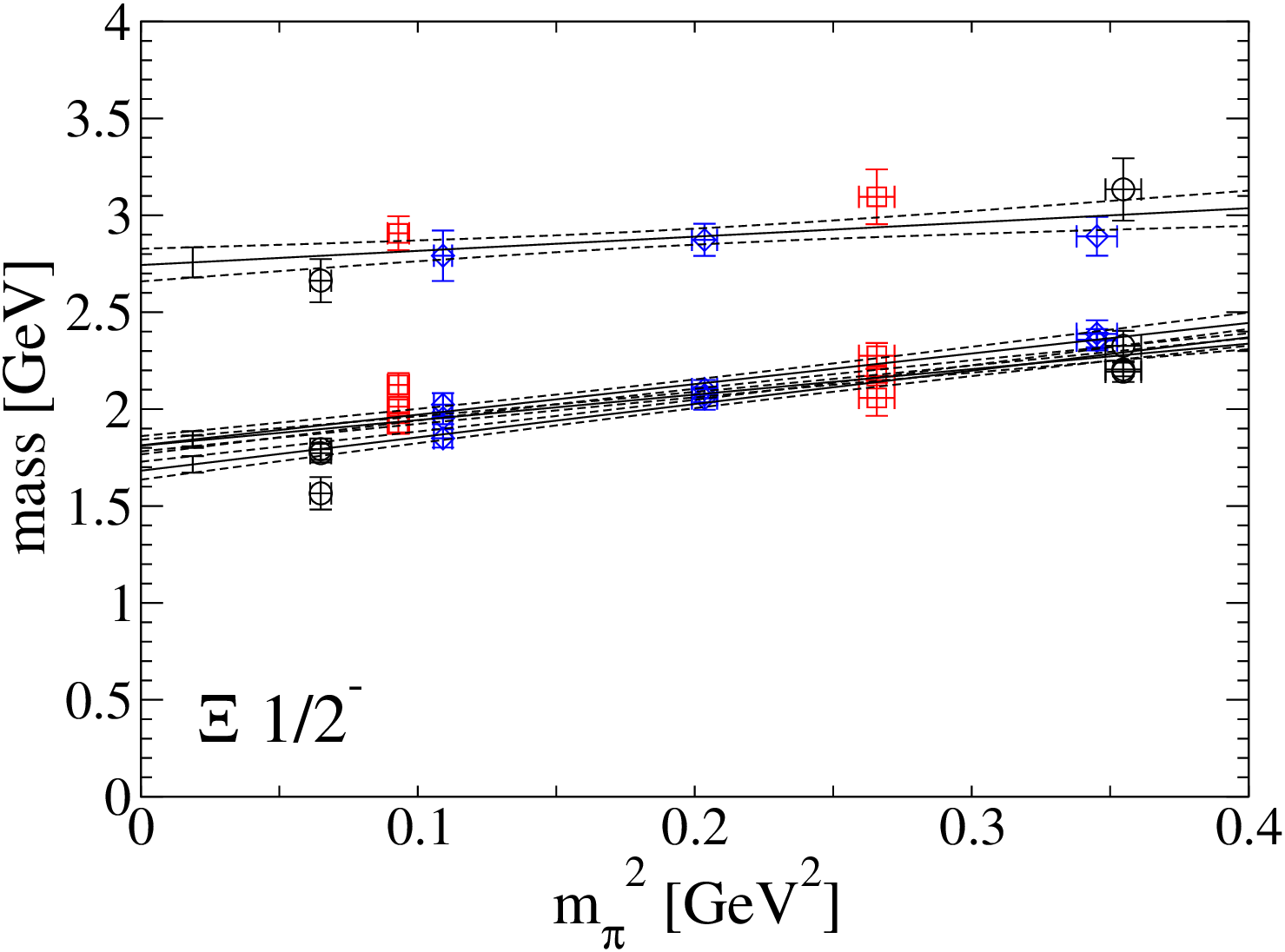}\\
\caption[Energy levels for $\Xi$ spin 1/2]{
Energy levels for $\Xi$ spin 1/2, positive (upper) and negative parity (lower).}
\label{fig:xi_1half}
\end{figure}
\begin{figure}
\noindent\includegraphics[width=\columnwidth,clip]{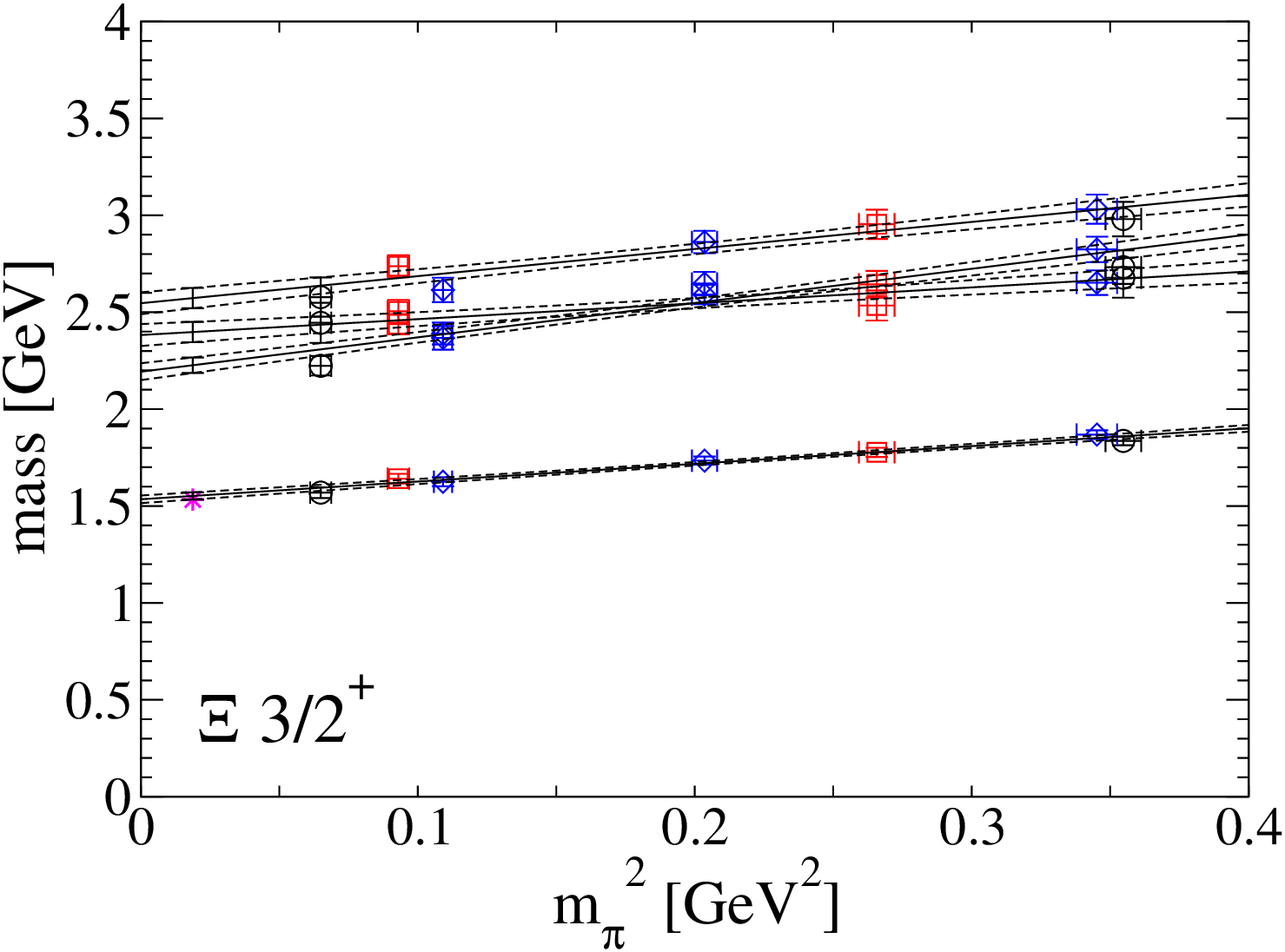}\\
\noindent\includegraphics[width=\columnwidth,clip]{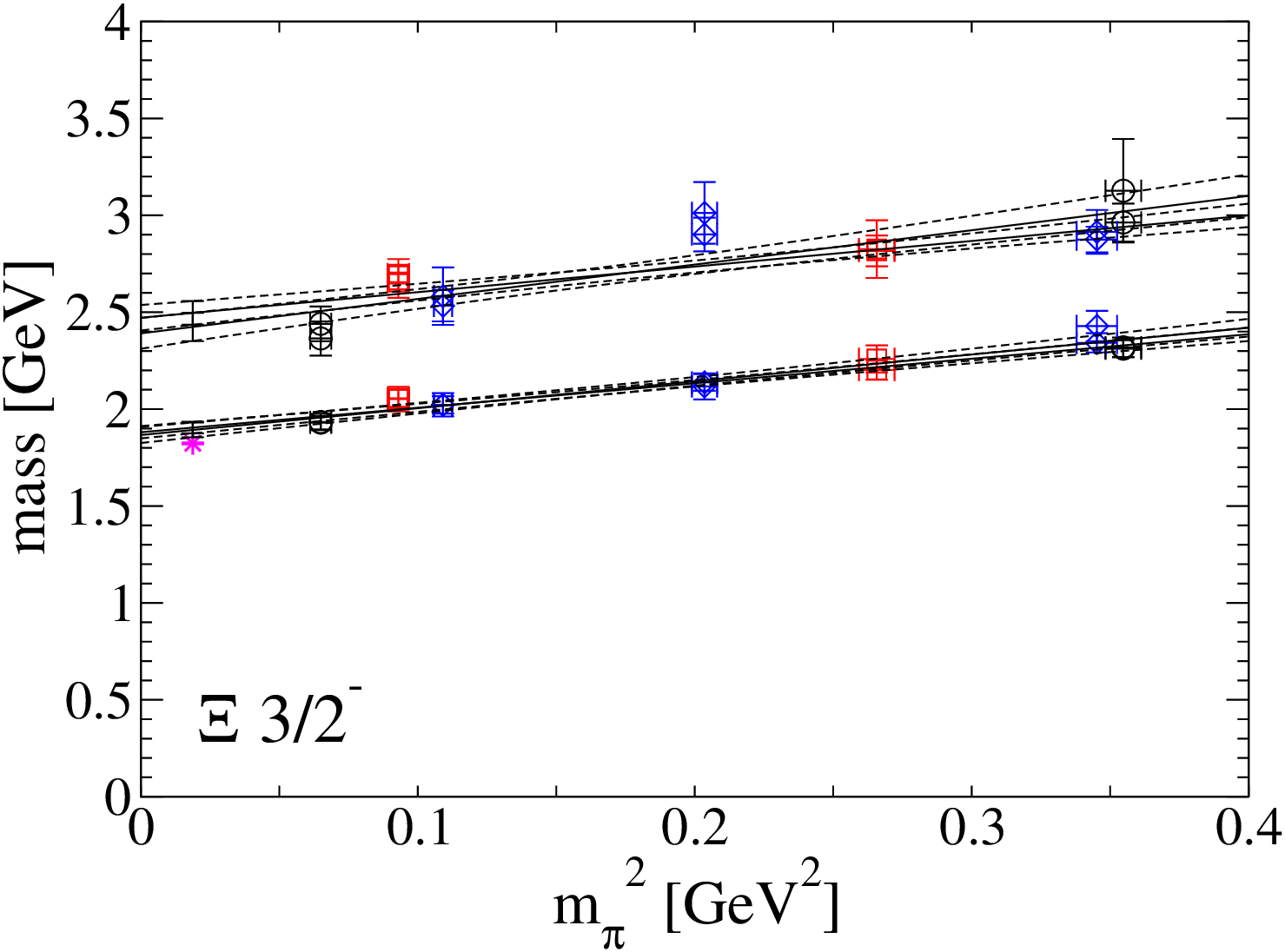}\\
\caption[Energy levels for $\Xi$ spin 3/2]{
Energy levels for $\Xi$ spin 3/2, positive (upper) and negative parity (lower).}
\label{fig:xi_3half}
\end{figure}

\myparagraph{\Xi: \;\; I(J^P)=\ot(\ot^+)}
Experimentally, only one resonance $\Xi(1322)$ is known in the $\Xi$ spin 1/2 positive parity channel.
We use interpolators (1,2,9,10,25,26) and extract the four lowest states. 
The ground state shows a fairly clean signal and its chiral extrapolation agrees nicely with the $\Xi(1322)$ (see Fig.~\ref{fig:xi_1half}).
The three excitations come out much higher and the results at the lightest pion mass may suggest a significant chiral curvature towards physical pion masses. 
This is also expressed in the poor $\chi^2/$d.o.f., which is above five for the first excitation (see Table 
\ref{tab:chi2baryons_pospar}).
Analyzing the eigenvectors, we find that -- within the finite basis used --
the ground state and the first excitation are strongly dominated by octet $\chi_1$. 
Of the third and the fourth excitation, one is dominated by decuplet and the other one by octet $\chi_2$ interpolators.
The mixing of octet and decuplet interpolators is found negligible in the range of simulated pion masses.

\myparagraph{\Xi: \;\; I(J^P)=\ot(\ot^-)}
No state is known in the $\Xi$ spin $1/2^-$ channel experimentally, and no low-lying 
state identified in quark model calculations like, e.g., \cite{Melde:2008yr}. 
Nevertheless, using interpolators (1,2,9,10,25,26), we identify four states in our simulations (see Fig.~\ref{fig:xi_1half}).
Of those, three are low lying and extrapolate to 1.7-1.9 GeV.
Note the poor $\chi^2/$d.o.f.~larger than three of the corresponding three chiral extrapolations.
The fourth state appears rather high at 2.7-2.9 GeV, but its extrapolation shows a nice $\chi^2/$d.o.f.~of order one.
From the eigenvectors we find that 
the ground state is dominated by octet $\chi_2$,
the first excitation by octet $\chi_1$,
the second excitation by decuplet and
the third excitation again by octet $\chi_1$ interpolators.
We emphasize the existence of a low lying state in this channel which is dominated by decuplet interpolators, 
analogous to the $\Sigma$ spin $1/2$ negative parity channel.

\myparagraph{\Xi: \;\;I(J^P)=\ot(\frac{3}{2}^+)}
In this channel, one state, $\Xi(1530)$, is experimentally known and well established.
We use interpolators (2,3,10,11,12) to extract four states from our simulation.
All four states show a stable signal  and the ground state energy level nicely extrapolates to the experimental $\Xi(1530)$ (see Fig.~\ref{fig:xi_3half}). 
The second and third energy levels appear rather close to each other and are compatible with a level crossing picture within pion masses of 300-500 MeV. 
Within the finite basis used, the ground state is dominated by decuplet interpolators, which agrees with quark model calculations.
At light pion masses, the first excitation is dominated by octet and the second by decuplet interpolators.
The third excitation is again dominated by decuplet interpolators.

\myparagraph{\Xi: \;\;I(J^P)=\ot(\frac{3}{2}^-)}
The Particle Data Group \cite{Beringer:1900zz}
lists one (established) state, $\Xi(1820)$, which is expected to be dominated by octet 
interpolators according to quark model calculations \cite{Beringer:1900zz}.
Using interpolators (2,3,10,11,12), we extract four energy levels in this channel.
We find two low lying states, the energy levels of which extrapolate close to the experimental $\Xi(1820)$ (see Fig.~\ref{fig:xi_3half}).
Analyzing the eigenvectors, we find that
of the two low lying states, one is dominated by octet and the other one by decuplet interpolators.
The third state is dominated by octet and the fourth state by decuplet interpolators.
Compared to the other $\Xi$ channels, there appears a small but measurable mixing of octet and decuplet interpolators.

\subsection{Omega}
\begin{figure}
\noindent\includegraphics[width=\columnwidth,clip]{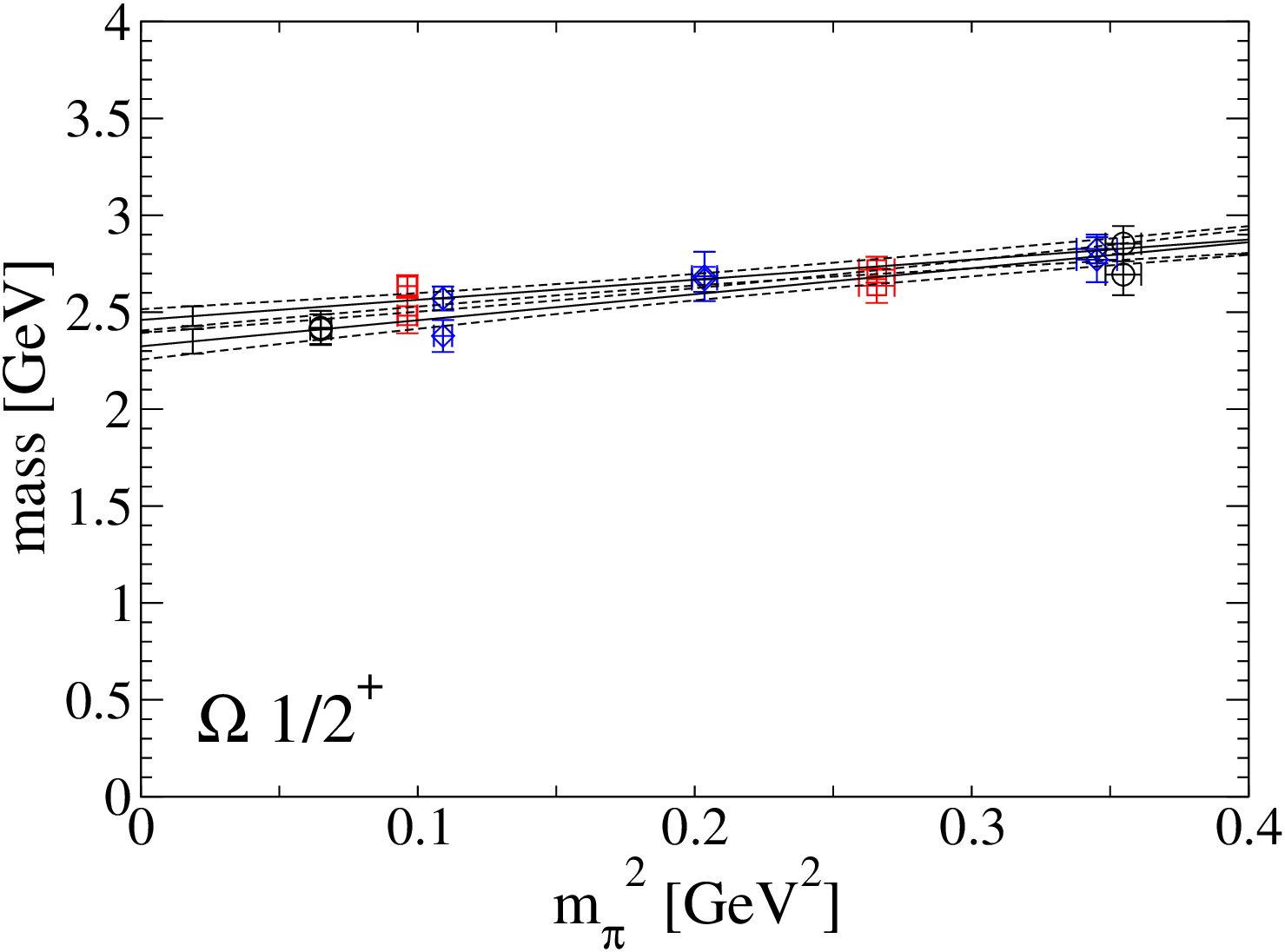}\\
\noindent\includegraphics[width=\columnwidth,clip]{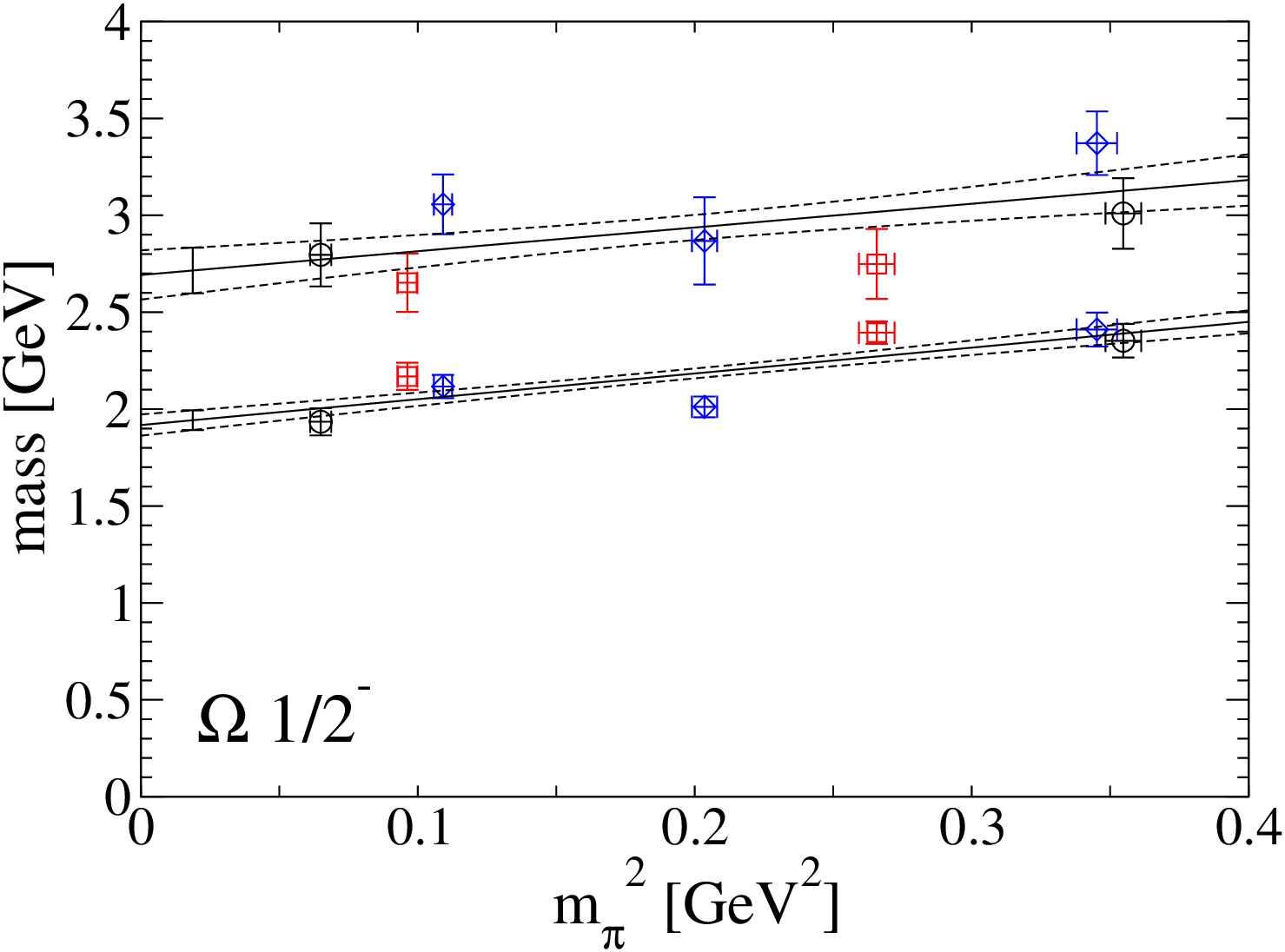}\\
\caption[Energy levels for $\Omega$ spin 1/2]{
Energy levels for $\Omega$ spin 1/2, positive (upper) and negative parity (lower).}
\label{fig:omega_1half}
\end{figure}
\begin{figure}
\noindent\includegraphics[width=\columnwidth,clip]{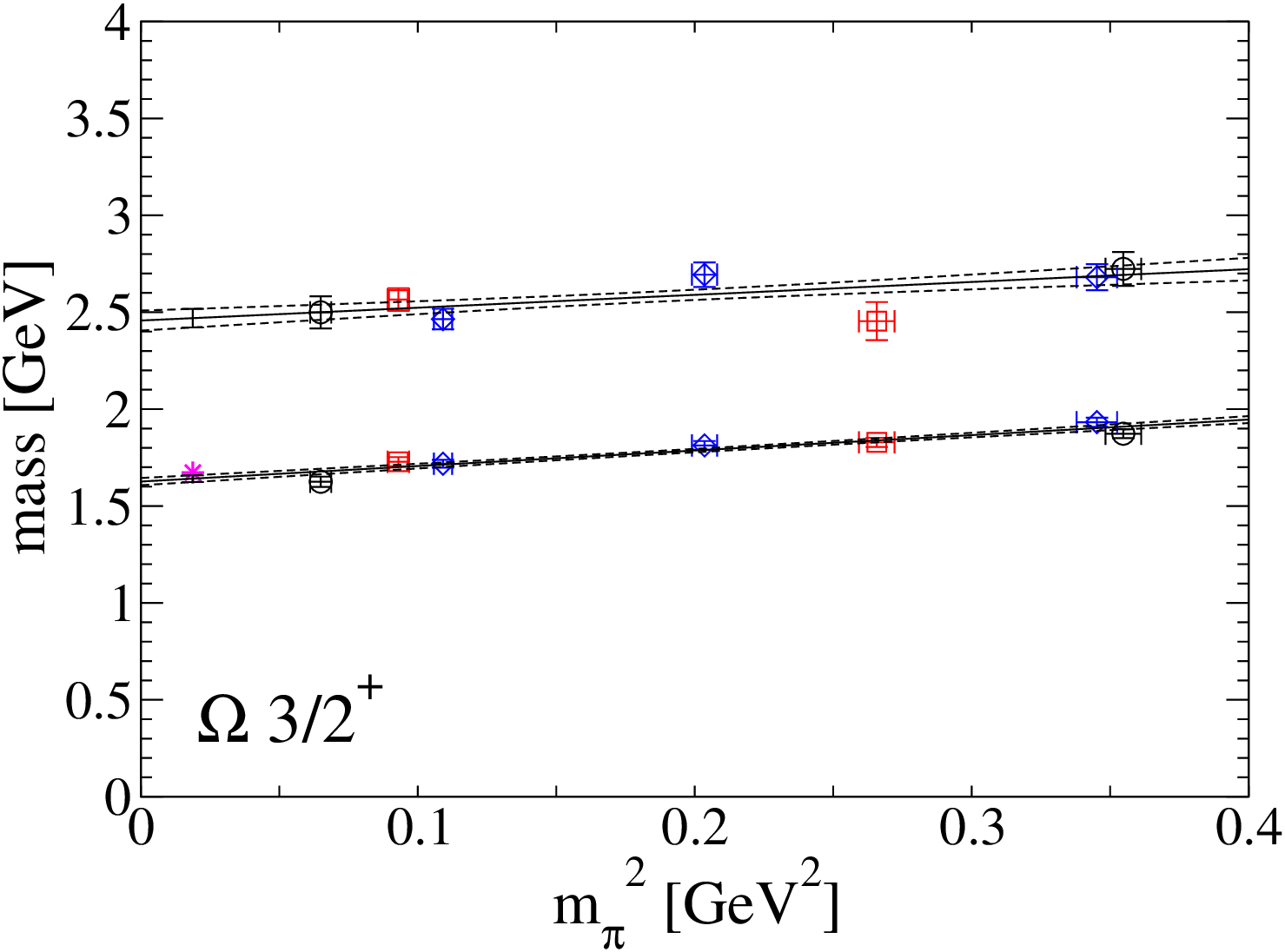}\\
\noindent\includegraphics[width=\columnwidth,clip]{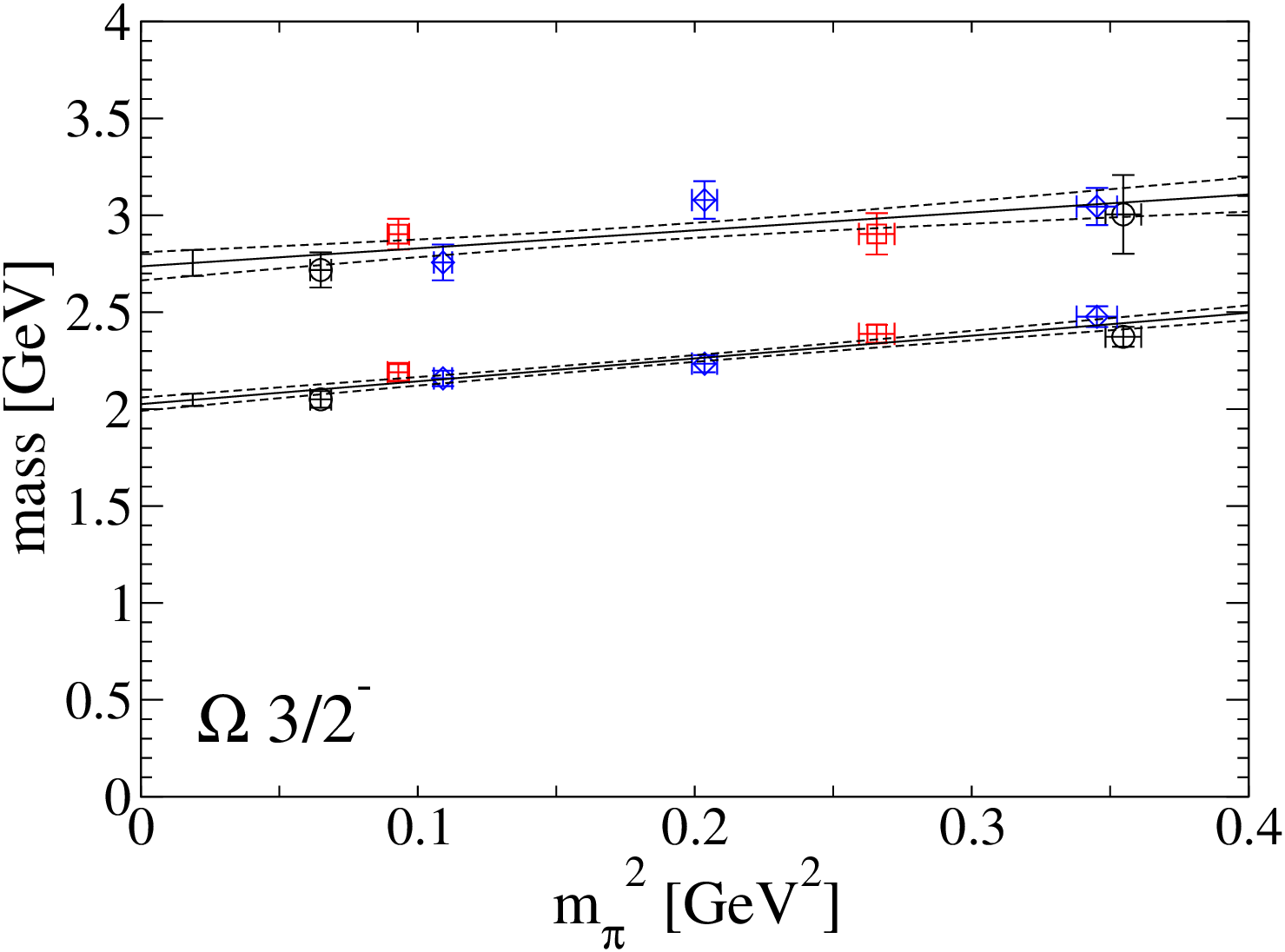}\\
\caption[Energy levels for $\Omega$ spin 3/2]{
Energy levels for $\Omega$ spin 3/2, positive (upper) and negative parity (lower).}
\label{fig:omega_3half}
\end{figure}
\myparagraph{\Omega: \;\;I(J^P)=0(\ot^+)}
Experimentally, the $\Omega$ baryons have been investigated little.
No state is known in the $J^P=1/2^+$ channel.
Using the same interpolators as in the corresponding $\Delta$ channel, we find two states, whose energy levels are close for all simulated pion masses (see Fig.~\ref{fig:omega_1half}).
Both predicted resonances lie between 2.3 and 2.6 GeV. 

\myparagraph{\Omega: \;\;I(J^P)=0(\ot^-)}
Again, there is no experimental experience in the $J^P=\ot^-$ channel of the $\Omega$ baryons.
We extract two states, where the excitation comes with some noise.
The chiral extrapolation of the ground state predicts a resonance around 2 GeV (see Fig.~\ref{fig:omega_1half}).
Note the corresponding poor $\chi^2/$d.o.f. larger than four (see Table \ref{tab:chi2baryons_negpar}); its main contribution comes 
from the light energy level of one ensemble (C72). 
Since this behavior is not systematically observed in other channels, we assume the deviation to be due to statistical fluctuations. 

\myparagraph{\Omega: \;\;I(J^P)=0(\frac{3}{2}^+)}
The $\Omega(1672)$ in the $J^P=3/2^+$ channel is known experimentally to very high accuracy.
This is one of the reasons why this state is often used to define the strange quark mass parameters.
This approach is pursued also in our setup.
The determination of the parameters has been performed along a different scheme of 
scale setting compared to \ref{Engel:2011aa}.
The Sommer parameter was identified with the experimental value for each ensemble, without extrapolation to physical pion masses.
In that scheme the lowest energy level in the $\Omega$ $J^P=3/2^+$ channel was identified with the experimental $\Omega(1672)$ for each ensemble. 
This identification used preliminary data on the $16^3\times32$ lattices
only.
Here we present results relying on another scheme of scale setting \cite{Engel:2011aa}.
Thus, the results shown here for the ground state serve as an additional cross check for the final setup of the simulation.

The ground state energy level extrapolates close to the experimental $\Omega(1672)$, undershooting it slightly (see Figure \ref{fig:omega_3half}). 
The corresponding $\chi^2/$d.o.f.~is around two (see Table \ref{tab:chi2baryons_pospar}),
half of it contributed by ensemble A66.
Using our final dataset and revisiting the tuning, we find that the strange
quark mass of ensemble A66 is  slightly too light while the mass from ensemble
C64 is slightly too heavy. This creates a slope in the chiral extrapolation which
causes the  $\Omega(1672)$ (and to a lesser extent all baryons involving one or
more strange quarks) to be lighter than a proper tuning would imply. A thorough
discussion is difficult since also other systematics enter.
We will provide some further discussion, also considering finite volume effects, in Section \ref{sec:results_vol}.

\myparagraph{\Omega: \;\;I(J^P)=0(\frac{3}{2}^-)}
In the $J^P=3/2^-$ channel of the $\Omega$ baryons there is no experimental evidence.
We find two states, both with a fairly good signal, in our simulations.
The chiral extrapolation of the ground state energy level predicts a resonance slightly above 2 GeV (see Figure \ref{fig:omega_3half}).
\section{Volume Dependence of Baryon Energy Levels}\label{sec:results_vol}

For resonance states in large volumes, there are two leading mechanisms of finite volume effects. 
For one, the spectral density of scattering states depends on the volume and distorts the energy spectrum through avoided level crossings. 
This mechanism is very important for the determination of resonance properties \cite{Luscher:1986pf,Luscher:1991cf}. 
The expected distortion from this effect is of $\mathcal O(\Gamma)$, where
$\Gamma$ is the width of the resonance. Notice that the resonance width is
expected to be quite a bit smaller than the physical one at unphysical pion masses. This
justifies identifying the pattern of energy levels qualitatively with the
spectrum of resonances. Therefore, this kind of finite volume effect is
discussed only qualitatively for particular observables. 
A second volume effect comes from virtual pion  exchange with the mirror image.
The so-called ``pion wrapping around the universe'' causes an exponential
correction to the energy level of the hadron \cite{Luscher:1985dn}. 
This mechanism can be discussed to higher orders in Chiral Perturbation Theory 
\cite{Colangelo:2005gd,Colangelo:2005cg,Meissner:2005ba}. 
Here we follow a fit form successfully applied in \cite{Durr:2008zz},
\be
E_h(L)	= E_h(L=\infty) + c_h(m_\pi) \E^{-m_\pi L} (m_\pi L)^{-3/2} \;,
\label{eq:vol}
\ee
where $E_h$ is the energy level of the hadron at linear size $L$ of the lattice.
It was suggested that $c_h(m_\pi)=c_{h,0} m_\pi^2$, which implies two fit parameters for each observable: $E_h(L=\infty)$ and $c_{h,0}$. 
The parameter $c_{h,0}$ is shared among different ensembles, which we exploit to make combined fits.
We remark that the fit form used is a fairly simple one, however, considering the small number of different volumes, we have to rely on a method which uses few parameters. 

Due to the exponential behavior, finite volume effects are expected to become non-negligible for $m_\pi L\lesssim 4$.
This region is entered in particular for the ensembles with small pion masses. 
Eq.~\eq{eq:vol} is valid only for asymptotically large volumes, power-like corrections are expected for $m_\pi L\lesssim 3$ and already earlier for higher excitations. 
For ensemble C77 ($m_\pi=330$ MeV) we generated data on configuration size $12^3\times 24$, $16^3\times 32$, and $24^3\times 48$, for 
ensemble A66 ($m_\pi=255$ MeV) we have data for size $16^3\times 32$ and $24^3\times 48$.
All these ensembles, $2.7<m_\pi L<6$, where the pion cloud exchange should have a measurable effect described by Eq.~\eq{eq:vol}.
We apply Eq.~\eq{eq:vol} separately to each observable. 
The data of sets A66 and C77 enter a combined fit, and the resulting parameters are used to extrapolate the data of all ensembles (for that observable) to infinite volume.
Finally, the results are extrapolated to the physical light-quark mass.

We focus on narrow or stable states with a good signal where 
clear finite volume effects can be expected.
This is the case in particular for the ground states of the positive parity baryon channels. 
As mentioned in the previous section, the results for strange baryons are
affected by our imperfect strange quark tuning. The tuning is of
acceptable quality for 5 out of the 7 ensembles of size $16^3\times32$.
We therefore omit
the data from C64 and A66 for our final chiral fits for baryons with
strangeness. As our tuning was done in finite volume the resulting value for the
$\Omega(1672)$ will still deviate from the physical value. Assuming a simple
dependence on the number of strange quarks, this deviation can be translated to
other states and we provide this simple estimate as a second uncertainty when
citing final values for the baryon masses. These values are also listed in Table
\ref{tab:chi2_vol}.

\subsection{Nucleon}\label{sec:results_vol:baryons:nucleon}

\myparagraph{N:\;\;I(J^P)=\ot(\ot^+)}
\begin{figure}[tb]
\centering
\includegraphics[width=\mylenC,clip]{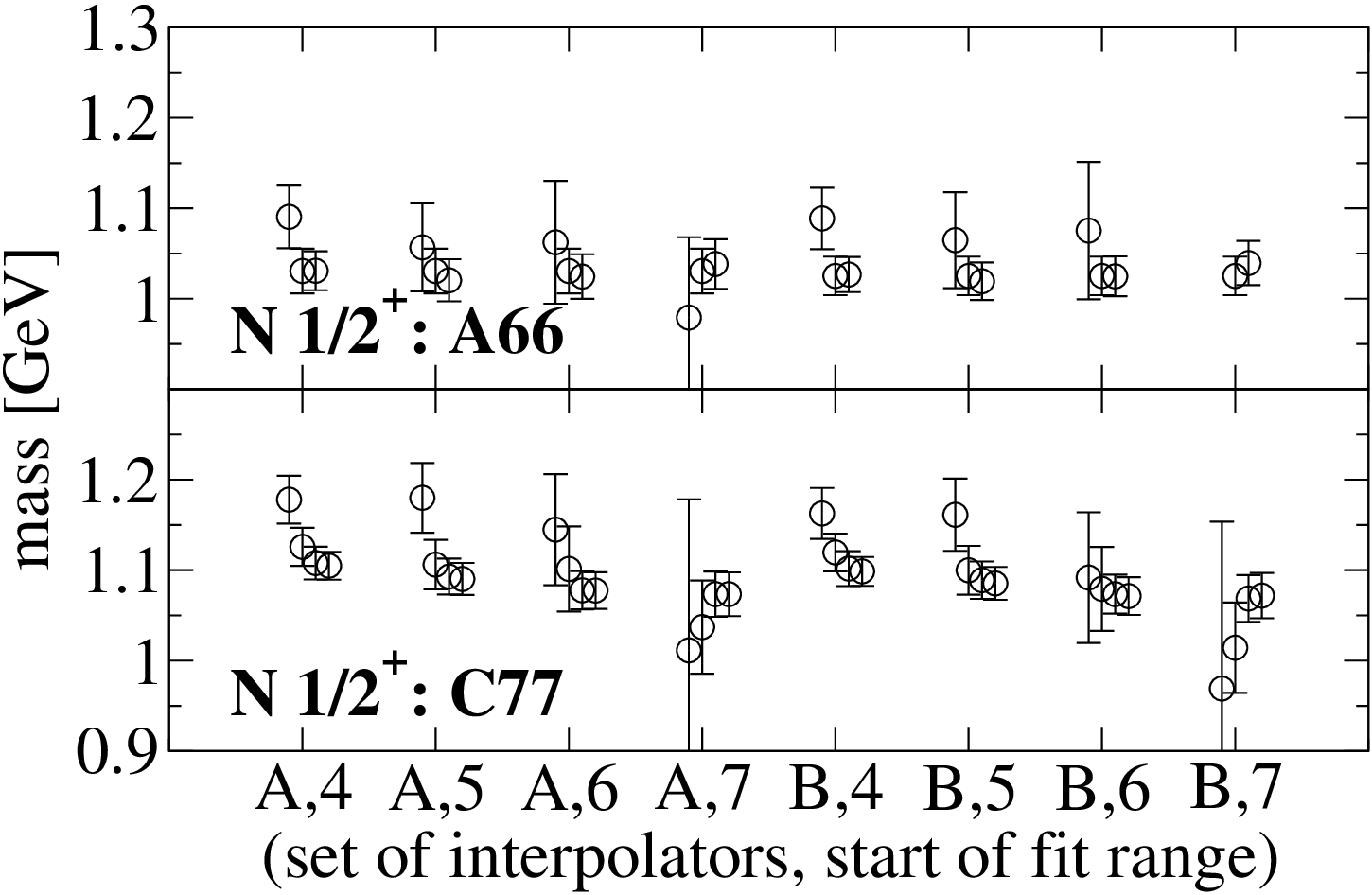}
\caption[Systematic error of the nucleon mass]
{Systematic error of the nucleon ground state energy level.
The levels are shown for different choices of interpolators and
fit ranges, labeled on the horizontal axis. E.g., ``A4'' denotes the set of
interpolator ``A'' and a fit range for the eigenvalues from $t=4a$ to
the onset of noise.
``A'' denotes set of interpolators (1,2,9,10,19,20), ``B'' denotes (3,4,11,12,19,20).
For each set of interpolator and fit range, results for small to large lattices 
(spatial size 16, 24 for ensemble A66, and 12,16, 24 for C77) are shown from left to right, the corresponding infinite volume limit rightmost.
}
\label{fig:nucleon_1half_pospar_vol_syserr}
\end{figure}
\begin{figure}[tb]
\centering
\includegraphics[width=\mylenC,clip]{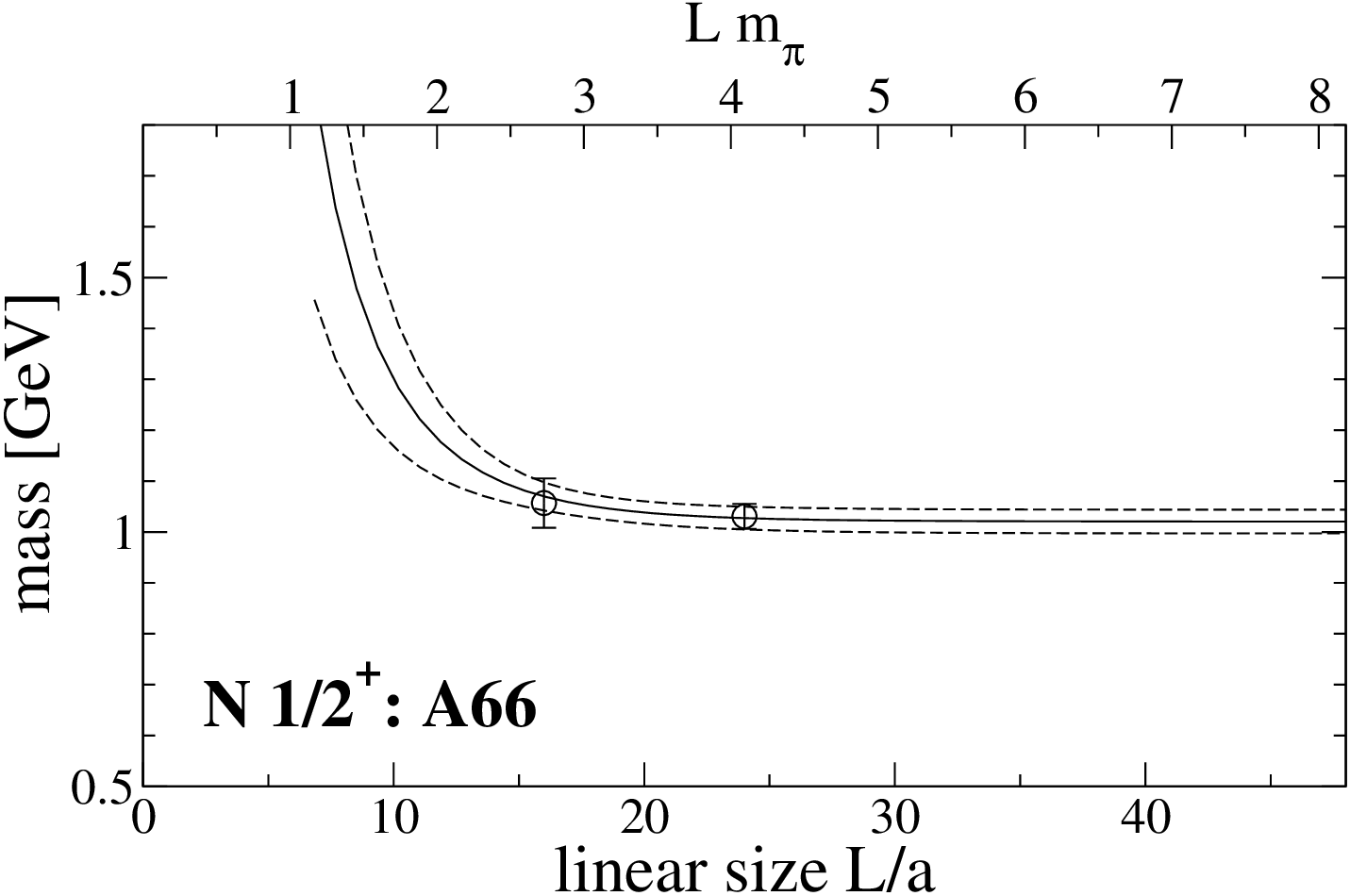}\\
\vspace{5mm}
\includegraphics[width=\mylenC,clip]{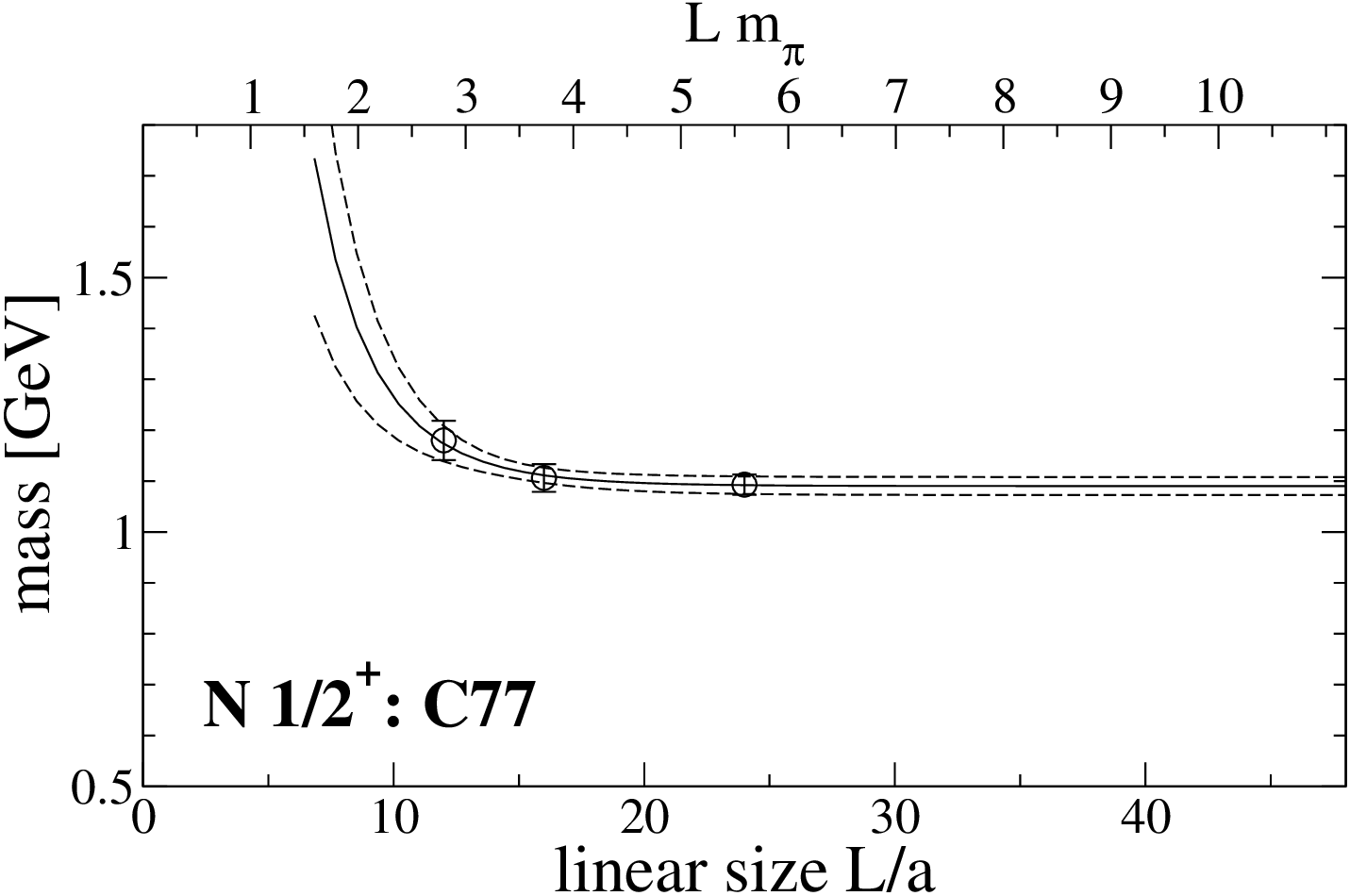}
\caption[Volume dependence of the nucleon mass]
{Volume dependence of the nucleon mass for the set of interpolators  (1,2,9,10,19,20) and $t_{\text{min}}=5a$ ((A,5) of Fig.~\ref{fig:nucleon_1half_pospar_vol_syserr}).
}
\label{fig:nucleon_1half_pospar_vol}
\end{figure}
\begin{figure}[htb]
\includegraphics[width=\mylenC,clip]{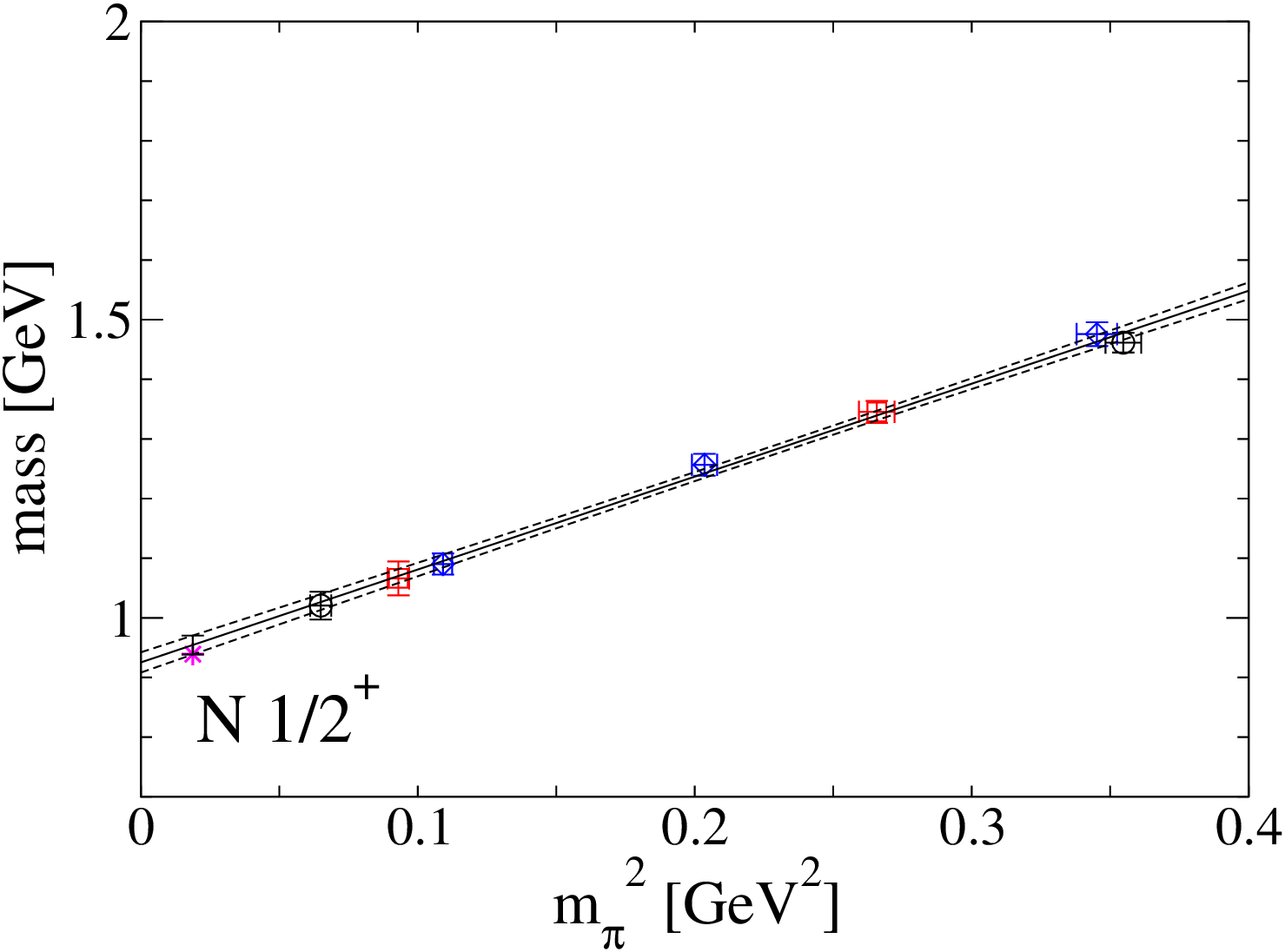}\\
\includegraphics[width=\mylenC,clip]{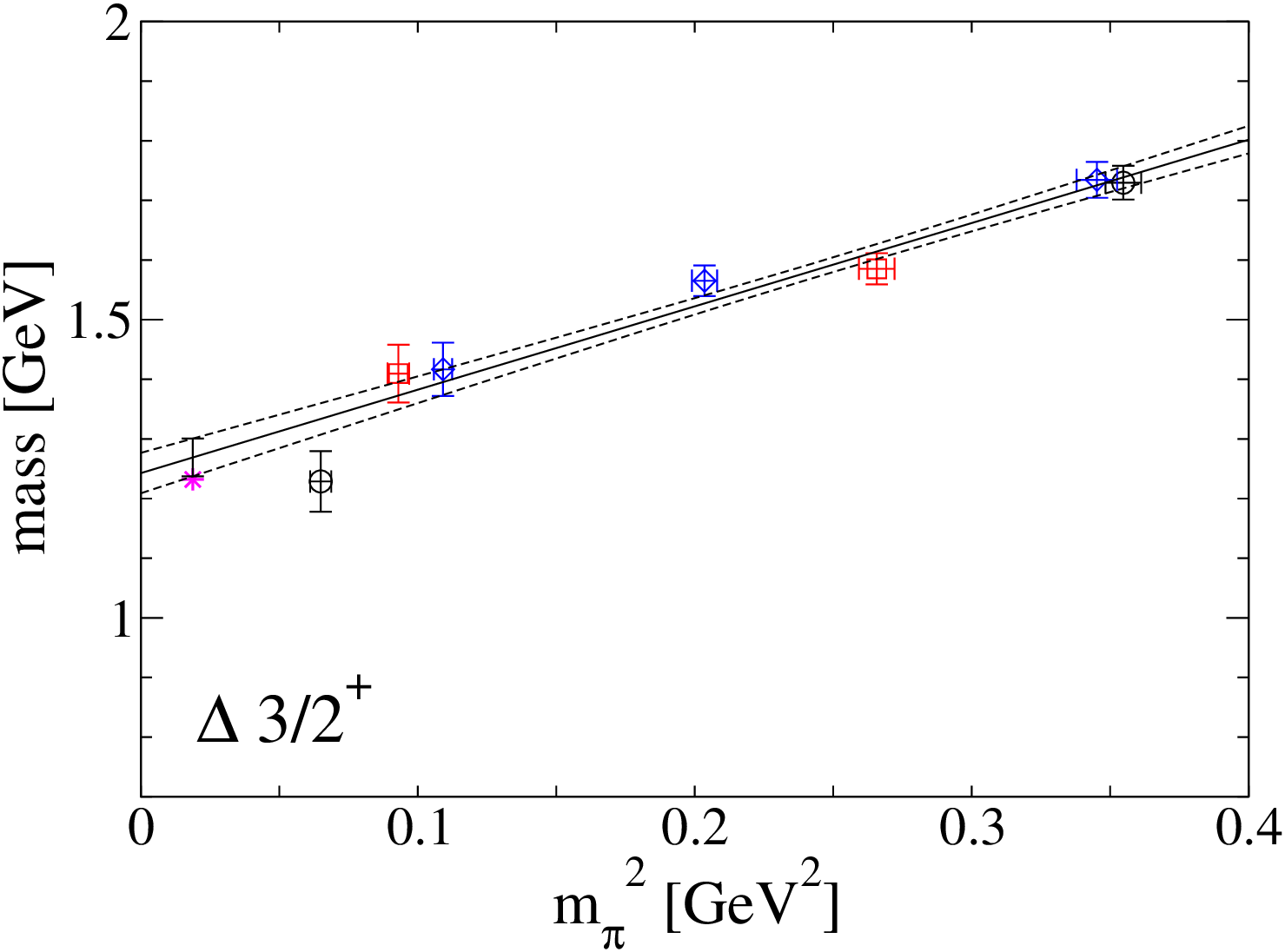}
\caption[Energy levels for the $N\,\ot^+$ and $\Delta\,\frac{3}{2}^+$ in the infinite volume limit]
{Energy levels for the nucleon spin $\ot^+$ (upper) and $\Delta$ spin $\frac{3}{2}^+$ (lower) in the infinite volume limit.
After infinite volume extrapolation ((A,5) of Fig.~\ref{fig:nucleon_1half_pospar_vol_syserr} resp.~Fig.~\ref{fig:delta_3half_pospar_vol_syserr}), we extrapolate to physical pion masses. 
We obtain $m_N$=954(16) MeV and $m_\Delta$=1268(32) MeV, which both match the experimental values within roughly 1$\sigma$.
}
\label{fig:nucleondelta_pospar_infvol}
\end{figure}

The nucleon spin $1/2^+$ ground state shows a very clean signal.
Our result for the finite box of roughly 2.2~fm  deviates significantly from experiment (see Fig.~\ref{fig:nucleon_1half}).
In order to estimate the systematic error we compare two sets of interpolators A=(1,2,9,10,19,20) and B=(3,4,10,11,19,20).
Furthermore, we consider different starting values for the fit range for the eigenvalues.
The results for the different ensembles and the corresponding infinite volume extrapolations are shown in Fig.~\ref{fig:nucleon_1half_pospar_vol_syserr}.
Note that the result for (B,7) of ensemble A66 lies outside the plotted region. 
We conclude that for small volumes late starts of the fit have to be avoided.

We find a clear dependence of the nucleon energy level on the lattice volume.
For definiteness, we choose the set of interpolators A and $t_{\text{min}}=5a$ and the corresponding infinite volume extrapolation, which is shown in Fig.~\ref{fig:nucleon_1half_pospar_vol}.
After infinite volume extrapolation of all ensembles with the extrapolation parameters 
determined from A66 and C77, 
we extrapolate to the physical pion mass, shown in Fig.~\ref{fig:nucleondelta_pospar_infvol} (upper).
Our final result is $m_N=954(16)$ MeV (error is statistical only), which agrees with the experimental $N(939)$ within 1$\sigma$.

\myparagraph{N:\;\;I(J^P)=\ot(\ot^-)}
\begin{figure}[tb]
\centering
\includegraphics[width=\mylenC,clip]{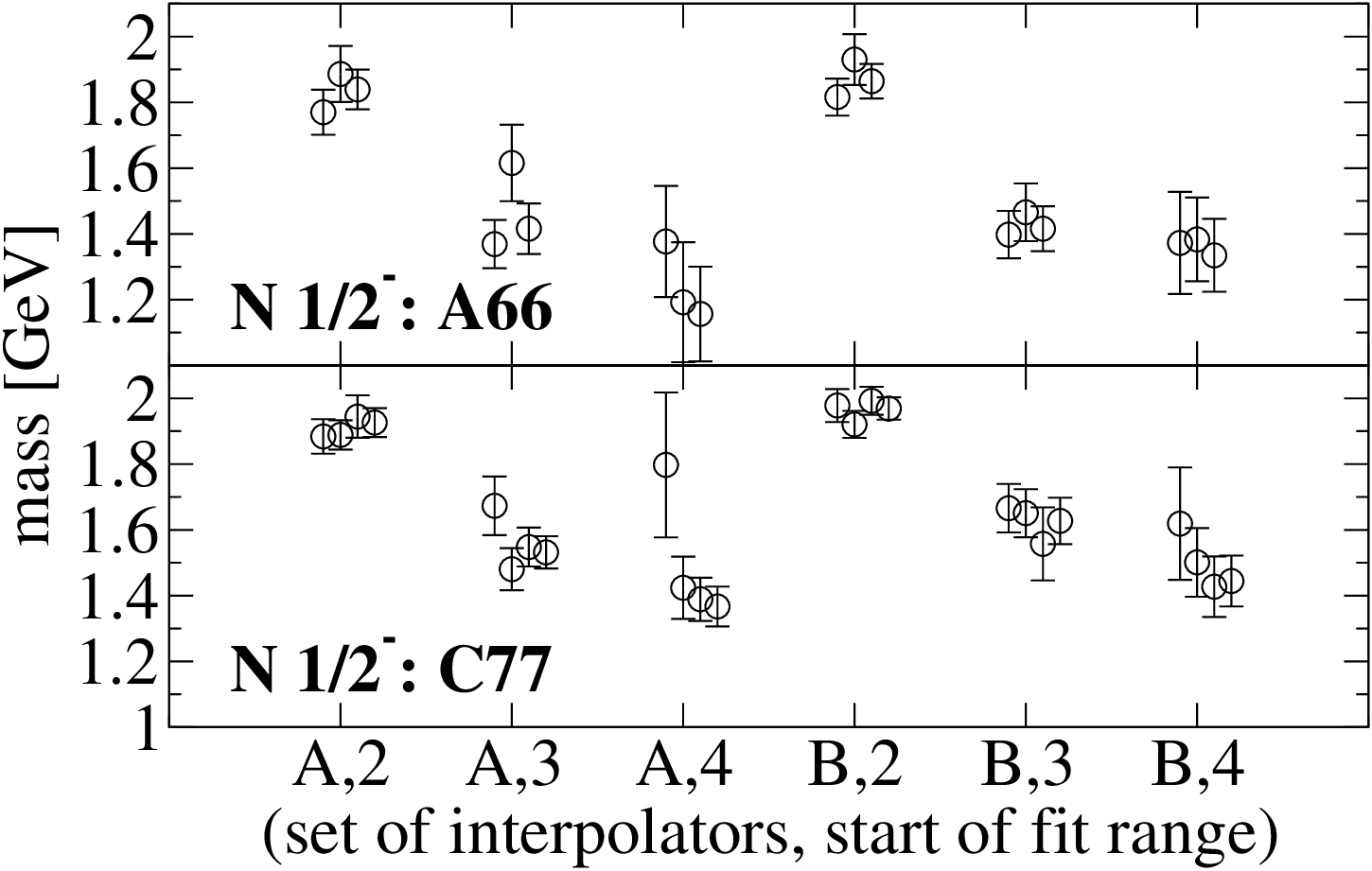}
\caption[Systematic error of the nucleon $1/2^-$ ground state mass]
{Systematic error of the nucleon spin $1/2^-$ ground state mass, analogous to Fig.~\ref{fig:nucleon_1half_pospar_vol_syserr}. 
``A'' denotes set of interpolators (5,11,17), ``B'' denotes (1,2,9,10,17,18).
For each set of interpolator and fit range, results for small to large lattices are shown from left to right, the corresponding infinite volume limit rightmost.
}
\label{fig:nucleon_1half_negpar_vol_syserr}
\end{figure}
\begin{figure}[tb]
\centering
\includegraphics[width=\mylenC,clip]{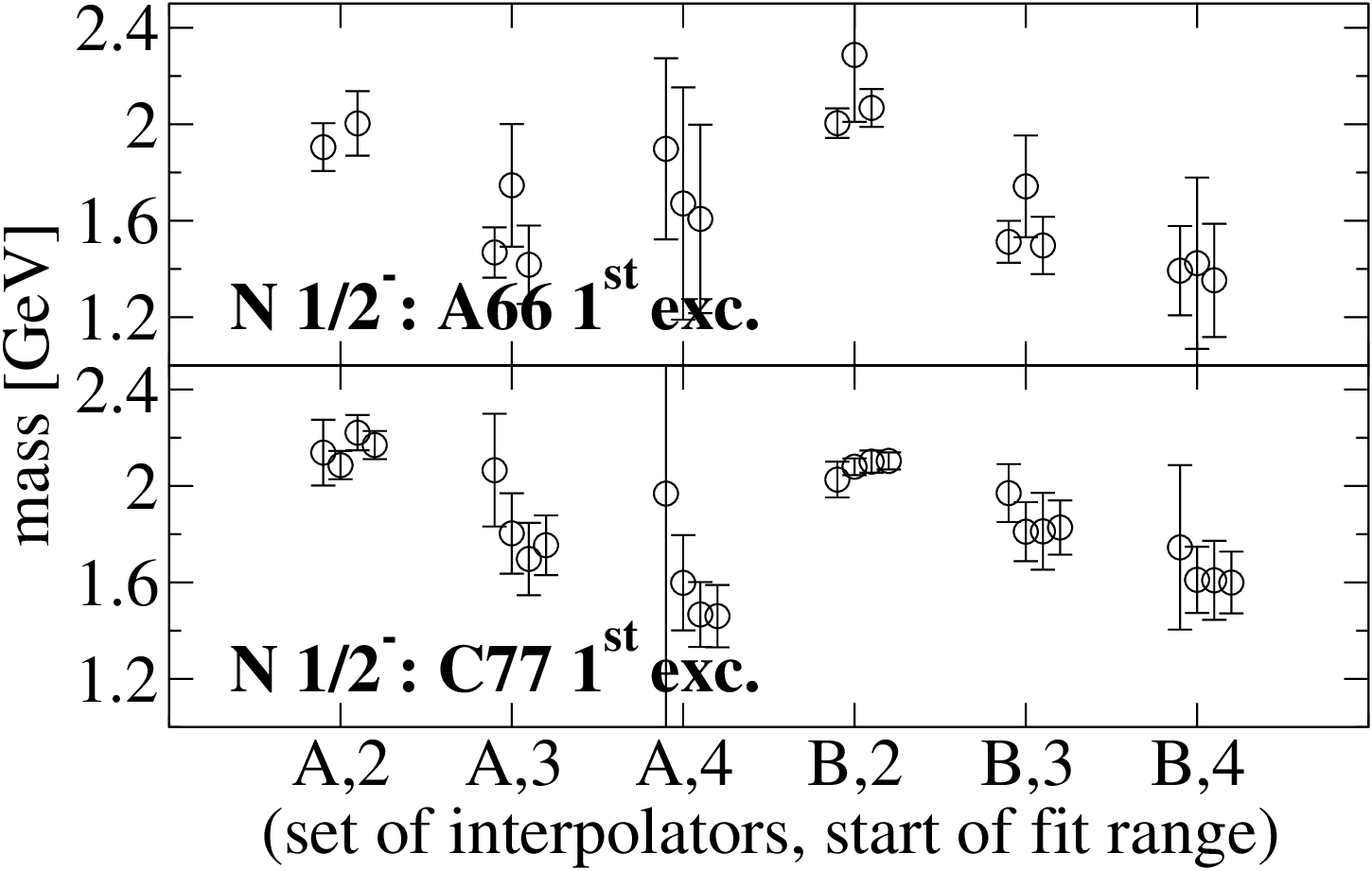}
\caption[Systematic error of the nucleon $1/2^-$ first excited mass]
{Systematic error of the nucleon spin $1/2^-$ first excited energy level, analogous to Fig.~\ref{fig:nucleon_1half_pospar_vol_syserr}. 
``A'' denotes set of interpolators (5,11,17), ``B'' denotes (1,2,9,10,17,18).
For each set of interpolator and fit range, results for small to large lattices are shown from left to right, the corresponding infinite volume limit rightmost.
}
\label{fig:nucleon_1half_negpar_1E_vol_syserr}
\end{figure}

In the nucleon spin $1/2^-$ channel we analyze the finite volume effects of the two lowest energy levels. 
Our results for the finite box of roughly 2.2 fm are a bit low compared to experiment (see Fig.~\ref{fig:nucleon_1half}).
We show results for  different volumes and infinite volume extrapolations for the ground state in Fig.~\ref{fig:nucleon_1half_negpar_vol_syserr} and for the first excitation in Fig.~\ref{fig:nucleon_1half_negpar_1E_vol_syserr}.
Note that in some cases the data suggest negative finite volume corrections to the energy level.
Such are compatible with an attractive $s$ wave scattering state $\pi N$.
However, the pattern is not systematically observed in A66 and C77, neither with nor without assuming a level crossing (with changing pion mass). 
Hence the finite volume analysis does not provide clear information on the particle content of the 
two lowest energy levels in the nucleon spin $1/2^-$ channel. 

In fact, as has been shown recently in a study which includes meson-baryon
interpolators \cite{Lang:2012db}, the spectrum should exhibit a sub-threshold
energy level in addition to two levels close the the resonance position.
Comparison of these results with the energy levels obtained here leads one to
interpret the present eigenstates as superpositions of those states.

\subsection{Delta Baryons}\label{sec:results_vol:baryons:delta}

\begin{figure}[htb]
\centering
\includegraphics[width=\mylenC,clip]{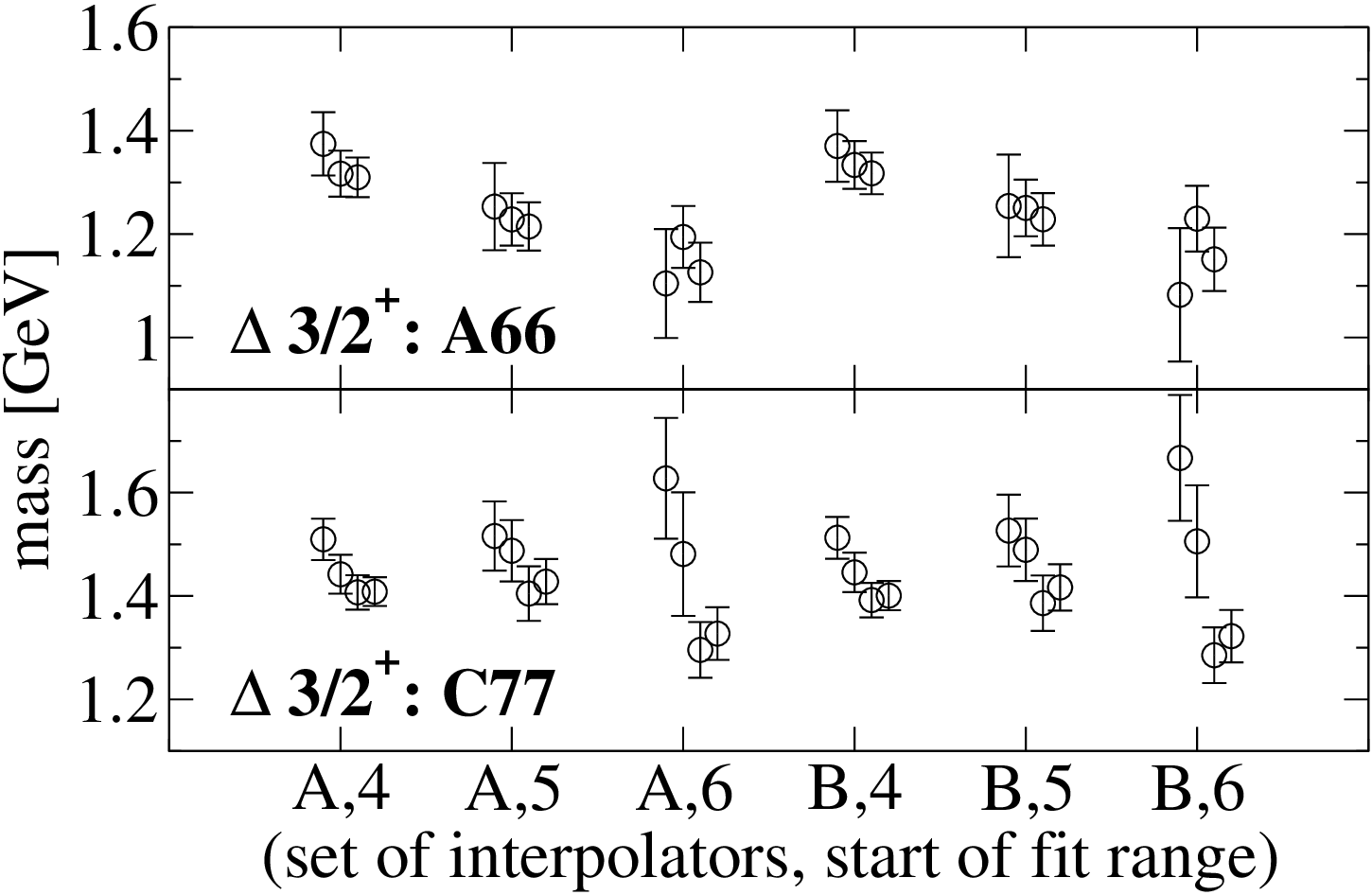}
\caption[Systematic error of the $\Delta\,3/2^+$  mass]
{Systematic error of the $\Delta$ spin $3/2^+$ mass, analogous to Fig.~\ref{fig:nucleon_1half_pospar_vol_syserr}. 
``A'' denotes set of interpolators (1,4,5), ``B'' denotes (1,5,8).
For each set of interpolator and fit range, results for small to large lattices are shown from left to right, the corresponding infinite volume limit rightmost.
}
\label{fig:delta_3half_pospar_vol_syserr}
\end{figure}

We show results and infinite volume extrapolations for different sets of interpolators and different fit ranges for the $\Delta$ spin $3/2^+$ ground state in Fig.~\ref{fig:delta_3half_pospar_vol_syserr}.
Compared to the nucleon, the fit ranges of the eigenvalues are short, correspondingly, and the results tend to fluctuate a bit more. 
The volume dependence appears to be the strongest of all observables considered. 
For definiteness, we choose the set of interpolators A and $t_{\text{min}}=5\,a$ and the corresponding infinite volume extrapolation, and note that the systematic error is of the order of the statistical error, or slightly larger. 
After infinite volume extrapolation of all ensembles, we extrapolate to the physical pion mass as shown in Fig.~\ref{fig:nucleondelta_pospar_infvol}.
Our final result is $m_\Delta=1268(32)$ MeV, 
which agrees with the experimental $\Delta(1232)$ within roughly 1$\sigma$.
We remark that the energy level  in ensemble A66 appears low compared to other ensembles.
This degrades the $\chi^2/$d.o.f.~of the chiral fit (see Table \ref{tab:chi2_vol}), but improves the comparison with experiment. 

\subsection{Omega Baryons}\label{sec:results_vol:baryons:omega}

\begin{figure}[htb]
\centering
\includegraphics[width=\mylenC,clip]{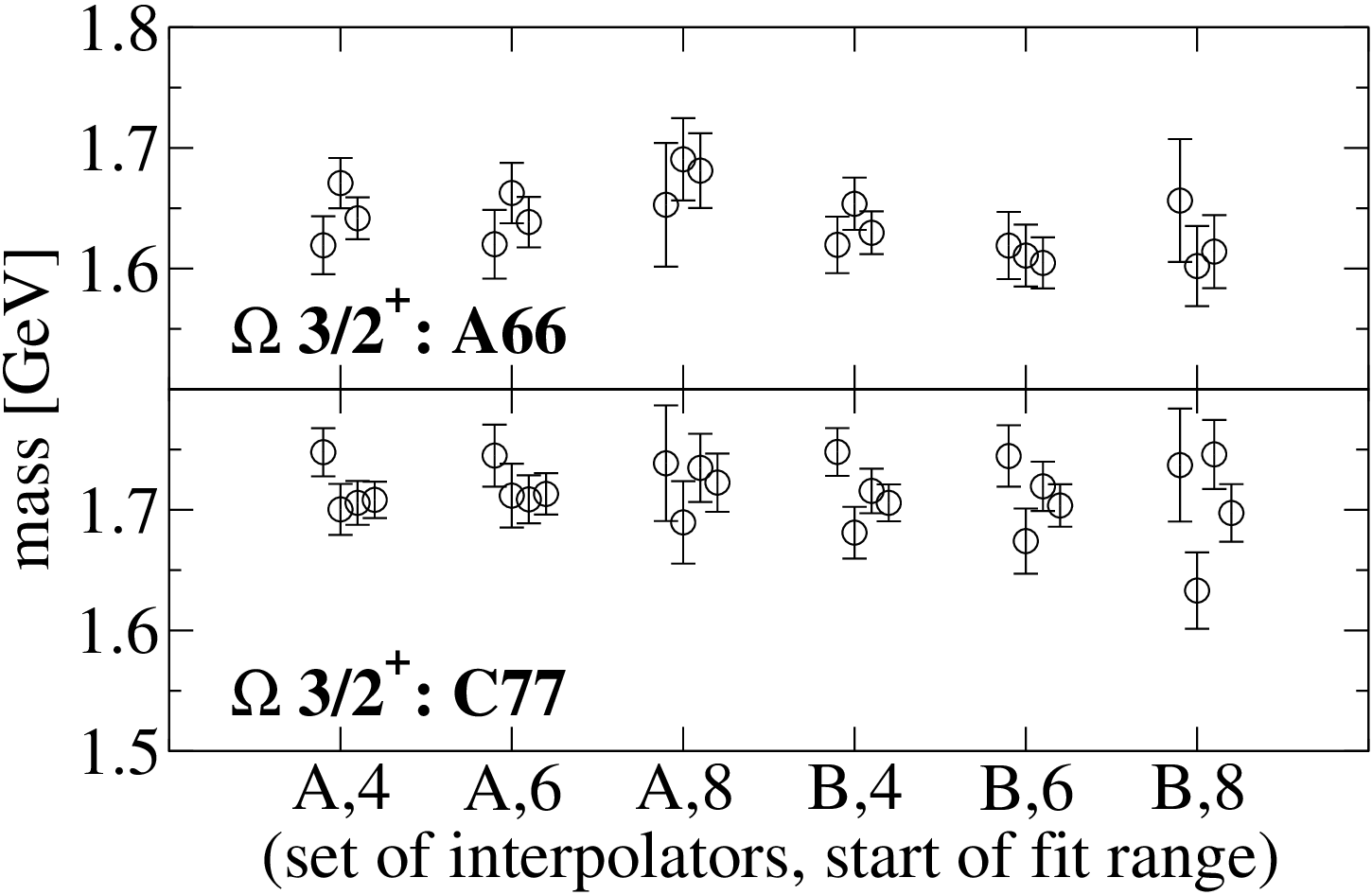}
\caption[Systematic error of the $\Omega$ $3/2^+$ mass]
{Systematic error of the $\Omega$ spin $3/2^+$ mass, analogous to Fig.~\ref{fig:nucleon_1half_pospar_vol_syserr}. 
``A'' denotes set of interpolators (1,5,8), ``B'' denotes (1,3,4).
For each set of interpolator and fit range, results for small to large lattices are shown from left to right, the corresponding infinite volume limit rightmost.
For definiteness we choose (B,4).
}
\label{fig:omega_3half_pospar_vol_syserr}
\end{figure}

\begin{figure}[htb]
\centering
\includegraphics[width=\mylenC,clip]{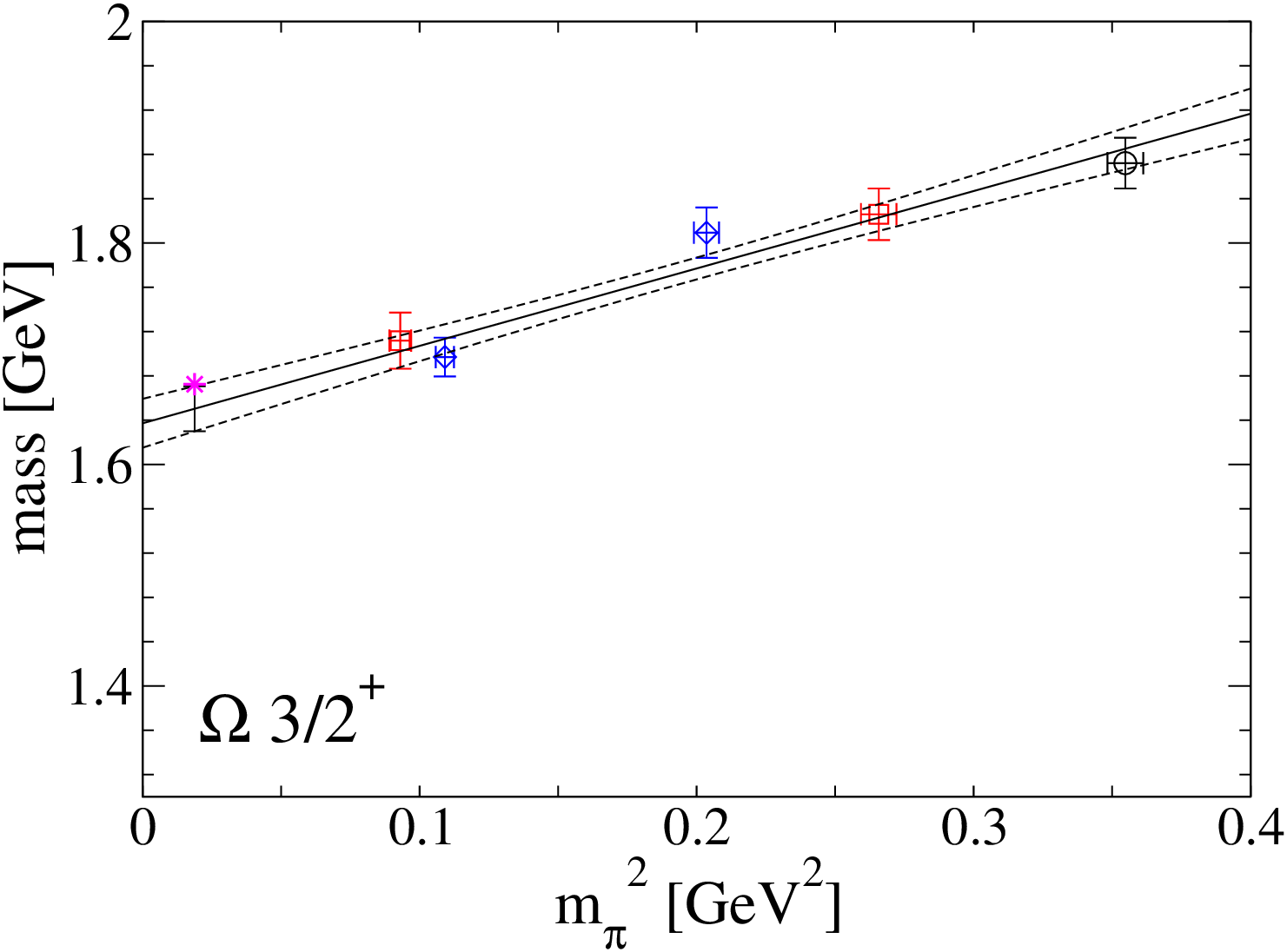}
\caption[Energy levels for $\Omega$ in the infinite volume limit]
{Energy levels for $\Omega$ spin $3/2^+$ in the infinite volume limit.
 After infinite volume extrapolation we extrapolate to physical pion masses, obtaining $m_\Omega$=1650(20) MeV.
 For discussion please refer to the text.
}
\label{fig:omega_3half_pospar_infvol}
\end{figure}

The $\Omega$ mass was used in the first place to define the strange quark mass parameter. 
We consider different sets of interpolators and fit ranges of the eigenvalues in order to estimate the corresponding systematic error. 
Figure \ref{fig:omega_3half_pospar_vol_syserr} shows some of the corresponding results. 
Here, we choose for definiteness interpolators (1,3,4) and a fit range starting from $t_{\text{min}}=4a$ 
for the ensembles with letter C and  $t_{\text{min}}=6a$ for the ensembles 
with letter A; we note that the corresponding systematic error appears to be somewhat smaller than the statistical one. 
We extrapolate the energy levels of all ensembles to infinite volume.
In the final extrapolation to physical light-quark masses (see Fig.~\ref{fig:omega_3half_pospar_infvol}), 
we omit ensemble A66 and C64, because they show a slight mistuning in the strange quark mass. 
This strategy is also pursued for other strange baryons in the infinite volume limit. 
We obtain $m_\Omega=1650(20)$ MeV, which agrees with the experimental $\Omega(1672)$ 
within 1.1 $\sigma$. The slight deviation originates from the quark mass tuning in finite volume using only partial statistics.


\begin{figure}[htb]
\centering
\includegraphics[width=\mylenC,clip]{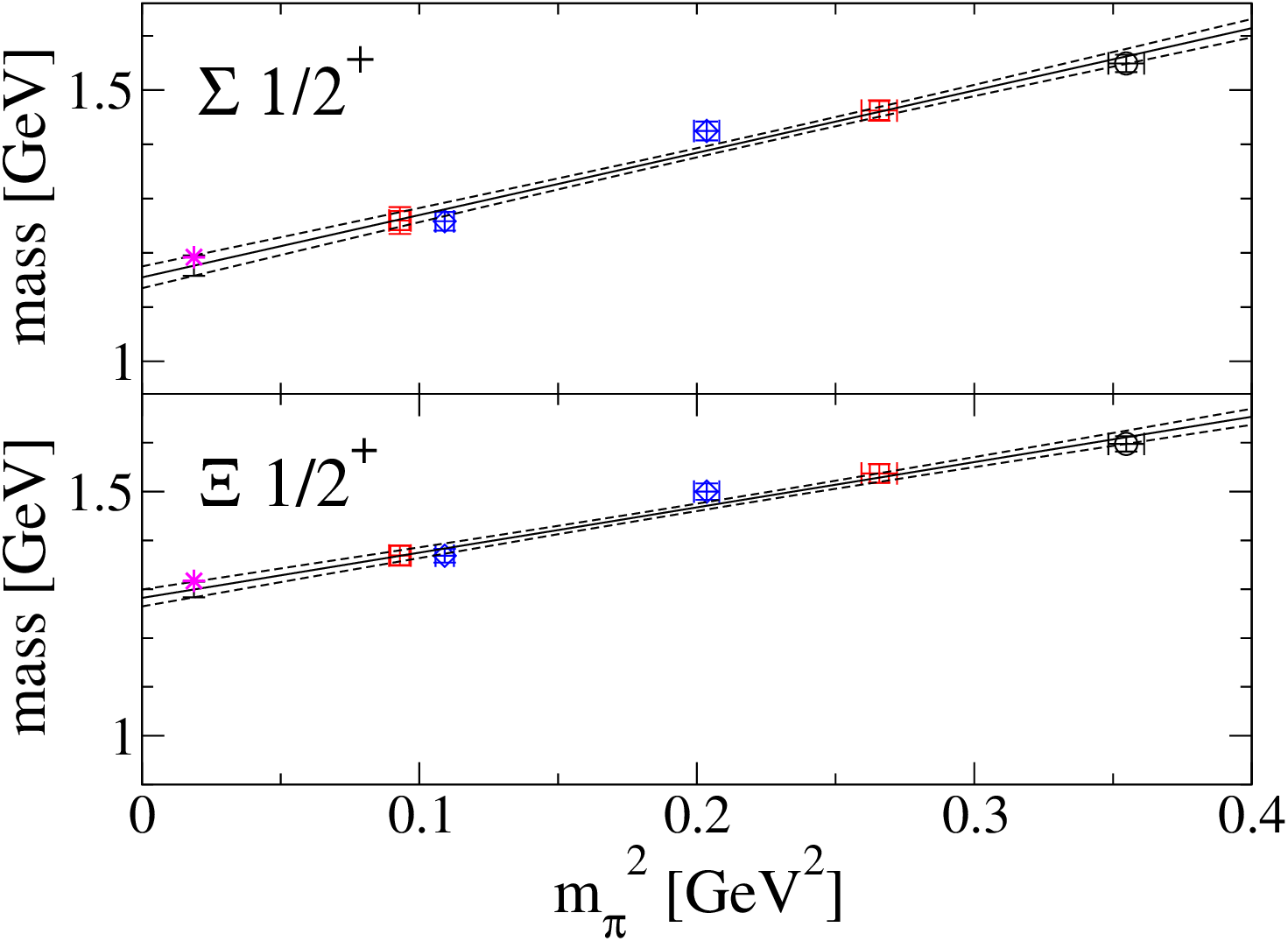}
\caption[Energy levels for $\Sigma$ $\ot^+$ (upper pane) and $\Xi$ $\ot^+$ (lower pane) in the infinite volume limit]
{Energy levels for the $\Sigma$ spin $\ot^+$ (upper pane) and the $\Xi$ spin $\ot^+$ (lower pane) ground states in the infinite volume limit.
After infinite volume extrapolation 
we extrapolate to physical pion masses. 
We obtain $m_\Sigma$=1176(19)(+07)~MeV and $m_\Xi=$1299(16)(+15)~MeV.
}
\label{fig:sigmaxi_1half_pospar_infvol}
\end{figure}

\subsection{Sigma Baryons}\label{sec:results_vol:baryons:sigma}

\begin{figure}[htb]
\centering
\includegraphics[width=\mylenC,clip]{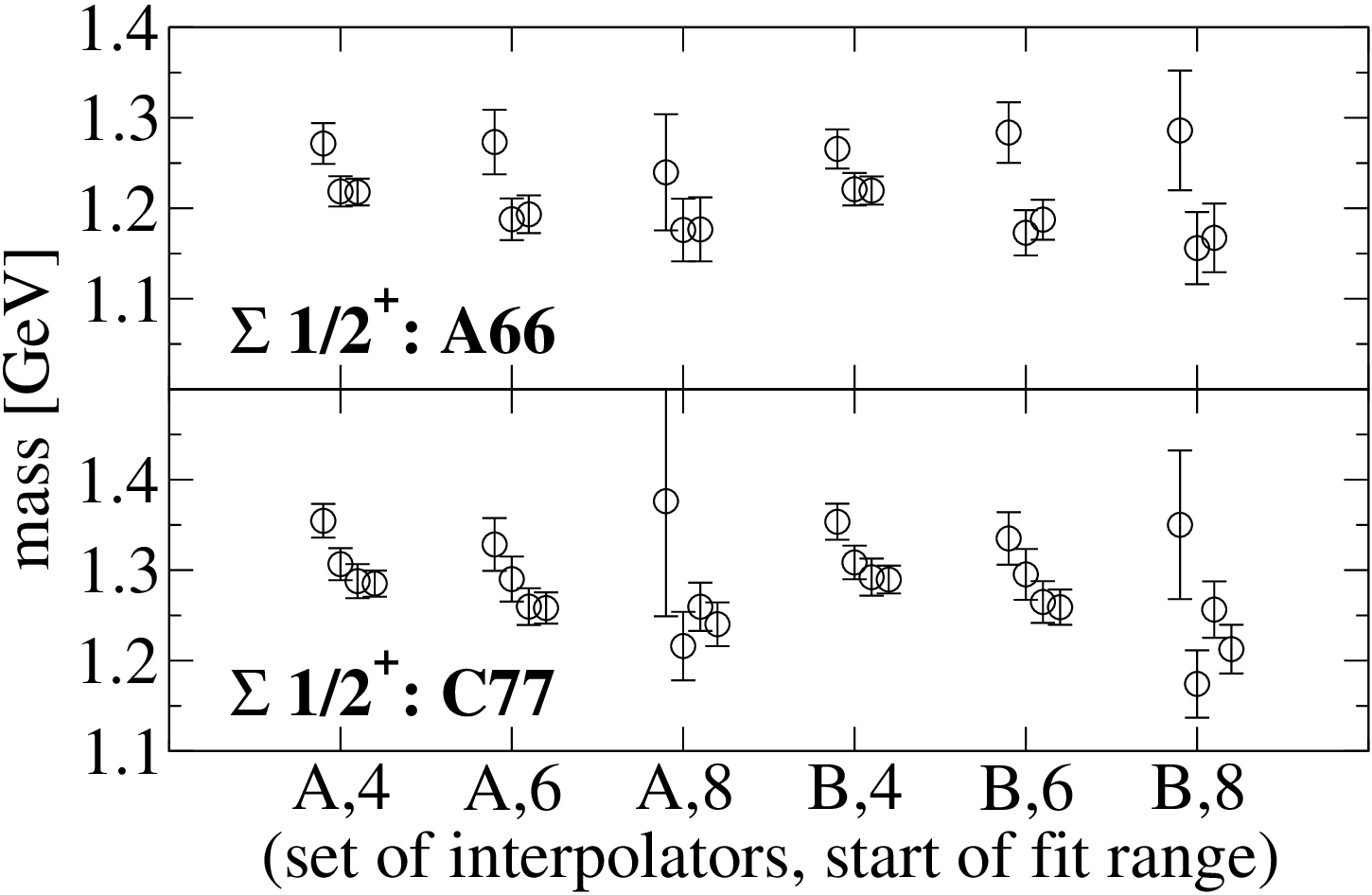}
\caption[Systematic error of the $\Sigma$ $1/2^+$ mass]
{Systematic error of the $\Sigma$ spin $\ot^+$ mass, analogous to Fig.~\ref{fig:nucleon_1half_pospar_vol_syserr}. 
``A'' denotes set of interpolators (1,2,9,10,25,26), ``B'' denotes (2,3,10,11,19,20,26,27).
For each set of interpolator and fit range, results for small to large lattices are shown from left to right, the corresponding infinite volume limit rightmost.
}
\label{fig:sigma_1half_pospar_vol_syserr}
\end{figure}

In the $\Sigma$ spin $1/2^+$ channel we apply the sets of interpolators A=(1,2,9,10,25,26) and B=(2,3,10,11,19,20,26,27) and different fit ranges to discuss the volume dependence of the ground state (see Fig.~\ref{fig:sigma_1half_pospar_vol_syserr}).
The volume dependence is found to be comparable in size to the one of the nucleon ground state energy level.
Towards larger fit ranges the results start to scatter; nevertheless, they are conclusive and the systematic error is of the order of the statistical one.
We choose interpolators A and $t_{\text{min}}=6a$, and show the results in the infinite volume limit in the upper pane of Fig.~
\ref{fig:sigmaxi_1half_pospar_infvol}.
Our final result is $m_\Sigma=1176(19)(+07)$ MeV (second error is a correction  estimate based on the slight mistuning of the strange quark mass), which is compatible with the experimental $\Sigma$ around 1193 MeV.

In the $\Sigma$ spin $3/2^+$ channel we again use interpolators 
(2,3,10,11,12). The results are shown in the upper pane of 
Fig.~\ref{fig:sigmaxi_3half_pospar_infvol}. 
Here our final result is $m_{\Sigma}=1431(25)(+07)$MeV 
which is somewhat larger than the experimental value of 1384 MeV.

\subsection{Xi Baryons}\label{sec:results_vol:baryons:xi}

\begin{figure}[htb]
\centering
\includegraphics[width=\mylenC,clip]{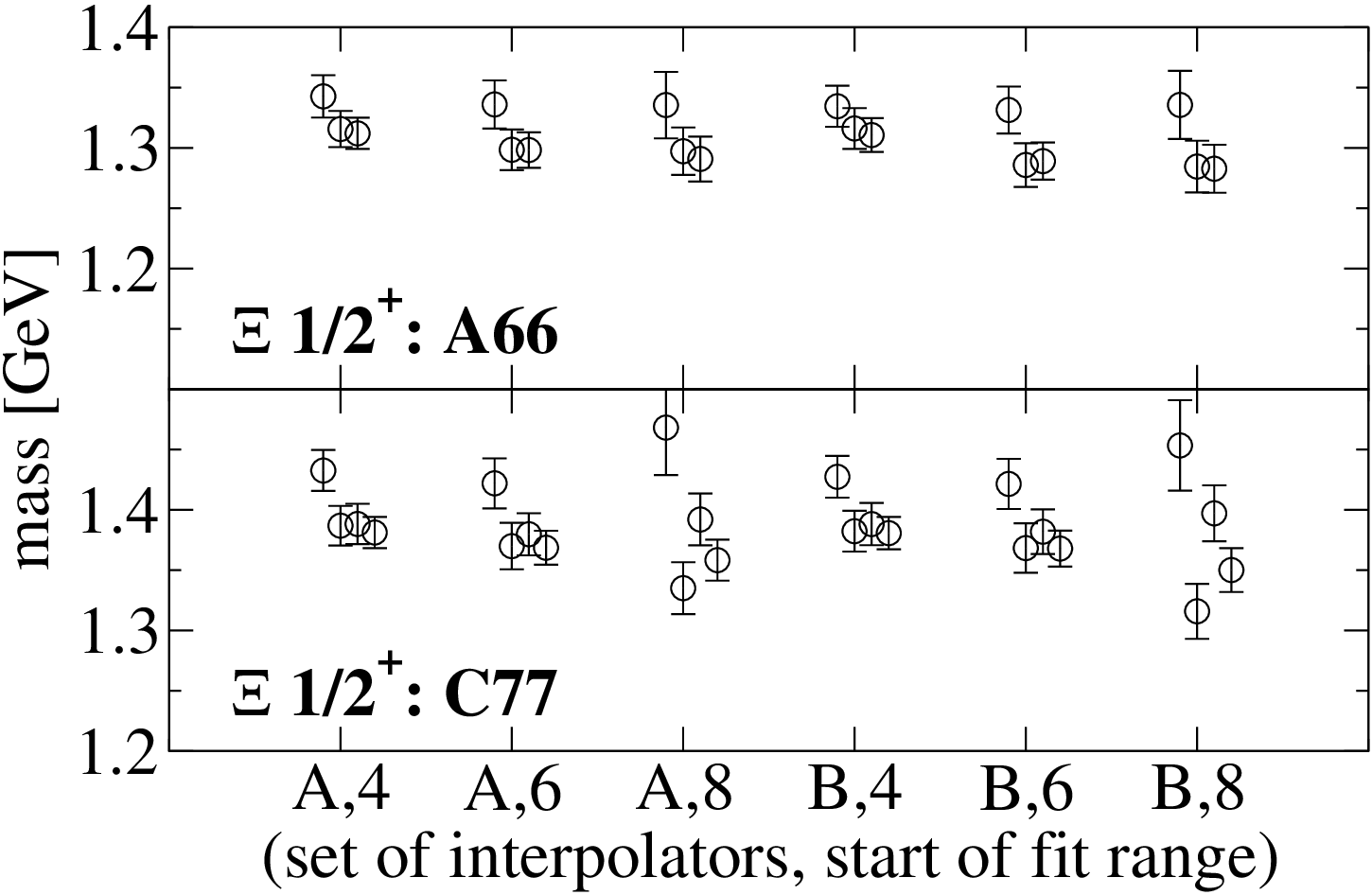}
\caption[Systematic error of the $\Xi$ $1/2^+$ mass]
{Systematic error of the $\Xi$ spin $1/2^+$mass, analogous to Fig.~\ref{fig:nucleon_1half_pospar_vol_syserr}. 
``A'' denotes set of interpolators (1,2,9,10,25,26), ``B'' denotes (2,3,10,11,19,20,26,27).
For each set of interpolator and fit range, results for small to large lattices are shown from left to right, the corresponding infinite volume limit rightmost.
}
\label{fig:xi_1half_pospar_vol_syserr}
\end{figure}

\begin{figure}[htb]
\centering
\includegraphics[width=\mylenC,clip]{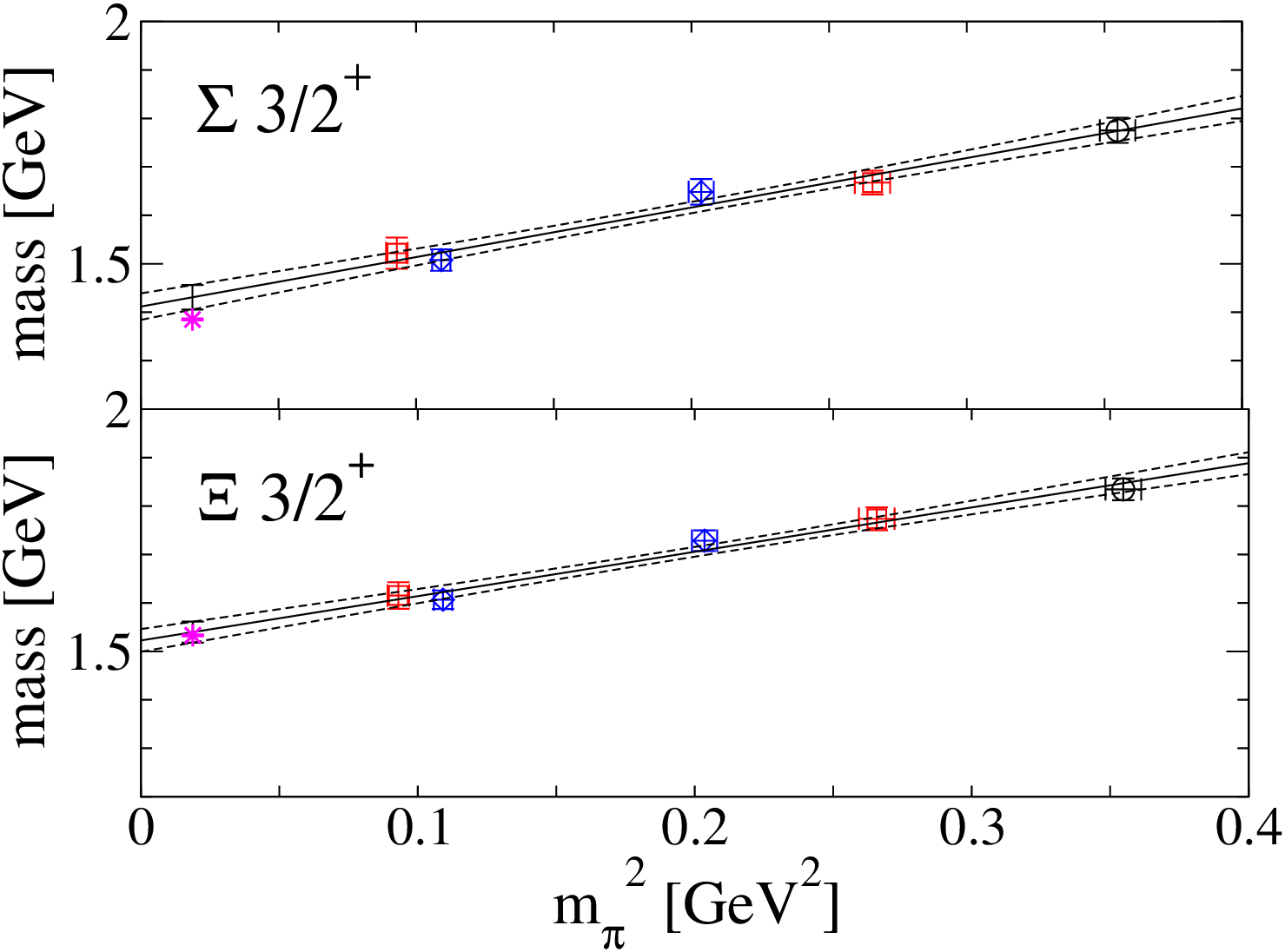}
\caption[Energy levels for $\Sigma$ $3/2^+$ (upper pane) and $\Xi$ $3/2^+$ (lower pane) in the infinite volume limit]
{Energy levels for $\Sigma$ spin $3/2^+$ (upper pane) and $\Xi$ spin $3/2^+$ (lower pane) ground states in the infinite volume limit.
After infinite volume extrapolation we extrapolate to physical pion masses.
We obtain $m_{\Sigma}$=1431(25)(+07) MeV and $m_{\Xi}$=1540(22)(+15) MeV.
}
\label{fig:sigmaxi_3half_pospar_infvol}
\end{figure}

We consider the sets of interpolators A=(1,2,9,10,25,26) and B=(2,3,10,11,19,20,26,27) and different fit ranges to discuss the volume dependence of the $\Xi$ spin $1/2^+$ ground state (see Fig.~\ref{fig:xi_1half_pospar_vol_syserr}). Again, the results are conclusive, and the systematic error is well bounded.
We choose interpolators A and $t_{\text{min}}=6a$, and show the results for infinite volume in the upper pane of Fig.~\ref{fig:sigmaxi_3half_pospar_infvol}.
Our final result is $m_\Xi=1299(16)(+15)$ MeV which is again slightly lower than the experimental $\Xi$ around 1317 MeV.

For the $\Xi$ spin $3/2^+$ ground state we use interpolators (2,3,10,11,12). The infinite volume results are shown in the right pane of Fig.~\ref{fig:sigmaxi_3half_pospar_infvol}. 
Our result is $m_{\Xi}=1540(22)(+15)$MeV which is slightly larger than the experimental value 1532MeV.

\section{Summary\label{summary}}
\begin{figure}
\noindent\includegraphics[width=\columnwidth,clip]{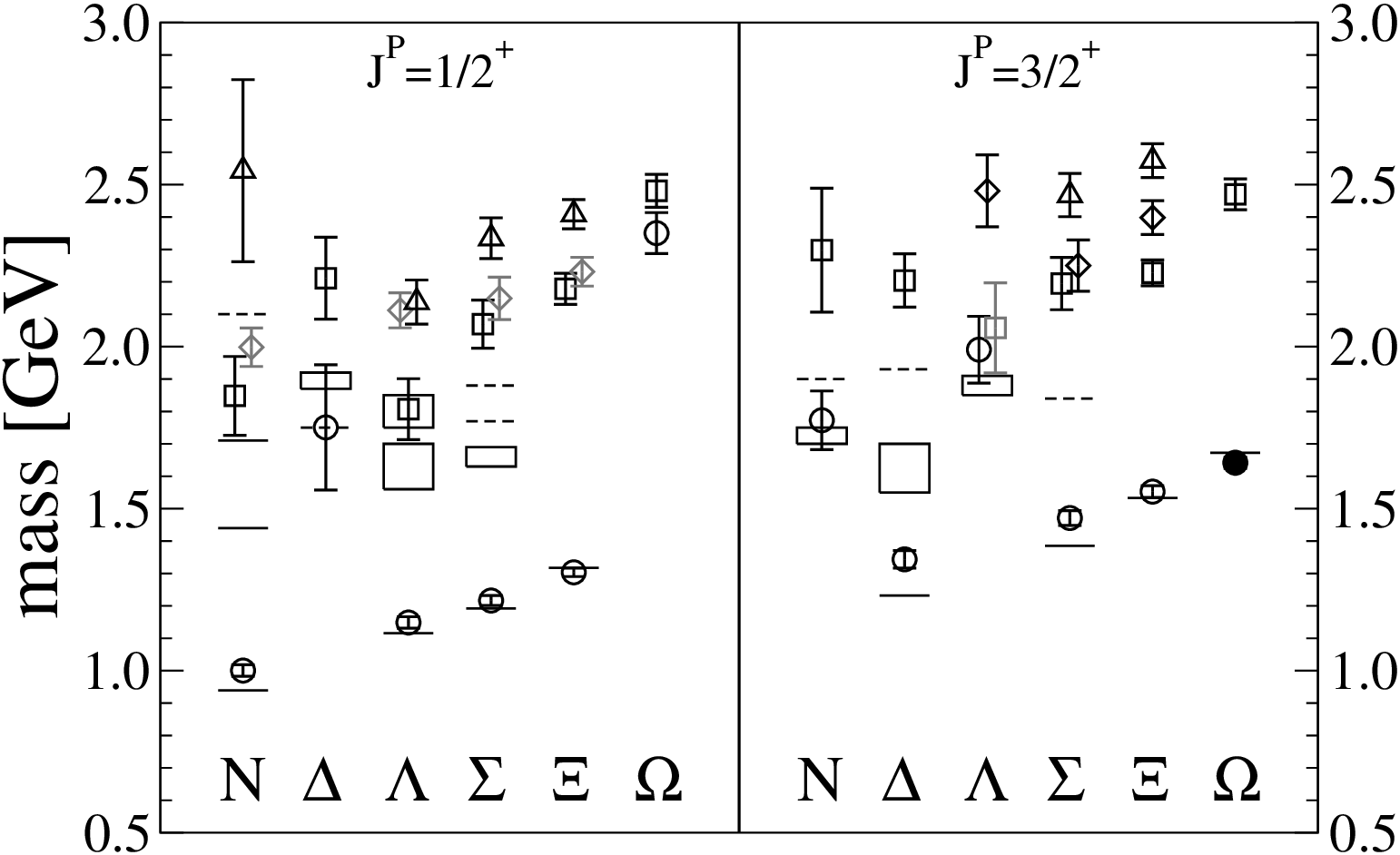}\\
\noindent\includegraphics[width=\columnwidth,clip]{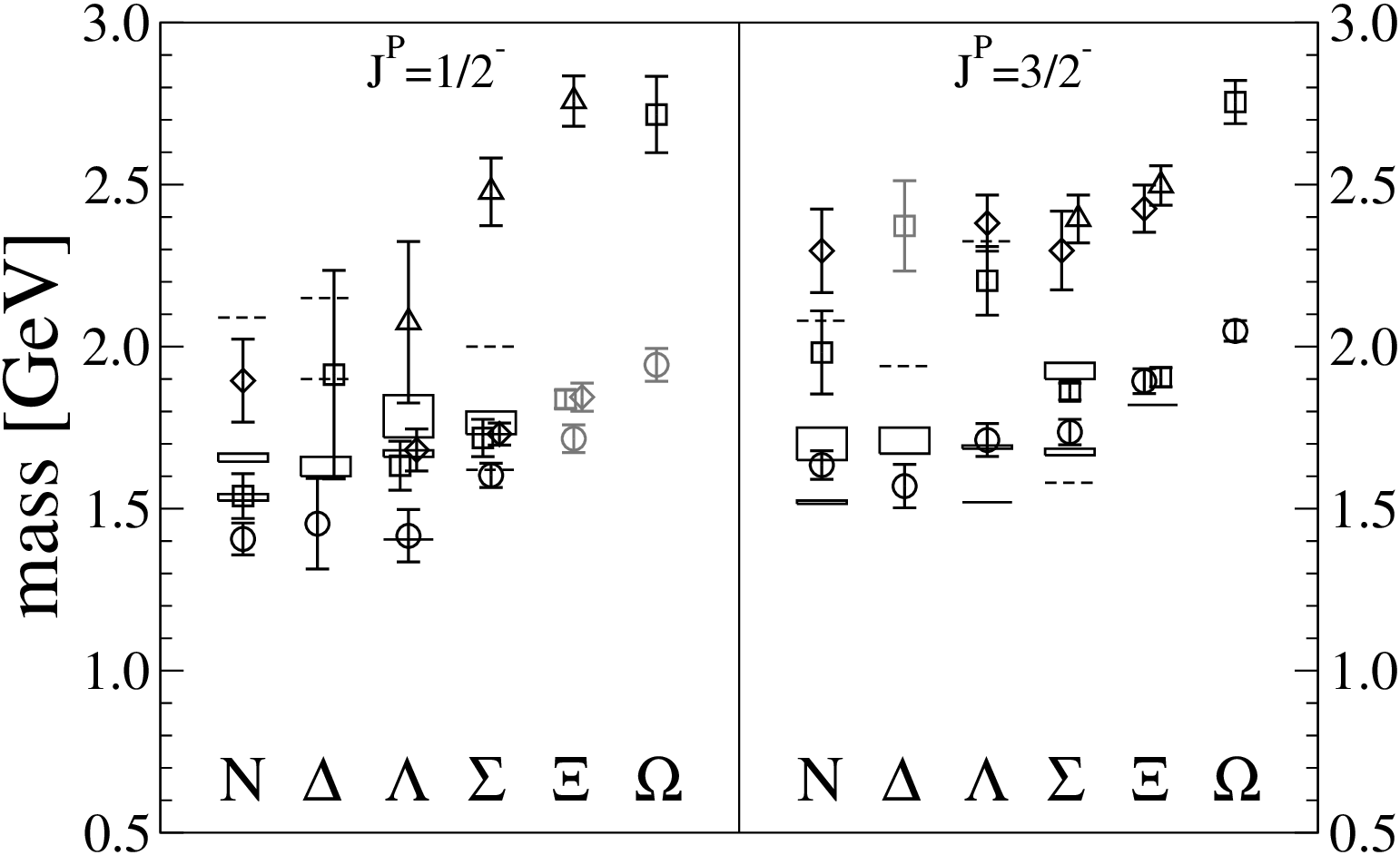}
\caption[Energy levels for baryons: Summary]{
Energy levels for positive parity (top) and negative parity baryons (bottom).
All values are obtained by chiral extrapolation linear in the pion mass squared.
Horizontal lines or boxes represent experimentally known states, 
dashed lines indicate poor evidence, according to \cite{Beringer:1900zz}.
The statistical uncertainty of our results is indicated by bands of 1$\sigma$, 
that of the experimental values by boxes of 1$\sigma$.
The strange quarks are implemented in valence approximation.
Grey symbols denote a poor $\chi^2$/d.o.f.~of the chiral fits (see Tables \ref{tab:chi2baryons_pospar} and \ref{tab:chi2baryons_negpar}).
}
\label{fig:baryons_summary}
\end{figure}

\begin{figure}[htb]
\centering
\includegraphics[width=\columnwidth,clip]{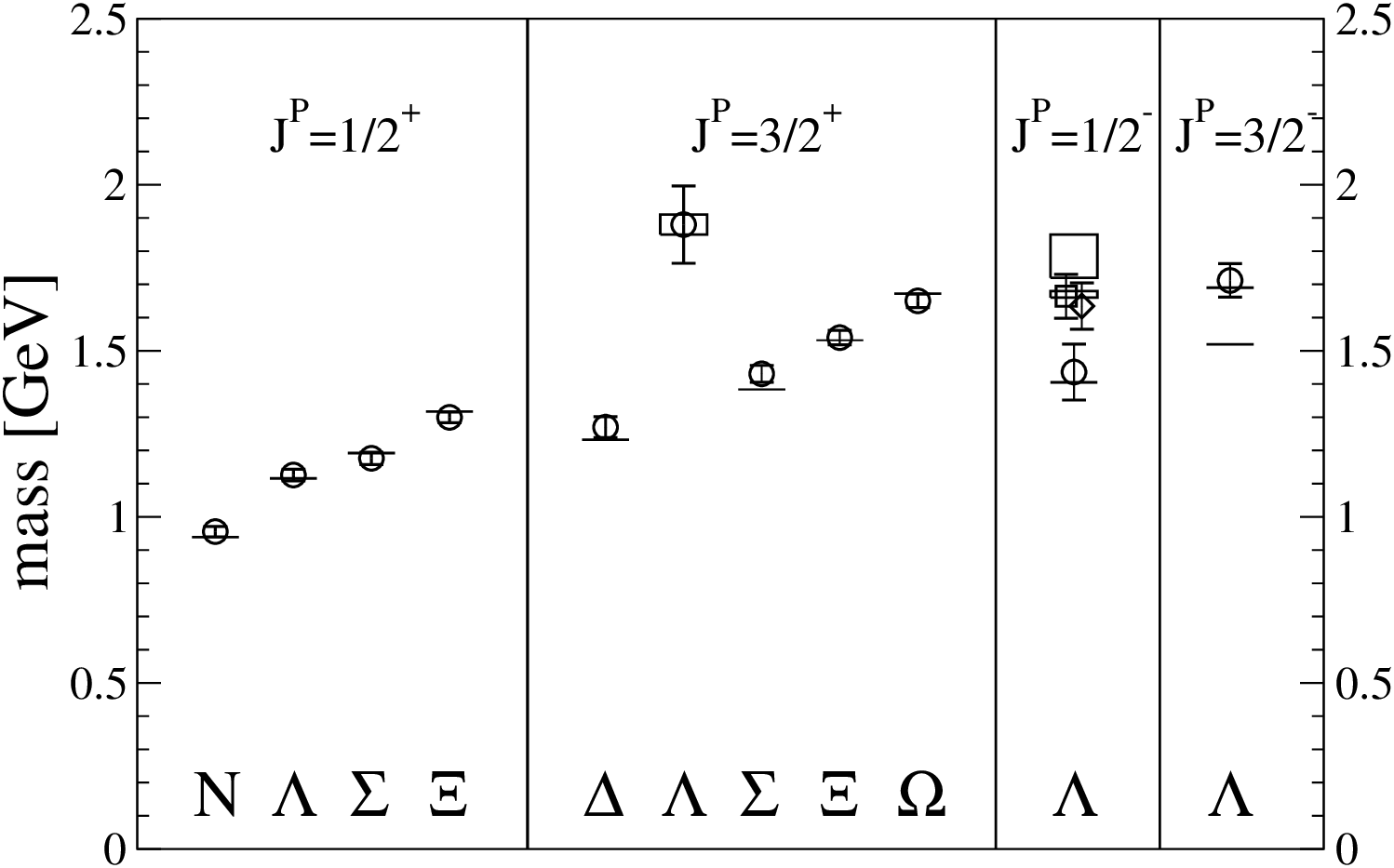}
\caption[Energy levels in infinite volume limit: Summary]
{Energy levels of baryons in the infinite volume limit at physical pion mass. 
Horizontal lines and boxes represent experimentally known states \cite{Beringer:1900zz}.
The statistical uncertainty of our results is indicated by bands of 1$\sigma$.
}
\label{fig:infvol_summary}
\end{figure}

We have derived results for the low lying energy levels in all baryon channels (spin 1/2 and 3/2, both parities) for baryons with
light and strange valence quark content. The light quarks were included as dynamical quarks in the generation of 
gauge configurations by the Hybrid Monte Carlo method. The quarks were implemented as Chirally Improved quarks, the pion masses range from
255 to 596 MeV. 

Figure \ref{fig:baryons_summary} shows our results for the extrapolation (leading order Chiral Perturbation Theory linear in $m_\pi^2$)
of the finite volume energy levels to physical pion mass.
We find good agreement of the ground state energy levels with the experimental values, where available. In some cases (e.g., in the $\Omega$ 
and the $\Xi$ sectors) our results suggest  the
existence of yet unobserved resonance states. 
We use 3-quark interpolators for the baryons throughout and find no signal for a coupling to dynamically
generated meson-baryon states in $p$- and $d$-wave channels. This is not so clear for the $s$ wave channels.
These show several energy levels close to ground states in the $\ot^-$ channels. 
In these cases there could be mixing with the $s$ wave meson-baryon sectors.

We want to mention that for all our ensembles (i.e., over the whole pion
mass range) the Gell-Mann--Okubo formula \cite{GellMann:1961ky,Okubo:1961jc} is fulfilled with high precision.
The values of the combination of the spin 1/2 positive parity octet ground state
masses obey 
\be
\left| \frac{2 M_N+2 M_\Xi - M_\Sigma-3 M_\Lambda 
}{2 M_N+2 M_\Xi + M_\Sigma+3 M_\Lambda} \right|<0.03
\ee
for all pion masses studies here.

We analyze the flavor symmetry content by identifying the singlet/octet/decuplet contributions. For the ground states agreement with the expectations from the quark model is found. In the $\ot^+$ nucleon channel the first excitation is considerably
higher than the Roper resonance and one possible interpretation is, that the physical state 
couples very weakly to our interpolators. This
may be also the case in the  $\Lambda\ot^{+}$ channel, where the first excitation is dominated by singlet interpolators 
matching the $\Lambda(1810)$ (singlet in the quark model) and the Roper-like  $\Lambda(1600)$ (octet
in the quark model) seems to be missing. 

We study the systematic errors due to the final choice of interpolator sets and fit ranges and we also perform
infinite volume extrapolations for the lowest energy levels. 
Because a slight mistuning of the strange quark mass is identified in two of the ensembles, we omit them in the final extrapolation to the physical pion mass.
Remaining small deviations are expected to stem from systematic effects which cannot be
identified uniquely given our limited dataset at a single lattice spacing with 2
dynamical quark flavors.	
In general, however, our results in the infinite volume limit compare favorably with experiment, as shown in Fig.~\ref{fig:infvol_summary}.

\acknowledgments
We would like to thank Elvira Gamiz, Christof Gattringer, Leonid Y.~Glozman, Markus Limmer, 
Willibald Plessas, Helios Sanchis-Alepuz, Mario Schr\"ock and Valentina Verduci
for valuable discussions.  The calculations have been performed on the SGI Altix
4700 of the Leibniz-Rechenzentrum Munich and on local clusters at UNI-IT at the
University of Graz. We thank these institutions for providing support.
G.P.E.~was partially supported by the MIUR--PRIN contract 20093BM-NPR. 
D.M.~acknowledges support by Natural Sciences and Engineering
Research Council of Canada (NSERC) and  G.P.E.~and A.S.~acknowledge support by
the DFG project SFB/TR-55. Fermilab is operated by Fermi Research Alliance, LLC under Contract No. De-AC02-07CH11359 with the United States Department of Energy.

\begin{appendix}
\section{Tables of Baryon Interpolators}
\label{sec:app_interpol}

All interpolators are projected to definite parity using the projector
\be
P^\pm=\frac{1}{2}(\mathds{1} \pm \gamma_t) \;.
\label{eq:parproj}
\ee
The resulting correlation matrices of positive and negative parity ($\pm$),
\be
C^\pm_{ij}(t)	= \pm Z_{ij}^\pm \E^{-tE^\pm} \pm Z_{ij}^\mp \E^{-(T-t)E^\mp} ,
\ee
are combined to the correlation matrices
\be
C(t)	= \frac{1}{2} \left( C^+(t) - C^-(T-t) \right) \; ,
\ee
which are then used in the variational method. 

All Rarita-Schwinger fields (spin 3/2 interpolators of Table \ref{tab:baryon:interpol:1}) are projected to definite spin 3/2 using the continuum formulation of the Rarita-Schwinger projector \cite{Lurie:1968}
\be
P^{3/2}_{\mu \nu} (\vec{p}) = \delta_{\mu \nu} - \frac{1}{3} \gamma_{\mu} \gamma_{\nu} - \frac{1}{3p^2} ( \gamma \cdot p \, \gamma_{\mu} p_{\nu} + p_{\mu}\gamma_{\nu} \gamma \cdot p) \;.
\label{eq:RaritaSchwinger}
\ee

The baryon interpolators  used in this work are detailed in Tables
\ref{tab:baryon:interpol:1}, \ref{tab:baryon:interpol:2} and
\ref{tab:baryon:interpol:3}. Table \ref{tab:baryon:interpol:1} shows the flavor
structure for all interpolators.  For the spin 1/2 channels of the nucleon, $\Sigma$,
$\Xi$ and $\Lambda$, we use the three different Dirac structures 
$\chi^{(i)}=(\Gamma_1^{(i)},\Gamma_2^{(i)}),(i=1,2,3)$, listed in Table
\ref{tab:baryon:interpol:2}. Details about the quark smearings in the
interpolators are found in Table \ref{tab:baryon:interpol:3}. The name convention
of all baryon interpolators is determined by Tables \ref{tab:baryon:interpol:2}
and \ref{tab:baryon:interpol:3}. In the $\Lambda$ channels, singlet and octet
interpolators are collected in one set.  We assign to the first octet
interpolator the number after the last singlet interpolator, and continue to
count for the remaining octet interpolators.   In the $\Sigma$ and $\Xi$
channels, the same holds for octet and decuplet interpolators. 

In the continuum, the actual number of independent fields is reduced by Fierz identities. In
particular, there are no non-vanishing point-like interpolators for $\Delta(\ot)$
and singlet $\Lambda(\frac{3}{2})$. However, using differently smeared quarks in
the construction of interpolators, we do access independent
information and find good signals for the singlet  $\Lambda(\frac{3}{2})$
propagation. 

\begin{table*}[!]
\begin{ruledtabular}
\begin{tabular}{cccc}
Spin		& Flavor channel 	& Name				& Interpolator \\
\hline
$\frac{1}{2}$	& Nucleon 		& $N_{1/2}^{(i)}$		& $\epsilon_{abc}\, \Gamma_1^{(i)}\, u_a\, \big( u_b^T\, \Gamma_2^{(i)}\, d_c - d_b^T\, \Gamma_2^{(i)}\, u_c \big)	$ \\
$\frac{1}{2}$	& Delta			& $\Delta_{1/2}$		& $\epsilon_{abc}\, \gamma_i \gamma_5 u_a\, \big(u_b^T\, C\, \gamma_i\, u_c \big) 					$ \\
$\frac{1}{2}$	& Sigma	octet		& $\Sigma_{1/2}^{(8,i)}$	& $\epsilon_{abc}\, \Gamma_1^{(i)}\, u_a\, \big( u_b^T\, \Gamma_2^{(i)}\, s_c - s_b^T\, \Gamma_2^{(i)}\, u_c \big) 	$ \\
$\frac{1}{2}$	& Sigma	decuplet	& $\Sigma_{1/2}^{(10,i)}$	& $\epsilon_{abc}\, \gamma_i \gamma_5 u_a\, \big(u_b^T\, C\, \gamma_i\, s_c - s_b^T\, C\, \gamma_i\, u_c \big) 		$ \\
$\frac{1}{2}$	& Xi octet		& $\Xi_{1/2}^{(8,i)}$		& $\epsilon_{abc}\, \Gamma_1^{(i)}\, s_a\, \big( s_b^T\, \Gamma_2^{(i)}\, u_c - u_b^T\, \Gamma_2^{(i)}\, s_c \big) 	$ \\
$\frac{1}{2}$	& Xi decuplet		& $\Xi_{1/2}^{(10,i)}$		& $\epsilon_{abc}\, \gamma_i \gamma_5 s_a\, \big(s_b^T\, C\, \gamma_i\, u_c - u_b^T\, C\, \Gamma_i\, s_c \big) 		$ \\
$\frac{1}{2}$	& Lambda singlet	& $\Lambda_{1/2}^{(1,i)}$	& $\epsilon_{abc} \Gamma^{(i)}_1 u_a ( d_b^T \Gamma^{(i)}_2 s_c - s_b^T \Gamma^{(i)}_2 d_c) 				$ \\ 
		&			& 				& $\,+ \, \mbox{cyclic permutations of}\;  u, d, s  									$ \\
$\frac{1}{2}$	& Lambda octet		& $\Lambda_{1/2}^{(8,i)}$	& $\epsilon_{abc} \Big[ \Gamma^{(i)}_1 s_a ( u_b^T  \Gamma^{(i)}_2 d_c - d_b^T  \Gamma^{(i)}_2 u_c ) 			$ \\
		&			& 				& $\, + \; \Gamma^{(i)}_1 u_a ( s_b^T  \Gamma^{(i)}_2 d_c) - \Gamma^{(i)}_1 d_a ( s_b^T  \Gamma^{(i)}_2 u_c) \Big]	$ \\
$\frac{1}{2}$	& Omega			& $\Omega_{1/2}$		&  $\epsilon_{abc}\, \gamma_i \gamma_5 s_a\, \big(s_b^T\, C\, \gamma_i\, s_c \big) 					$ \\
\hline
$\frac{3}{2}$	& Nucleon 		& $N_{3/2}^{(i)}$		& $\epsilon_{abc}\, \gamma_5 \, u_a\, \big( u_b^T\, C \gamma_5 \gamma_i\, d_c - d_b^T\, C \gamma_5 \gamma_i\, u_c \big)	$ \\
$\frac{3}{2}$	& Delta			& $\Delta_{3/2}^{(i)}$		& $\epsilon_{abc}\, u_a\, \big(u_b^T\, C\, \gamma_i\, u_c \big) 							$ \\
$\frac{3}{2}$	& Sigma	octet		& $\Sigma_{3/2}^{(8,i)}$	& $\epsilon_{abc}\, \gamma_5 \, u_a\, \big( u_b^T\, C \gamma_5 \gamma_i\, s_c - s_b^T\, C \gamma_5 \gamma_i\, u_c \big)	$ \\
$\frac{3}{2}$	& Sigma decuplet	& $\Sigma_{3/2}^{(10,i)}$	& $\epsilon_{abc}\, u_a\, \big( u_b^T\, C \gamma_i\, s_c - s_b^T\, C \gamma_i\, u_c \big)	$ \\
$\frac{3}{2}$	& Xi octet		& $\Xi_{3/2}^{(8,i)}$		& $\epsilon_{abc}\, \gamma_5 \, s_a\, \big( s_b^T\, C \gamma_5 \gamma_i\, u_c - u_b^T\, C \gamma_5 \gamma_i\, s_c \big)	$ \\
$\frac{3}{2}$	& Xi decuplet		& $\Xi_{3/2}^{(10,i)}$		& $\epsilon_{abc}\, s_a\, \big( s_b^T\, C \gamma_i\, u_c - u_b^T\, C \gamma_i\, s_c \big)	$ \\
$\frac{3}{2}$	& Lambda singlet	& $\Lambda_{3/2}^{(1,i)}$	& $\epsilon_{abc} \gamma_5 u_a ( d_b^T C \gamma_5 \gamma_i s_c - s_b^T C \gamma_5 \gamma_i d_c) 			$ \\ 
		&			& 				& $\,+ \, \mbox{cyclic permutations of}\;  u, d, s  									$ \\
$\frac{3}{2}$	& Lambda octet		& $\Lambda_{3/2}^{(8,i)}$	& $\epsilon_{abc} \Big[ \gamma_5 s_a ( u_b^T C \gamma_5 \gamma_i d_c - d_b^T  C \gamma_5 \gamma_i u_c ) 		$ \\
		&			& 				& $\, + \; \gamma_5 u_a ( s_b^T  C \gamma_5 \gamma_i d_c) - \gamma_5 d_a ( s_b^T  C \gamma_5 \gamma_i u_c\Big]		$ \\
$\frac{3}{2}$	& Omega			& $\Omega_{3/2}^{(i)}$		& $\epsilon_{abc}\, s_a\, \big(s_b^T\, C\, \gamma_i\, s_c \big) 							$ \\
\end{tabular}
\end{ruledtabular}
\caption[Baryon interpolators: Flavor structure]{
Baryon interpolators: Flavor structure.
The possible choices for the Dirac matrices $\Gamma_{1,2}^{(i)}$ in the spin 1/2 channels are listed in Table \ref{tab:baryon:interpol:1}. 
All interpolators are projected to definite parity according to Eq.~\eq{eq:parproj}.
All spin 3/2 interpolators include the Rarita-Schwinger projector, according to Eq.~\eq{eq:RaritaSchwinger}, which is suppressed for clarity in the table.
$C$ denotes the charge conjugation matrix, $\gamma_i$ the spatial Dirac matrices and $\gamma_t$ the Dirac matrix in time direction.
Spin 1/2 and spin 3/2 channels are separated by a solid line.
Summation convention applies for repeated indices, and in the case of spin 3/2 observables, the open Lorentz index (after spin projection) is summed after taking the expectation value of correlation functions.
}
\label{tab:baryon:interpol:1}
\end{table*}

\begin{table}[!]
\begin{ruledtabular}
\begin{tabular}{ccccc}
i	& $\Gamma^{(i)}_1$	& $\Gamma^{(i)}_2$	& \multicolumn{2}{c}{Numbering of associated interpolators} \\
	&			&			& $N_{1/2},\Lambda_{1/2}^1,\Sigma_{1/2}^8,\Xi_{1/2}^8$	& $\Lambda_{1/2}^8,\Sigma_{1/2}^{10},\Xi_{1/2}^{10}$	\\
\hline
$1$ 	& $\mathds{1}$  	& $C\gamma_5$         	& 1-8		& 25-32	\\
$2$ 	& $\gamma_5$    	& $C$                 	& 9-16		& 33-40	\\
$3$ 	& $i\mathds{1}$ 	& $C\gamma_t\gamma_5$ 	& 17-24		& 41-48	\\
\end{tabular}
\end{ruledtabular}
\caption[Baryon interpolators: Dirac structure]{
Baryon interpolators: Dirac structures used for the spin 1/2 nucleon, $\Lambda$, $\Sigma$ and $\Xi$  interpolators, according to Table \ref{tab:baryon:interpol:1}.
The naming convention (numbering) for associated interpolators in the different channels is given as well.
The subscripts denote the spin, the superscripts the flavor irreducible representation. 
}
\label{tab:baryon:interpol:2}
\end{table}

\begin{table}[!]
\begin{ruledtabular}
\begin{tabular}{c|cccc}
quark		& \multicolumn{4}{c}{Numbering of associated interpolators} \\
smearing	& ~$\Delta_{1/2},\Delta_{3/2}$~	&$\Lambda_{3/2}^8,$			& ~$N_{1/2},\Lambda_{1/2}^1,$~		& $\Lambda_{1/2}^8,$	\\
& ~$\Omega_{1/2},\Omega_{3/2},$~			&$\Sigma_{3/2}^{10}$,		& $\Sigma_{1/2}^8,\Xi_{1/2}^8$		& ~$\Sigma_{1/2}^{10},\Xi_{1/2}^{10}$		\\
		& $N_{3/2},\Lambda_{3/2}^1$		&$\Xi_{3/2}^{10}$	& \\ 
& $\Sigma_{3/2}^8,\Xi_{3/2}^8$		&&& \\
\hline
(nn)n		& 1								& ~9					& 1,9,17				& 25,33,41	\\
(nn)w		& 2								& 10					& 2,10,18				& 26,34,42	\\
(nw)n		& 3 								& 11					& 3,11,19				& 27,35,43	\\
(nw)w		& 4								& 12					& 4,12,20				& 28,36,44	\\
(wn)n		& 5								& 13					& 5,13,21				& 29,37,45	\\
(wn)w		& 6								& 14					& 6,14,22				& 30,38,46	\\
(ww)n		& 7								& 15					& 7,15,23				& 31,39,47	\\
(ww)w		& 8								& 16					& 8,16,24				& 32,40,48	\\
\end{tabular}
\end{ruledtabular}
\caption[Baryon interpolators: Quark smearing types]{
Baryon interpolators: Quark smearing types and naming convention for the interpolators in the different channels.
The subscripts denotes the spin, the superscripts the flavor irreducible representation. 
The brackets in the first row symbolize the diquark part. 
Due to Fierz identities, some of the interpolators may be linearly dependent. 
}
\label{tab:baryon:interpol:3}
\end{table}

\section{Tables of Energy Levels and $\chi^2$}\label{sec:app_chisq}

We give the results of our extrapolation (linear in $m_\pi^2$) to
the physical pion mass together with the associated value of 
$\chi^2$/d.o.f.~in Tables~\ref{tab:chi2baryons_pospar} to \ref{tab:chi2_vol}.

\begin{table}
\begin{center}
\begin{tabular}{lcc}
\hline
\hline
Baryon: $I(J^P)$	& Energy level [MeV]	& $\chi^2$/d.o.f. 	\\
\hline
$N:\,1/2(1/2^+)$	& 1000(18)		& 2.16/5		\\
$N:\,1/2(1/2^+)$	& 1848(120)		& 3.61/5		\\
$N:\,1/2(1/2^+)$	& 1998(59)		& 18.31/5		\\
$N:\,1/2(1/2^+)$	& 2543(280)		& 1.96/3		\\
$\Delta:\,3/2(1/2^+)$	& 1751(190)		& 1.58/5		\\
$\Delta:\,3/2(1/2^+)$	& 2211(126)		& 1.15/5		\\
$\Lambda:\,0(1/2^+)$	& 1149(18)		& 1.89/3		\\ 
$\Lambda:\,0(1/2^+)$	& 1807(94)		& 4.63/5		\\
$\Lambda:\,0(1/2^+)$	& 2112(54)		& 20.27/5		\\
$\Lambda:\,0(1/2^+)$	& 2137(68)		& 1.50/5		\\
$\Sigma:\,1(1/2^+)$	& 1216(15)		& 6.94/5		\\
$\Sigma:\,1(1/2^+)$	& 2069(74)		& 3.41/5		\\
$\Sigma:\,1(1/2^+)$	& 2149(66)		& 20.37/5		\\
$\Sigma:\,1(1/2^+)$	& 2335(63)		& 2.09/5		\\
$\Xi:\,1/2(1/2^+)$	& 1303(13)		& 8.31/5		\\
$\Xi:\,1/2(1/2^+)$	& 2178(48)		& 7.51/5		\\
$\Xi:\,1/2(1/2^+)$	& 2231(44)		& 26.53/5		\\
$\Xi:\,1/2(1/2^+)$	& 2408(45)		& 10.37/5		\\
$\Omega:\,0(1/2^+)$	& 2350(63)		& 4.14/5		\\
$\Omega:\,0(1/2^+)$	& 2481(51)		& 4.35/5		\\
\hline
$N:\,1/2(3/2^+)$	& 1773(91)		& 8.35/5		\\
$N:\,1/2(3/2^+)$	& 2298(191)		& 3.79/5		\\
$\Delta:\,3/2(3/2^+)$	& 1344(27)		& 6.13/5		\\
$\Delta:\,3/2(3/2^+)$	& 2204(82)		& 6.23/5		\\
$\Lambda:\,0(3/2^+)$	& 1991(103)		& 3.56/3		\\
$\Lambda:\,0(3/2^+)$	& 2058(139)		& 23.04/5		\\
$\Lambda:\,0(3/2^+)$	& 2481(111)		& 4.26/5		\\
$\Sigma:\,1(3/2^+)$	& 1471(23)		& 2.52/5		\\
$\Sigma:\,1(3/2^+)$	& 2194(81)		& 4.78/5		\\
$\Sigma:\,1(3/2^+)$	& 2250(79)		& 7.05/5		\\
$\Sigma:\,1(3/2^+)$	& 2468(67)		& 4.22/5		\\
$\Xi:\,1/2(3/2^+)$	& 1553(18)		& 3.78/5		\\
$\Xi:\,1/2(3/2^+)$	& 2228(40)		& 6.99/5		\\
$\Xi:\,1/2(3/2^+)$	& 2398(52)		& 7.03/5		\\
$\Xi:\,1/2(3/2^+)$	& 2574(52)		& 4.26/5		\\
$\Omega:\,0(3/2^+)$	& 1642(17)		& 10.86/5		\\
$\Omega:\,0(3/2^+)$	& 2470(49)		& 8.14/5		\\
\hline
\hline
\end{tabular}
\end{center}
\caption{
Energy levels at the physical pion mass and corresponding $\chi^2$/d.o.f.~for the chiral fits of 
the positive baryon energy levels reported in this work.
Sources of large $\chi^2$/d.o.f.~($\geq 3$) are discussed in the text.
Spin 1/2 and spin 3/2 baryons are separated by a line.
Given errors are statistical only.
}
\label{tab:chi2baryons_pospar}
\end{table}

\begin{table}
\begin{center}
\begin{tabular}{lcc}
\hline
\hline
Baryon: $I(J^P)$ 	& Energy level [MeV]	& $\chi^2$/d.o.f. 	\\
\hline
$N:\,1/2(1/2^-)$	& 1406(49)		& 6.51/5		\\
$N:\,1/2(1/2^-)$	& 1539(69)		& 8.72/5		\\
$N:\,1/2(1/2^-)$	& 1895(128)		& 6.35/5		\\
$N:\,1/2(1/2^-)$	& 1918(211)		& 5.94/5		\\
$\Delta:\,3/2(1/2^-)$	& 1454(140)		& 11.16/5		\\
$\Delta:\,3/2(1/2^-)$	& 1914(322)		& 3.24/5		\\
$\Lambda:\,0(1/2^-)$	& 1416(81)		& 1.25/3		\\
$\Lambda:\,0(1/2^-)$	& 1546(110)		& 0.57/3		\\
$\Lambda:\,0(1/2^-)$	& 1713(116)		& 3.49/3		\\
$\Lambda:\,0(1/2^-)$	& 2075(249)		& 13.56/5		\\
$\Sigma:\,1(1/2^-)$	& 1603(38)		& 7.45/5		\\
$\Sigma:\,1(1/2^-)$	& 1718(58)		& 12.78/5		\\
$\Sigma:\,1(1/2^-)$	& 1730(34)		& 10.79/5		\\
$\Sigma:\,1(1/2^-)$	& 2478(104)		& 11.94/5		\\
$\Xi:\,1/2(1/2^-)$	& 1716(43)		& 19.10/5		\\
$\Xi:\,1/2(1/2^-)$	& 1837(28)		& 20.25/5		\\
$\Xi:\,1/2(1/2^-)$	& 1844(43)		& 15.75/5		\\
$\Xi:\,1/2(1/2^-)$	& 2758(78)		& 5.61/5		\\
$\Omega:\,0(1/2^-)$	& 1944(56)		& 20.48/5		\\
$\Omega:\,0(1/2^-)$	& 2716(118)		& 8.58/5		\\
\hline
$N:\,1/2(3/2^-)$	& 1634(44)		& 14.75/5		\\
$N:\,1/2(3/2^-)$	& 1982(128)		& 7.40/5		\\
$N:\,1/2(3/2^-)$	& 2296(129)		& 9.59/5		\\
$\Delta:\,3/2(3/2^-)$	& 1570(67)		& 4.01/5		\\
$\Delta:\,3/2(3/2^-)$	& 2373(140)		& 17.97/5		\\
$\Lambda:\,0(3/2^-)$	& 1751(41)		& 1.42/3		\\
$\Lambda:\,0(3/2^-)$	& 2203(106)		& 3.97/5		\\
$\Lambda:\,0(3/2^-)$	& 2381(87)		& 6.48/5		\\
$\Sigma:\,1(3/2^-)$	& 1861(26)		& 6.33/5		\\
$\Sigma:\,1(3/2^-)$	& 1736(40)		& 2.25/5		\\
$\Sigma:\,1(3/2^-)$	& 2394(74)		& 9.73/5		\\
$\Sigma:\,1(3/2^-)$	& 2297(122)		& 3.90/5		\\
$\Xi:\,1/2(3/2^-)$	& 1906(29)		& 3.12/5		\\
$\Xi:\,1/2(3/2^-)$	& 1894(38)		& 3.19/5		\\
$\Xi:\,1/2(3/2^-)$	& 2497(61)		& 8.53/5		\\
$\Xi:\,1/2(3/2^-)$	& 2426(73)		& 7.60/5		\\
$\Omega:\,0(3/2^-)$	& 2049(32)		& 7.32/5		\\
$\Omega:\,0(3/2^-)$	& 2755(67)		& 5.68/5		\\
\hline
\hline
\end{tabular}
\end{center}
\caption{
Same as Table \ref{tab:chi2baryons_pospar}, but for negative parity baryons.
Spin 1/2 and spin 3/2 baryons are separated by a line.
}
\label{tab:chi2baryons_negpar}
\end{table}

\begin{table}
\begin{center}
\begin{tabular}{lccc}
\hline
\hline
Hadron		& $I(J^P)$ 	& Energy level [MeV]	& $\chi^2$/d.o.f. 	\\
\hline
$N$		&$1/2(1/2^+)$	& 954(16)		& 2.26/5		\\
$\Lambda$	&$0(1/2^+)$	& 1126(17)(+07)		& 2.74/3		\\
$\Sigma$	&$1(1/2^+)$	& 1176(19)(+07)		& 6.67/3		\\
$\Xi$		&$1/2(1/2^+)$	& 1299(16)(+15)		& 5.05/3		\\
\hline
$\Delta$	&$3/2(3/2^+)$	& 1268(32)		& 8.67/5		\\
$\Lambda$	&$0(3/2^+)$	& 1880(116)(+07)		& 2.38/3		\\
$\Sigma$	&$1(3/2^+)$	& 1431(25)(+07)		& 2.29/3		\\
$\Xi$	        &$1/2(3/2^+)$	& 1540(22)(+15)		& 2.05/3		\\
$\Omega$	&$0(3/2^+)$	& 1650(20)(+22)		& 3.10/3		\\
\hline
$\Lambda $	&$0(1/2^-)$	& 1436(84)(+07)		& 1.25/3		\\
$\Lambda $	&$0(1/2^-)$	& 1635(70)(+07)		& 4.93/3		\\
$\Lambda $	&$0(1/2^-)$	& 1664(66)(+07)		& 3.49/3		\\
\hline
$\Lambda $	&$0(3/2^-)$	& 1712(51)(+07)		& 2.92/3		\\
\hline
\hline
\end{tabular}
\end{center}
\caption{
Same as Table \ref{tab:chi2baryons_pospar}, but for hadrons after the infinite volume extrapolation.
The horizontal line separates different parity and spin. 
Notice that the $\Omega$ mass is not a prediction of our calculation.
The second errors given are na\"ive estimates for the systematic error from a mistuning of the strange quark mass.}
\label{tab:chi2_vol}
\end{table}

\end{appendix}

\clearpage

%

\end{document}